\documentclass[sigconf]{acmart}

\usepackage{multirow}
\usepackage{bbding}
\usepackage{subfigure}
\usepackage{algorithm}
\usepackage{algorithmic}
\usepackage[table]{xcolor}
\usepackage{bbold}
\usepackage{makecell}

\AtBeginDocument{%
  }

\setcopyright{acmlicensed}
\copyrightyear{2018}
\acmYear{2018}
\acmDOI{XXXXXXX.XXXXXXX}
\acmConference[Conference acronym 'XX]{Make sure to enter the correct
  conference title from your rights confirmation email}{June 03--05,
  2018}{Woodstock, NY}
\acmISBN{978-1-4503-XXXX-X/2018/06}

\acmSubmissionID{123-A56-BU3}

\begin{document}

\title[Adaptive Semantic-aware Watermarking for Embedding-as-a-Service Copyright Protection]{From Essence to Defense: Adaptive Semantic-aware Watermarking for Embedding-as-a-Service Copyright Protection}


\author{Hao Li}
\affiliation{%
  \institution{Institute of Information Engineering, Chinese Academy of Sciences}
  \institution{School of Cyber Security, University of Chinese Academy of Sciences}
  \state{Beijing}
  \country{China}
}
\email{lihao1998@iie.ac.cn}

\author{Yubing Ren}
\authornote{Corresponding author}
\affiliation{%
  \institution{Institute of Information Engineering, Chinese Academy of Sciences}
  \institution{School of Cyber Security, University of Chinese Academy of Sciences}
  \state{Beijing}
  \country{China}
}
\email{renyubing@iie.ac.cn}

\author{Yanan Cao}
\affiliation{%
  \institution{Institute of Information Engineering, Chinese Academy of Sciences}
  \institution{School of Cyber Security, University of Chinese Academy of Sciences}
  \state{Beijing}
  \country{China}
}

\author{Yingjie Li}
\affiliation{%
  \institution{Institute of Information Engineering, Chinese Academy of Sciences}
  \institution{School of Cyber Security, University of Chinese Academy of Sciences}
  \state{Beijing}
  \country{China}
}

\author{Fang Fang}
\affiliation{%
  \institution{Institute of Information Engineering, Chinese Academy of Sciences}
  \institution{School of Cyber Security, University of Chinese Academy of Sciences}
  \state{Beijing}
  \country{China}
}

\author{Xuebin Wang}
\affiliation{%
  \institution{Institute of Information Engineering, Chinese Academy of Sciences}
  \institution{School of Cyber Security, University of Chinese Academy of Sciences}
  \state{Beijing}
  \country{China}
}

\renewcommand{\shortauthors}{Hao Li et al.}

\begin{abstract}
Benefiting from the superior capabilities of large language models in natural language understanding and generation, Embeddings-as-a-Service (EaaS) has emerged as a successful commercial paradigm on the web platform. However, prior studies have revealed that EaaS is vulnerable to imitation attacks. Existing methods protect the intellectual property of EaaS through watermarking techniques, but they all ignore the most important properties of embedding: semantics, resulting in limited harmlessness and stealthiness. To this end, we propose SemMark, a novel semantic-based watermarking paradigm for EaaS copyright protection. SemMark employs locality-sensitive hashing to partition the semantic space and inject semantic-aware watermarks into specific regions, ensuring that the watermark signals remain imperceptible and diverse. In addition, we introduce the adaptive watermark weight mechanism based on the local outlier factor to preserve the original embedding distribution. Furthermore, we propose Detect-Sampling and Dimensionality-Reduction attacks and construct four scenarios to evaluate the watermarking method. Extensive experiments are conducted on four popular NLP datasets, and SemMark achieves superior verifiability, diversity, stealthiness, and harmlessness.
\end{abstract}


\keywords{Embedding-as-a-Service Watermark, Semantic-aware Watermarking, Copyright Protection
}

\received{20 February 2007}
\received[revised]{12 March 2009}
\received[accepted]{5 June 2009}

\maketitle

\section{Introduction}
Large Language Models (LLMs) currently demonstrate remarkable capabilities in natural language understanding and generation. With the rapid growth of web platforms and commercial cloud services, an increasing number of LLM owners provide Embeddings-as-a-Service (EaaS), such as OpenAI \cite{openai2024embedding}, Google \cite{google2024grounding}, and Mistral AI \cite{mistral2024embedding}. EaaS provides APIs to generate high-quality embeddings to help users extract features and complete various natural language processing (NLP) tasks, such as text generation \cite{ram-etal-2023-context, shi-etal-2024-replug}, information retrieval \cite{10.1145/3394486.3403305, kamalloo-etal-2023-evaluating-embedding}, and information extraction \cite{wan-etal-2023-gpt, li2025bridging}. However, \citet{liu2022stolenencoder} demonstrates that EaaS is vulnerable to imitation attacks. Attackers train their models by querying the victim model at a much lower cost and computing resources than the victim. This seriously infringes the intellectual property of service providers and hinders the development of the EaaS community.

\begin{figure}[t]
\centering
\includegraphics[width=1.0\linewidth]{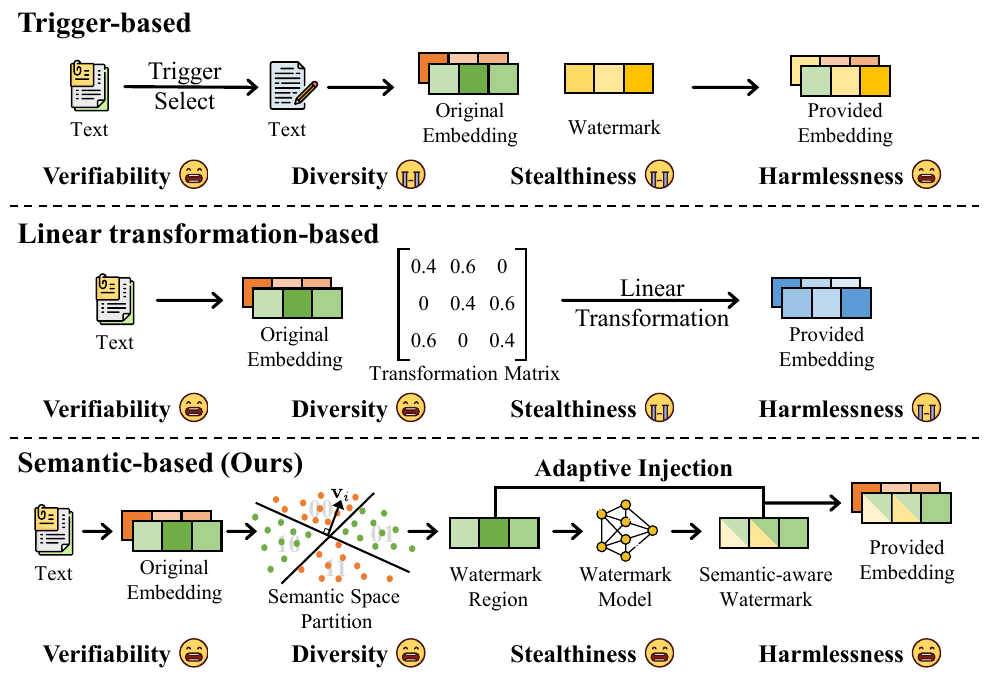}
\caption{Paradigm comparison between our semantic-based watermark SemMark and existing trigger-based/linear transformation-based watermarks.}
\label{fig: example}
\centering
\end{figure}

To defend against imitation attacks, recent studies have introduced watermarking techniques for EaaS, which can be broadly categorized as trigger-based and transformation-based. Trigger-based methods \cite{liu2022stolenencoder, peng-etal-2023-copying, shetty-etal-2024-warden, wang-etal-2025-robust} embed watermark signals by associating them with specific trigger words. During verification, these triggers are queried to expose ownership. While conceptually simple, such methods depend on fixed trigger sets and handcrafted queries, which are easy for adversaries to detect and filter, and their limited diversity further reduces robustness. In contrast, transformation-based methods such as WET \cite{shetty-etal-2025-wet} avoid explicit triggers by applying a linear transformation to the entire embedding space. However, this approach introduces noticeable distortion and makes watermarking heavily reliant on the chosen transformation matrix, at the cost of stealthiness. Despite their differences, both categories share the same limitation: they treat watermarks as add-ons, disconnected from the semantic properties of embeddings.

Semantics are a crucial property of embeddings, as they encode high-dimensional representations that capture meaning and underpin the effectiveness of downstream NLP tasks \cite{DBLP:journals/corr/abs-1301-3781, chen2013expressive, conneau-etal-2018-cram}. Existing watermarking methods overlook these semantic properties, either inject signals that are unrelated to meaning or impose rigid global transformations. This disconnect degrades embedding quality and makes watermark patterns more conspicuous, reducing their reliability in practice. These observations suggest that watermarking for EaaS should be designed with greater sensitivity to the semantic characteristics of embeddings, ensuring that protection mechanisms remain both effective and unobtrusive.

Building on this intuition, we argue that watermarking for EaaS should be guided by four key desiderata: 1) \textbf{Verifiability}. The watermark can be reliably detected, allowing providers to distinguish benign models from stolen ones. 
2) \textbf{Diversity}. Watermark signals should exhibit sufficient variability across embeddings, making them resistant to removal attacks.
3) \textbf{Stealthiness}. The watermark should remain imperceptible during both insertion and verification, preventing adversaries from identifying it.
4) \textbf{Harmlessness}. The watermark should minimally affect the original embeddings, preserving semantic integrity and downstream task performance.

To this end, we introduce a novel semantic-based paradigm for EaaS copyright protection, which injects \textbf{Sem}antic-aware water\textbf{Mark}s into specific regions of the embedding space (\textbf{SemMark}). Specifically, the embedding space is partitioned using locality-sensitive hashing to define watermark regions. Semantic-aware watermarks are then injected into these regions via the watermark mapping model, ensuring that similar embeddings receive similar watermark signals while satisfying both verifiability and diversity. Furthermore, considering the original embedding distribution, we introduce the adaptive watermark weight mechanism based on the local outlier factor to assign watermark weights. During watermark verification, only normal texts across different regions are compared, making it difficult for adversaries to detect the watermark. To evaluate the stealthiness and robustness against dimensionality changes, we propose the Detect-Sampling and Dimensionality-Reduction attack, and conduct extensive experiments on four widely used NLP datasets: SST2, MIND, AGNews, and Enron Spam. Experimental results demonstrate that SemMark is both efficient and robust across multiple attack scenarios.

The contributions are summarized as follows:
\begin{itemize}
    \item We conduct an in-depth analysis of the existing watermarking methods for EaaS and introduce a novel watermarking paradigm from the semantic perspective.
    \item We propose SemMark that injects semantic-aware watermarks with adaptive weights, achieving superior verifiability, diversity, stealthiness, and harmlessness.
    \item We introduce two new watermark attacks to evaluate the watermarking method. Extensive experiments demonstrate that SemMark achieves excellent efficiency and robustness.\footnote{Code and data are available at \url{https://github.com/hlee-top/SemMark}.}
\end{itemize}

\section{Related Work}
\subsection{Imitation Attacks}
Imitation Attacks are also known as ``model stealing'' or ``model extraction'' \cite{10.5555/3241094.3241142, orekondy2019knockoff, DBLP:conf/iclr/KrishnaTPPI20, wallace-etal-2020-imitation}, which replicate or steal the model's functionality by querying the victim model without the authorization of service providers. Imitation attacks are currently widespread in multiple fields and tasks, including natural language processing tasks \cite{Krishna2020Thieves, wallace-etal-2020-imitation, xu-etal-2022-student} and vision tasks \cite{orekondy2019knockoff, sanyal2022towards, beetham2023dual}. It does not access internal parameters, architecture, and training data, and attackers can easily imitate deployed models in reduced time with marginal computational resources. More critically, imitation models using advanced techniques can outperform the victim model in specific scenarios \cite{xu-etal-2022-student, 10.1609/aaai.v39i7.32734}.  For instance, \citet{xu-etal-2022-student} utilize the unsupervised domain adaptation and multi-victim integration technique to train the imitation models, enabling them to outperform the victim model in the sentiment analysis and machine translation fields.
Similarly, \citet{10.1609/aaai.v39i7.32734} propose the multimodal medical imitation attack, leveraging adversarial attacks for domain alignment and oracle models for report enrichment to generate diverse training data. Their imitation model achieves superior performance in radiology report generation compared to the victim model. In summary, imitation attacks allow adversaries to provide comparable or even superior services at a lower cost \cite{10.1145/3711896.3736573}, severely infringing on the intellectual property of the victim and enabling the illegal resale of services.

\begin{figure*}[t]
    \centerline{\includegraphics[width=1.0\linewidth]{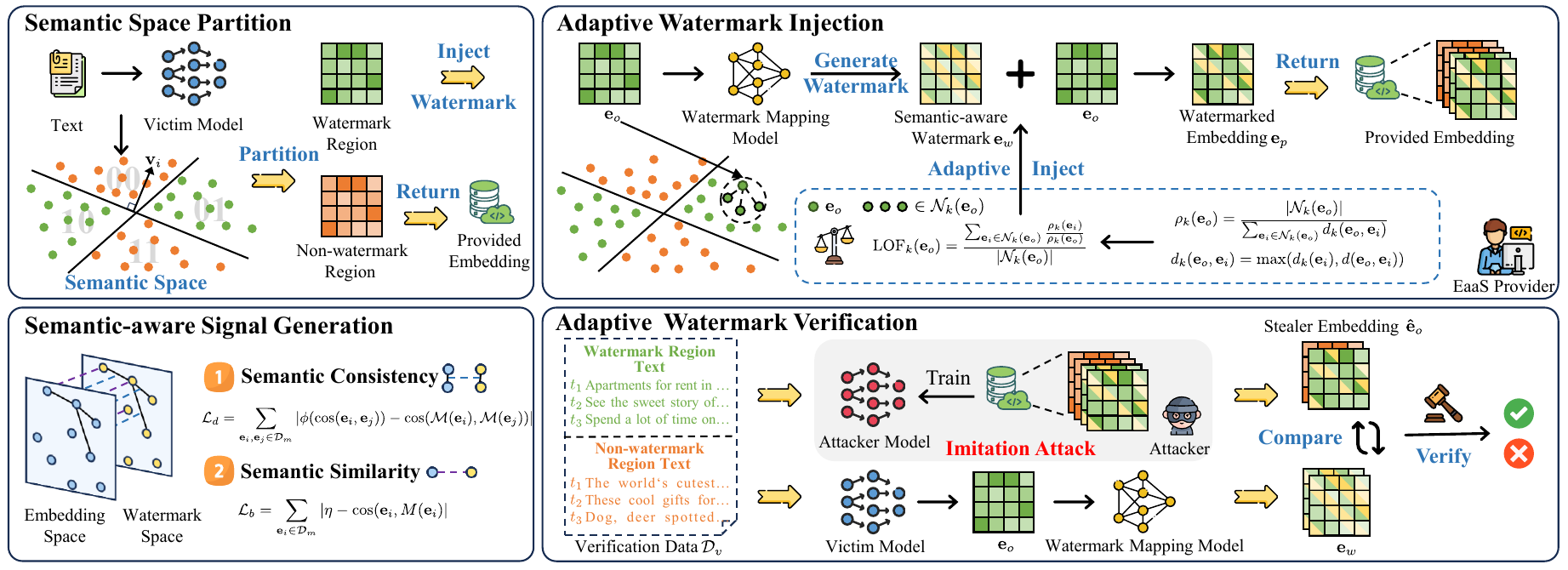}}
    \caption{Overall framework of our watermarking method SemMark. We use green and orange to distinguish the watermark region and non-watermark region embedding, and use gold to represent the inserted watermark signal.}
    \label{fig: model framework}
\end{figure*}

\subsection{EaaS Watermarks}
LLMs have demonstrated remarkable ability to generate high-quality, context-based embeddings, and many institutions currently provide Embeddings as a Service (EaaS), such as OpenAI \cite{openai2024embedding}, Google \cite{google2024grounding}, and Mistral AI \cite{mistral2024embedding}. Users can easily obtain high-quality embeddings via the service provider's API without expensive computational resources. However, such convenience also renders EaaS vulnerable to imitation attacks. As an efficient approach for copyright protection, preliminary studies have proposed leveraging watermarking techniques for EaaS.
These methods can be broadly categorized into trigger-based and linear transformation-based.
EmbMarker \cite{peng-etal-2023-copying} is the first trigger-based watermarking method to protect EaaS against imitation attacks. It uses moderate-frequency words as triggers and inserts watermark embeddings into the text embeddings containing the trigger words. 
\citet{shetty-etal-2024-warden} further propose the CSE (Clustering, Selection, Elimination) attack that can remove the backdoor watermark while preserving the quality of embeddings. To defend against CSE attack, WARDEN \cite{shetty-etal-2024-warden} strengthens EmbMarker by utilizing multiple trigger sets and watermark embeddings, thereby injecting multiple watermarks that are more resilient to CSE attack.
Despite these improvements, methods like EmbMarker and WARDEN insert the same watermark components into each embedding via linear interpolation, making them relatively easy to detect and eliminate. EspeW \cite{wang-etal-2025-robust} addresses this limitation by inserting watermarks only in a small subset of the original embedding dimensions, reducing shared components across embeddings.
In contrast, transformation-based methods, such as WET \cite{shetty-etal-2025-wet}, construct a transformation matrix to apply linear transformations across all embeddings during watermark injection. During verification, the suspected embedding is inversely transformed and compared with the original embedding to detect the watermark.

\section{Methodology}
In this section, we first introduce the preliminaries of EaaS copyright protection (\S\ref{sec: preliminary}). Next, the semantic space is partitioned into watermark and non-watermark regions (\S\ref{sec: partition}), and the watermark mapping model is trained to generate semantic-aware watermark signals (\S\ref{sec: generation}). These signals are then injected into the watermark region with adaptive weights (\S\ref{sec: injection}). Finally, watermark verification is conducted by comparing the embeddings across watermark and non-watermark regions (\S\ref{sec: verification}). The overall framework of SemMark is illustrated in Figure \ref{fig: model framework}, and the watermark injection and verification processes are detailed in Algorithms \ref{algo: inject} and \ref{algo: verif}, respectively.

\subsection{Preliminary}
\label{sec: preliminary}
\paragraph{Problem Definition}
EaaS provider $S_v$ provides the embedding service based on model $\Theta_v$ (called the victim model in imitation attack). When users query $S_v$ with sentence $t$, model $\Theta_v$ generates the corresponding original embedding $\mathbf{e}_o = \Theta_v(t)$. Due to the threat of imitation attacks, the copyright protection mechanism $\mathbf{e}_p = f(\mathbf{e}_o, \mathbf{e}_w)$ is applied. This mechanism injects the watermark signal $\mathbf{e}_w$ into the original embedding $\mathbf{e}_o$, and obtains the watermarked embedding $\mathbf{e}_p$, which is finally returned to the user.

\paragraph{Attacker}
The attacker's goal is to steal the model and provide a similar service $S_a$ at a much lower cost than training the model from scratch. The attacker queries the service $S_v$ to collect embeddings for their text dataset to train an attacker model $\Theta_a$, but is unaware of the details for the service $S_v$, including its model structure, algorithm details, and training data. In addition, the attacker can actively employ several strategies to evade EaaS copyright verification. 

\paragraph{Defender}
The defender possesses full knowledge about the service $S_v$ and has the capability to manipulate the original embeddings prior to returning them to users. Additionally, the defender maintains a verification dataset, which is used to query the potentially suspicious service $S_a$. By analyzing the returned embedding, the defender can verify whether $S_a$ originates from its own service $S_v$.

\subsection{Semantic Space Partition}
\label{sec: partition}
Semantics, as the expressive capacity in the high-dimensional vector space, naturally divides the embeddings representing different texts \cite{chen2013expressive, liu2024a}. Different from the existing methods that rely on backdoor trigger words to determine where to insert watermarks, we exploit this natural property to partition specific regions for watermark injection, thereby rendering the watermark algorithm imperceptible to potential attackers. Specifically, we utilize locality-sensitive hashing (LSH) \cite{10.1145/276698.276876, hou2023semstamp} to partition the semantic embedding space into watermark and non-watermark regions, where similar inputs are hashed to similar representations. Given the dimension $c$ of LSH, we apply the cosine-preserving method \cite{10.1145/509907.509965} to generate $c$ random normal vectors $\mathbf{v} = \{\mathbf{v}_1, ..., \mathbf{v}_c\}$ from the $h'$-dimensional Gaussian distribution.\footnote{Note that we reduce the dimensionality of the original embedding $\mathbf{e}_o$ to $h'$ dimensions by Principal Component Analysis (PCA) before hashing due to curse of dimensionality, as embedding distances tend to concentrate (e.g., The embedding dimension for OpenAI’s text-embedding-002 API is 1536).} These normal vectors serve as hyperplanes that partition the semantic space as follows:
\begin{align}
\label{eq: lsh}
    \operatorname{LSH}(\mathbf{e}_o) &= [\operatorname{LSH}_1(\mathbf{e}_o), ..., \operatorname{LSH}_c(\mathbf{e}_o)], \\
    \operatorname{LSH}_i(\mathbf{e}_o) &= \mathbb{1}(\mathbf{e}_o\cdot \mathbf{v}_i> 0).
\end{align}
where $\mathbb{1}$ represents the indicator function, and $\operatorname{LSH}(\mathbf{e}_o) \mapsto \{0,1\}^c$ represents $c$ binary representations to partition the semantic space into $2^c$ regions. Given the watermark proportion $\alpha$, $\alpha*2^c$ regions are randomly sampled as watermark regions $\mathcal{R}_w$, while the remaining regions are designated as non-watermark regions.

\begin{algorithm}[t]
\caption{Watermark Injection Process}
\begin{algorithmic}[1]
    \STATE \textbf{Input:} Text $t$, EaaS provider model $\Theta_v$, watermark mapping model $\mathcal{M}$, hash vector $\mathbf{v}$,  watermark regions $\mathcal{R}_w$, surrogate dataset $\mathcal{D}_s$
    \STATE \textbf{Output:} Returned embedding
    \STATE Encode text $t$ to generate original embedding $\mathbf{e}_o \leftarrow \Theta_v(t)$
    \STATE // Mapping embeddings to semantic region
    \STATE Reduce dimensionality $\mathbf{e}_o' \leftarrow \operatorname{PCA}(\mathbf{e}_o)$
    \STATE Determine the semantic region $r_o \leftarrow \mathbf{e}_o' \cdot \mathbf{v}$ \hfill Equation \ref{eq: lsh}
    \STATE // Original embedding in watermark region
    \IF{$r_o \in \mathcal{R}_w$}
        \STATE Generate watermark embedding $\mathbf{e}_w \leftarrow \mathcal{M}(\mathbf{e}_o)$
        \STATE Calculate local reachability density $\rho_k(\mathbf{e}_o)$ \hfill Equation \ref{eq: lrd}
        \STATE Calculate local outlier factor $L_o \leftarrow \operatorname{LOF}_k(\mathbf{e}_o)$ \hfill Equation \ref{eq: lof}
        \STATE Generate weight $\mathbf{u}_o \leftarrow \operatorname{Scale}(L_o, \mathcal{D}_s)$ \hfill Equation \ref{eq: scale}
        \STATE Inject watermark $\mathbf{e}_p \leftarrow \operatorname{Inject}(\mathbf{e}_o, \mathbf{e}_w , \mathbf{u}_o)$  \hfill Equation \ref{eq: inject}
        \STATE \textbf{Return:} $\mathbf{e}_p$
    \ELSE
        \STATE \textbf{Return:} $\mathbf{e}_o$
    \ENDIF
    
\end{algorithmic}
\label{algo: inject}
\end{algorithm}

\subsection{Semantic-aware Signal Generation}
\label{sec: generation}
Watermark signal generation is a critical component of watermarking methods, requiring the creation of diverse and verifiable watermarks with negligible impact on the original embedding. Previous methods \cite{peng-etal-2023-copying, shetty-etal-2024-warden, wang-etal-2025-robust, shetty-etal-2025-wet}  either arbitrarily select several fixed watermark signals or apply linear transformations to all embeddings, while disregarding their semantic information. Moreover, the watermarked embeddings generated by these approaches tend to share a common direction, resulting in limited diversity and heightened vulnerability to perturbations, which makes them easier to identify and eliminate. In contrast, we utilize a trainable neural network to generate the diverse watermark signal based on the semantics of embedding, thereby enhancing watermark diversity while maintaining essential semantic information. Complex high-dimensional semantic space and neural networks further increase the complexity of the watermark, making it substantially more difficult for adversaries to detect and reverse-engineer.

Specifically, we train the watermark mapping model $\mathcal{M}$ to transform the original embedding $\mathbf{e}_o$ to the watermark signal $\mathbf{e}_w = \mathcal{M}(\mathbf{e}_o)$, which is a multi-layer feed-forward neural network with residual connections. It ensures that the generated watermark signal satisfies the following properties:

(1) \textbf{Semantic Consistency}: Embeddings that are close in the semantic space exhibit highly correlated watermark signals. This property ensures consistency between the embeddings and their corresponding watermarks, enabling the watermark signals to tolerate slight perturbations and enhancing robustness. To this end, we formulate and minimize the consistency loss:
\begin{align}
    \mathcal{L}_d = \sum_{\mathbf{e}_i,\mathbf{e}_j \in \mathcal{D}_m} |\phi(\cos(\mathbf{e}_i,\mathbf{e}_j)) - \cos(\mathcal{M}(\mathbf{e}_i),\mathcal{M}(\mathbf{e}_j))|,
\end{align}
where $\mathcal{D}_m$ means the watermark mapping model training dataset, and $\cos$ represents the cosine similarity operation.
$\phi(x) = x + \tau(x-\bar{x}) $ is the scaling function based on the mean cosine similarity of the original embeddings, which makes similar embeddings generate more relevant watermark signals, and vice versa.

\begin{algorithm}[t]
\caption{Watermark Verification Process}
\begin{algorithmic}[1]
    \STATE \textbf{Input:} Watermark verification dataset $\mathcal{D}_v = \mathcal{D}_w \cup \mathcal{D}_n$, suspicious model $\Theta_a$, EaaS provider model $\Theta_v$, watermark mapping model $\mathcal{M}$
    \STATE \textbf{Output:} Evaluation metric p-value, $\Delta_{cos}$, $\Delta_{L_2}$
    
    \STATE Initialize list $Cos_w$, $Cos_n$, ${L_2}_w$, ${L_2}_n$
    \FOR{$t_i$ in $\mathcal{D}_v$}
    \STATE Encoded text $t_i$  by suspicious model $\mathbf{\hat{e}}_o^i \leftarrow \Theta_a(t_i)$
    \STATE Encoded text $t_i$  by EaaS provider model $\mathbf{e}_o^i \leftarrow \Theta_v(t_i)$
    \STATE Generate watermark embedding $\mathbf{e}_w^i \leftarrow \mathcal{M}(\mathbf{e}_o^i)$
    \STATE Calculate cosine similarity $\cos(\mathbf{\hat{e}}_o^i, \mathbf{e}_w^i)$ \hfill Equation \ref{eq: cos}
    \STATE Calculate $L_2$ distance $\operatorname{L_2}(\mathbf{\hat{e}}_o^i, \mathbf{e}_w^i)$ \hfill Equation \ref{eq: l2}
    \STATE // $t_i$ in watermark  region text set $\mathcal{D}_w$
    \IF{$t_i \in \mathcal{D}_w$}
        \STATE $Cos_w$.append($\cos(\mathbf{\hat{e}}_o^i, \mathbf{e}_w^i)$), ${L_2}_w$.append($\operatorname{L_2}(\mathbf{\hat{e}}_o^i, \mathbf{e}_w^i)$)
    \ELSE
        \STATE $Cos_n$.append($\cos(\mathbf{\hat{e}}_o^i, \mathbf{e}_w^i)$), ${L_2}_n$.append($\operatorname{L_2}(\mathbf{\hat{e}}_o^i, \mathbf{e}_w^i)$)
    \ENDIF
    \ENDFOR
    \STATE Kolmogorov-Smirnov test p-value $\leftarrow \operatorname{KS-test}(Cos_w, Cos_n)$
    \STATE Calculate $\Delta_{cos} \leftarrow \operatorname{mean}(Cos_w) - \operatorname{mean}(Cos_n)$ \hfill Equation \ref{eq: delta cos}
    \STATE Calculate $\Delta_{L_2} \leftarrow \operatorname{mean}({L_2}_w) - \operatorname{mean}({L_2}_n)$ \hfill Equation \ref{eq: delta l2}
    \STATE \textbf{Return:}  p-value, $\Delta_{cos}$, $\Delta_{L_2}$
    
\end{algorithmic}
\label{algo: verif}
\end{algorithm}

(2) \textbf{Semantic Similarity}: The watermark signal preserves relative similarity with the original embedding, ensuring semantic similarity while maintaining a non-trivial margin. This balance enables the watermark to remain semantic-aware and reliably verifiable. To this end, we formulate and minimize the similarity loss:
\begin{align}
    \mathcal{L}_b = \sum_{\mathbf{e}_i \in \mathcal{D}_m}|\eta  -\cos(\mathbf{e}_i,\mathcal{M}(\mathbf{e}_i))|,
\end{align}
where $\eta $ is a hyperparameter that constrains the similarity between the embedding and the watermark signal.

Considering all the properties, the loss function of the watermark mapping model is:
\begin{align}
    \mathcal{L} = \mathcal{L}_d + \lambda\mathcal{L}_b.
\end{align}

\subsection{Adaptive Watermark Injection}
\label{sec: injection}
Given the watermark signal $\mathbf{e}_w$, we inject it into the original embedding $\mathbf{e}_o$ to generate the watermarked embedding:
\begin{align}
\label{eq: inject}
    \mathbf{e}_p = \operatorname{Norm}((1-\mathbf{u}_o)*\mathbf{e}_o + \mathbf{u}_o * \mathbf{e}_w),
\end{align}
where $\operatorname{Norm}$ represents the $L_2$ normalization operation, and $\mathbf{u}_o$ represents the watermark weight. A common practice is to employ fixed weight parameters for watermark injection. However,  this strategy suffers from limited adaptability, particularly in watermarking schemes with low diversity. Considering the original embedding distribution, we aim to insert larger watermark weights for outliers instead of using fixed weight parameters to inject watermarks, which can better hide the watermark signal and maintain the overall quality of embeddings. To this end, we propose the adaptive watermark weight based on the local outlier factor (LOF) \cite{10.1145/342009.335388}. LOF is the density-based anomaly detection method used to identify data with significantly lower density than its neighbors.

Specifically, we construct the surrogate dataset $\mathcal{D}_s$ to estimate the local outlier factor for each embedding. We first compute the local reachability density, which quantifies the degree of density in the local neighborhood:
\begin{align}
\label{eq: lrd}
     \rho_k(\mathbf{e}_o) = \frac{|\mathcal{N}_k(\mathbf{e}_o)|}{\sum_{\mathbf{e}_i\in{\mathcal{N}_k(\mathbf{e}_o)}}d_k(\mathbf{e}_o,\mathbf{e}_i)},
\end{align}
where $\mathcal{N}_k(\mathbf{e}_o) = \{\mathbf{e}_i \in \mathcal{D}_s| \mathbf{e}_i \ne \mathbf{e}_o, d(\mathbf{e}_o,\mathbf{e}_i)\leq d_k(\mathbf{e}_o)\} $ represents the k-distance neighborhood, $d_k(\mathbf{e}_o)$ is the k-th distance of $\mathbf{e}_o$, and $d$ is the distance metric, e.g., Euclidean distance. $d_k(\mathbf{e}_o,\mathbf{e}_i) = \max(d_k(\mathbf{e}_i), d(\mathbf{e}_o,\mathbf{e}_i))$ represents the k-th reachable distance from $\mathbf{e}_o$ to $\mathbf{e}_i$. Then we compute the local outlier factor, which represents the relative ratio of local density and measures the density difference between the data and its neighbors:
\begin{align}
\label{eq: lof}
     L_o  = \operatorname{LOF}_k(\mathbf{e}_o) = \frac{\sum_{\mathbf{e}_i\in{\mathcal{N}_k(\mathbf{e}_o)}}\frac{\rho_k(\mathbf{e}_i)}{\rho_k(\mathbf{e}_o)}}{|\mathcal{N}_k(\mathbf{e}_o)|}.
\end{align}

Finally, we obtain the adaptive watermark weight based on the local outlier factor:
\begin{align}
\label{eq: scale}
    \mathbf{u}_o =
    \begin{cases}
        \delta, & L_o > L_{\max} \\
        \delta - \epsilon, & L_o < L_{\min} \\
        \delta - \epsilon + \frac{L_o - L_{\min}}{L_{\max} - L_{\min}}\epsilon, & L_{\min} \leq L_o \leq L_{\max}
    \end{cases}, \\
L_{\max}, L_{\min} = \max_{\mathbf{e}_i \in \mathcal{D}_s} \operatorname{LOF}_k(\mathbf{e}_i),\;
                    \min_{\mathbf{e}_i \in \mathcal{D}_s} \operatorname{LOF}_k(\mathbf{e}_i),
\end{align}
where $\delta$ and $\epsilon$ represent the maximum watermark strength and weight margin, respectively.

\subsection{Adaptive Watermark Verification}
\label{sec: verification}
To verify the copyright of EaaS, we construct the verification dataset $\mathcal{D}_v = \mathcal{D}_w \cup \mathcal{D}_n$ that includes the watermark region text set $\mathcal{D}_w$ and the non-watermark region text set $\mathcal{D}_n$:
\begin{align}
    \mathcal{D}_w &= \{[t_1,t_2,...,t_m]|\Theta_v(t_i) \in \mathcal{R}_w\}, \\
    \mathcal{D}_n &= \{[t_1,t_2,...,t_m]|\Theta_v(t_i) \notin \mathcal{R}_w\}.
\end{align}

By leveraging the semantic-based design, both text sets $\mathcal{D}_w$ and $\mathcal{D}_n$ consist solely of normal texts, obviating the need for crafted backdoor trigger words. Consequently, the watermark verification process closely mirrors standard API usage, thereby preserving the stealthiness of the watermarking scheme.

The model trained with watermark embeddings will inherit the watermark features and exhibit stronger alignment with the corresponding target watermark signal in watermarked regions than in non-watermarked regions. This characteristic is then leveraged to verify copyright. Specifically, we use cosine similarity and $L_2$ distance to measure the closeness:
\begin{align}
    \cos(\mathbf{e}_o, \mathbf{e}_w) &= \frac{\mathbf{e}_o \cdot \mathbf{e}_w}{||\mathbf{e}_o||\cdot||\mathbf{e}_w||}, \label{eq: cos}\\
    \operatorname{L_2}(\mathbf{e}_o, \mathbf{e}_w) &= ||\frac{\mathbf{e}_o}{||\mathbf{e}_o||}- \frac{\mathbf{e}_w}{||\mathbf{e}_w||}||^2. \label{eq: l2}
\end{align}

Then we use the cosine similarity and the square of $L_2$ distance as metrics to verify the copyright of EaaS:
\begin{align}
    \Delta_{cos} &= \frac{1}{|\mathcal{D}_w|}\sum_{\mathbf{\hat{e}}_o^i \in \mathcal{D}_w}\cos(\mathbf{\hat{e}}_o^i, \mathbf{e}_w^i) - \frac{1}{|\mathcal{D}_n|}\sum_{\mathbf{\hat{e}}_o^j \in \mathcal{D}_n}\cos(\mathbf{\hat{e}}_o^j, \mathbf{e}_w^j), \label{eq: delta cos}\\
    \Delta_{L_2} &= \frac{1}{|\mathcal{D}_w|} \sum_{\mathbf{\hat{e}}_o^i \in \mathcal{D}_w}\operatorname{L_2}(\mathbf{\hat{e}}_o^i, \mathbf{e}_w^i) - \frac{1}{|\mathcal{D}_n|} \sum_{\mathbf{\hat{e}}_o^j \in \mathcal{D}_n}\operatorname{L_2}(\mathbf{\hat{e}}_o^j, \mathbf{e}_w^j), \label{eq: delta l2}
\end{align}
where $\mathbf{\hat{e}}_o^i$ denotes the output embedding produced by the suspicious model $\Theta_a$, and $\mathbf{e}_w^i$ represents the watermark signal corresponding to the original embedding $\mathbf{e}_o^i$. Furthermore, we employ the Kolmogorov-Smirnov (KS) test \cite{berger2014kolmogorov} for hypothesis testing and compute the p-value as the third evaluation metric. The null hypothesis is defined as: The distribution of cosine similarity with the corresponding watermark signal in $\mathcal{D}_w$ and $\mathcal{D}_n$ is consistent. A lower p-value indicates stronger evidence against the null hypothesis.

\begin{figure}[t]
\centering
\includegraphics[width=1.0\linewidth]{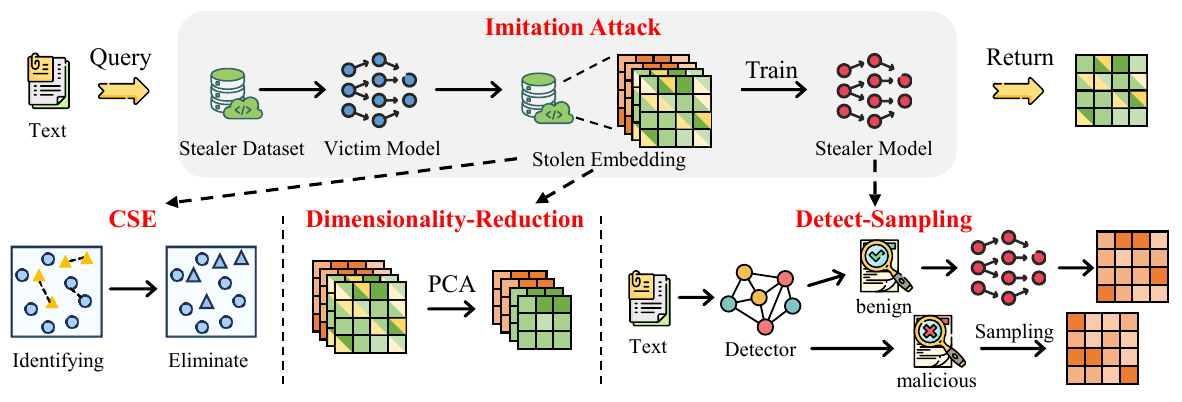}
\caption{Watermark attack details in the embedding stealing stage and watermark verification stage.}
\label{fig: wm attack}
\centering
\end{figure}

\section{Experiments}
\subsection{Experimental Settings}
\subsubsection{Datasets}  
We conduct experiments on four widely used natural language processing datasets: SST2 \cite{socher-etal-2013-recursive}, MIND \cite{wu-etal-2020-mind}, AGNews \cite{10.5555/2969239.2969312}, and Enron Spam \cite{metsis2006spam}. SST2 and Enron Spam cover 2 categories, AGNews covers 4 categories, and MIND covers 17 categories. Detailed data analysis and processing are listed in Appendix \ref{ap: dataset}.

\begin{table*}[ht]
\caption{Experimental results on four NLP datasets, where ``Verif'', ``Detect'', and  ``Dim'' are the abbreviations of Verifiability, Detect-Sampling, and Dimensionality-Reduction. In ``Verif'' column, successful verification (p-value < 0.05) is indicated by \Checkmark.}

\label{tab: main result}
\centering
\resizebox{\linewidth}{!}{
\begin{tabular}{llcccccccccc}
\toprule
\multirow{2}{*}{\textbf{Method}} & \multirow{2}{*}{\textbf{Attack}} & \multicolumn{2}{c}{\textbf{SST2}} & \multicolumn{2}{c}{\textbf{MIND}} & \multicolumn{2}{c}{\textbf{AGNews}} & \multicolumn{2}{c}{\textbf{Enron Spam}} & \multirow{2}{*}{\textbf{Verif}} \\
 \cmidrule(lr){3-4}  \cmidrule(lr){5-6}     \cmidrule(lr){7-8}  \cmidrule(lr){9-10}
& & \textbf{ACC} $\uparrow$ &  \textbf{p-value} $\downarrow$ & \textbf{ACC} $\uparrow$ &  \textbf{p-value} $\downarrow$ & \textbf{ACC} $\uparrow$ &  \textbf{p-value} $\downarrow$ & \textbf{ACC} $\uparrow$ &  \textbf{p-value} $\downarrow$ & & \\
\midrule

\textbf{Original} & \multicolumn{1}{c}{-} & $93.39 \pm 0.35$ & $>0.71$ & $77.21 \pm 0.05$ & $>0.46$ & $93.73 \pm 0.05$ & $>0.46$ & $94.92 \pm 0.05$ & $>0.17$ & \multicolumn{1}{c}{-} & \\
\midrule

\multirow{4}{*}{\textbf{EmbMarker}} & \textbf{No Attack} & $93.35 \pm 0.09$ & $<10^{-6}$ &  $ 77.26 \pm 0.06$  & $<10^{-5}$ & $93.72 \pm 0.17$ & $<10^{-10}$ & $94.63 \pm 0.19$ & $<10^{-6}$ & \Checkmark \\
& \textbf{CSE} &  $90.48 \pm 0.99$ & $>0.03$ &  $75.48 \pm 0.03$ & $>0.03$ & $92.80 \pm 0.16$ & $>0.34$ & $95.33 \pm 0.30$ & $<10^{-3}$ & \XSolidBrush \\
& \textbf{Detect} &  $93.00 \pm 0.19$ & $>0.34$ &  $77.22 \pm 0.05$ & $>0.83$ & $93.44 \pm 0.08$ & $>0.08$ & $94.77 \pm 0.27$ & $>0.34$ & \XSolidBrush \\
& \textbf{Dim} &  $93.12 \pm 0.16$  & $<10^{-9}$ & $77.24 \pm 0.09$ & $<10^{-9}$ & $93.96 \pm 0.13$ & $<10^{-10}$ & $95.85 \pm 0.08$ & $<10^{-9}$ & \Checkmark \\
\midrule

\multirow{4}{*}{\textbf{WARDEN}} & \textbf{No Attack} & $93.20 \pm 0.19$ & $<10^{-7}$ &  $77.21 \pm 0.05$  & $<10^{-10}$ & $93.55 \pm 0.11$ & $<10^{-10}$ & $94.70 \pm 0.35$ & $<10^{-7}$ & \Checkmark \\
& \textbf{CSE} &  $88.34 \pm 0.27$ & $<10^{-4}$ & $75.28 \pm 0.13$ & $<10^{-4}$ & $92.82 \pm 0.10$ & $<10^{-2}$ & $95.25 \pm 0.43$ & $<10^{-2}$ & \Checkmark  \\
& \textbf{Detect} &  $93.50 \pm 0.47$ & $>0.17$ &  $77.20 \pm 0.11$ & $>0.17$ & $93.60 \pm 0.02$ & $>0.83$ & $94.38 \pm 0.21$ & $>0.83$ & \XSolidBrush \\
& \textbf{Dim} &  $92.77 \pm 0.16$ & $<10^{-9}$ &  $77.33 \pm 0.10$ & $<10^{-10}$ & $93.86 \pm 0.11$ & $<10^{-10}$ & $95.88 \pm 0.12$ & $<10^{-9}$ & \Checkmark \\
\midrule

\multirow{4}{*}{\textbf{EspeW}} & \textbf{No Attack} & $93.54 \pm 0.38$ & $<10^{-9}$ &  $77.17 \pm 0.14$  & $<10^{-6}$ & $93.59 \pm 0.07$ & $<10^{-10}$ & $94.73 \pm 0.27$ & $<10^{-9}$ & \Checkmark \\
& \textbf{CSE} &  $86.81 \pm 0.43$  & $>0.34$ &  $75.48 \pm 0.08$ & $>0.03$ & $92.76 \pm 0.06$ & $>0.34$ & $95.27 \pm 0.06$ & $<10^{-2}$ & \XSolidBrush \\
& \textbf{Detect} &  $93.43 \pm 0.19$ & $>0.57$ &  $77.18 \pm 0.02$ & $>0.98$ & $93.47 \pm 0.15$ & $>0.83$ & $94.80 \pm 0.16$ & $>0.34$ & \XSolidBrush \\
& \textbf{Dim} &  $92.35 \pm 0.14$ & $<10^{-10}$ &  $77.25 \pm 0.08$ & $<10^{-9}$ & $93.80 \pm 0.12$ & $<10^{-10}$ & $95.92 \pm 0.13$ & $<10^{-9}$ & \Checkmark \\
\midrule

\multirow{4}{*}{\textbf{WET}} & \textbf{No Attack} & $93.08 \pm 0.39$ & $<10^{-10}$ &  $76.89 \pm 0.10$  & $<10^{-10}$ & $93.40 \pm 0.09$ & $<10^{-10}$ & $94.20 \pm 0.11$ & $<10^{-10}$ & \Checkmark \\
& \textbf{CSE} &  $86.81 \pm 0.61$ & $<10^{-10}$ & $75.37 \pm 0.09$ & $<10^{-10}$ & $92.78 \pm 0.12$ & $<10^{-10}$ & $95.38 \pm 0.34$ & $<10^{-10}$  & \Checkmark \\
& \textbf{Detect} &  $93.20 \pm 0.30$ & $<10^{-10}$ &  $76.83 \pm 0.02$ & $<10^{-10}$ & $93.40 \pm 0.09$ & $<10^{-9}$ & $94.50 \pm 0.11$ & $<0.02$ & \Checkmark \\
& \textbf{Dim} &  $93.35 \pm 0.43$ & $>0.57$ &  $75.94 \pm 0.10$ & $<0.02$ & $92.46 \pm 0.05$ & $>0.08$ & $92.47 \pm 0.50$ & $>0.98$ & \XSolidBrush \\
\midrule

\multirow{4}{*}{\makecell{\textbf{SemMark}\\ \textbf{(Ours)}}} & \textbf{No Attack} & $93.31 \pm 0.27$ & $<10^{-10}$ &  $77.25 \pm 0.04$  & $<10^{-10}$ & $93.45 \pm 0.05$ & $<10^{-10}$ & $94.45 \pm 0.07$ & $<10^{-10}$ & \Checkmark \\
& \textbf{CSE} &  $89.41 \pm 0.61$ & $<10^{-10}$ &  $75.49 \pm 0.03$ & $<10^{-6}$ & $93.14 \pm 0.20$ & $<10^{-10}$ & $95.42 \pm 0.51$ & $<10^{-10}$ & \Checkmark \\
& \textbf{Detect} &  $93.27 \pm 0.27$ & $<10^{-10}$ & $77.09 \pm 0.09$ & $<10^{-10}$ & $93.62 \pm 0.10$ & $<10^{-10}$ & $94.30 \pm 0.29$ & $<10^{-10}$  & \Checkmark \\
& \textbf{Dim} &  $93.43 \pm 0.14$ & $<10^{-10}$ & $77.29 \pm 0.01$ & $<10^{-10}$ & $93.81 \pm 0.05$ & $<10^{-10}$ & $95.40 \pm 0.08$ & $<10^{-10}$ & \Checkmark \\
\bottomrule
\end{tabular}
}
\end{table*}

\begin{table}[t]
\caption{Experimental results of similarity performance.}
\label{tab: sim result}
\centering
\resizebox{\linewidth}{!}{
\begin{tabular}{lcccc}
\toprule
\textbf{Method} & \textbf{SST2} & \textbf{MIND} & \textbf{AGNews} & 
\textbf{Enron Spam} \\
\rowcolor[gray]{0.9} \multicolumn{5}{c}{Low-Diversity Watermark} \\
\textbf{EmbMarker} & $98.04$ & $97.93$ & $98.08$ & $98.10$ \\
\textbf{WARDEN} & $97.27$ & $96.99$ & $96.88$ & $97.12$ \\
\textbf{EspeW} & $92.37$ & $92.18$ & $92.98$ & $92.39$ \\
\rowcolor[gray]{0.9} \multicolumn{5}{c}{High-Diversity Watermark} \\
\textbf{WET} & $28.48$ & $27.18$ & $28.73$ & $27.84$ \\
\textbf{SemMark (Ours)} & $96.40$ & $94.46$ & $96.82$ & $96.84$ \\

\bottomrule
\end{tabular}
}
\end{table}

\subsubsection{Evaluation Metrics} We employ the same evaluation metrics as the previous method \cite{peng-etal-2023-copying, shetty-etal-2024-warden, wang-etal-2025-robust, shetty-etal-2025-wet} to validate the effectiveness of the watermarking method, including:
\begin{itemize}
    \item \textbf{Task Performance}:
    We train a multi-layer perceptron (MLP) classifier using the provider’s embeddings as input. The classifier's accuracy (\textbf{ACC}) on the downstream task is used to measure the quality of the embeddings. Ideally, watermarked embeddings perform comparably to original embeddings on downstream tasks.
    \item \textbf{Similarity Performance} We measure the closeness of watermarked embeddings with original embeddings by reporting their cosine similarity (\textbf{Sim}). Ideally, the watermarked and original embeddings are highly similar.
    \item \textbf{Detection Performance}
    We employ three metrics to measure the watermark detection performance: the p-value of the KS test (\textbf{p-value}), the difference of cosine similarity (\textbf{$\Delta$ Cos}), and the difference of squared $L_2$ distance (\textbf{$\Delta$} $\boldsymbol{L_2}$).
\end{itemize}

\subsubsection{Baselines} We compare the performance of our method with the following baselines:
\begin{itemize}
    \item \textbf{EmbMarker} \cite{peng-etal-2023-copying}: EmbMarker uses moderate-frequency words as triggers and inserts watermarks based on the number of trigger words by linear interpolation. 
    \item \textbf{WARDEN} \cite{shetty-etal-2024-warden}: WARDEN strengthens EmbMarker against CSE attack by integrating multiple trigger sets with watermark embedding.
    \item \textbf{EspeW} \cite{wang-etal-2025-robust}: Inserting watermarks in partial dimensions to avoid sharing common components. 
    \item \textbf{WET} \cite{shetty-etal-2025-wet}: WET applies linear transformations to all original embeddings to generate watermarked embeddings.
\end{itemize}

\subsubsection{Imitation Attack Details} Following previous work \cite{peng-etal-2023-copying}, we use OpenAI's text-embedding-002 API (text-embedding-ada-002) as the original embedding of EaaS and BERT (bert-base-cased) \cite{devlin-etal-2019-bert} as the backbone model of the attacker to simulate the imitation attack. The attacker uses the original embedding as the input feature and connects two layers of MLP to train their model. The loss function is Mean Squared Error (MSE), optimized using AdamW optimizer \cite{DBLP:conf/iclr/LoshchilovH19} with a learning rate of $5\times10^{-5}$. The training epoch is set to 3.

\begin{figure*}[t]
   \centering
  \subfigure[No Attack] {
    \includegraphics[width=0.23\linewidth]{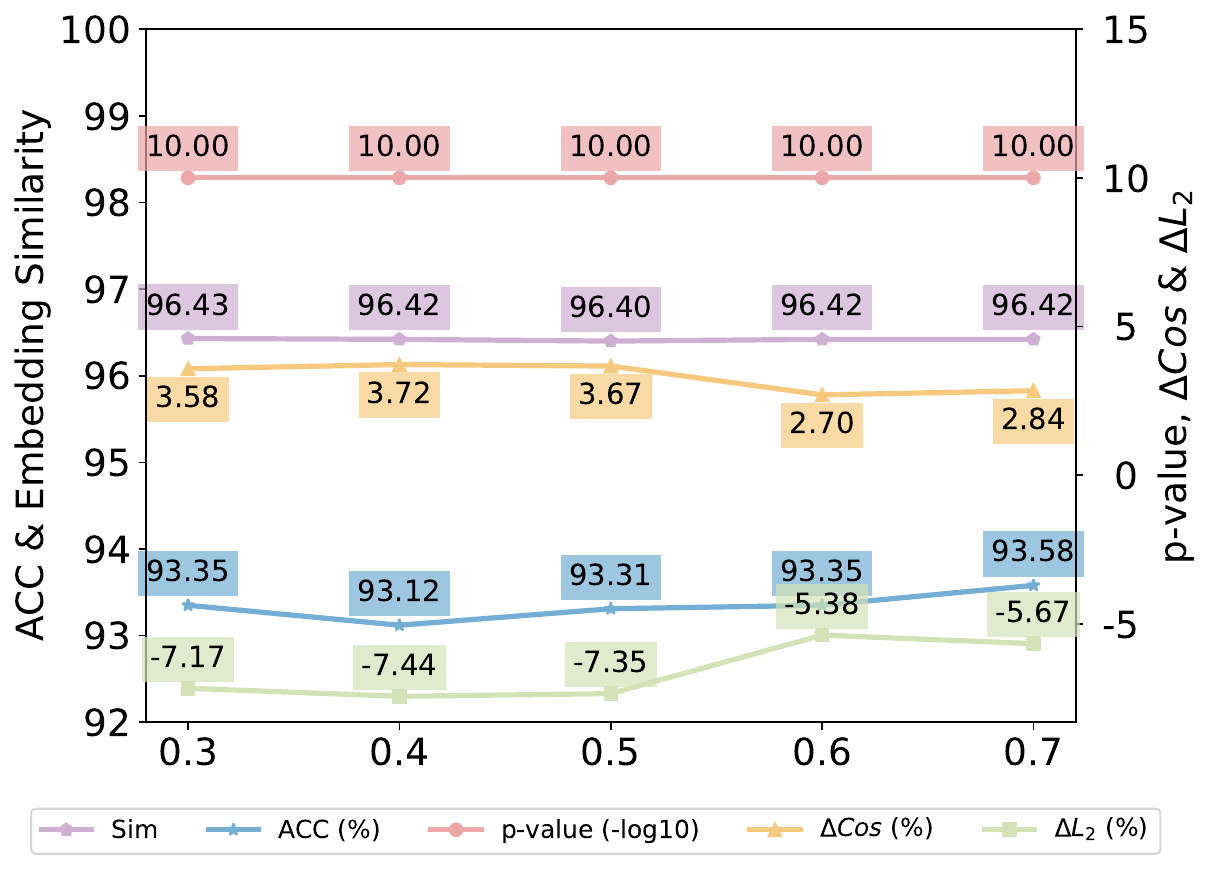}
    \label{fig: sst area noattack}
  }
  \subfigure[CSE] {
    \includegraphics[width=0.23\linewidth]{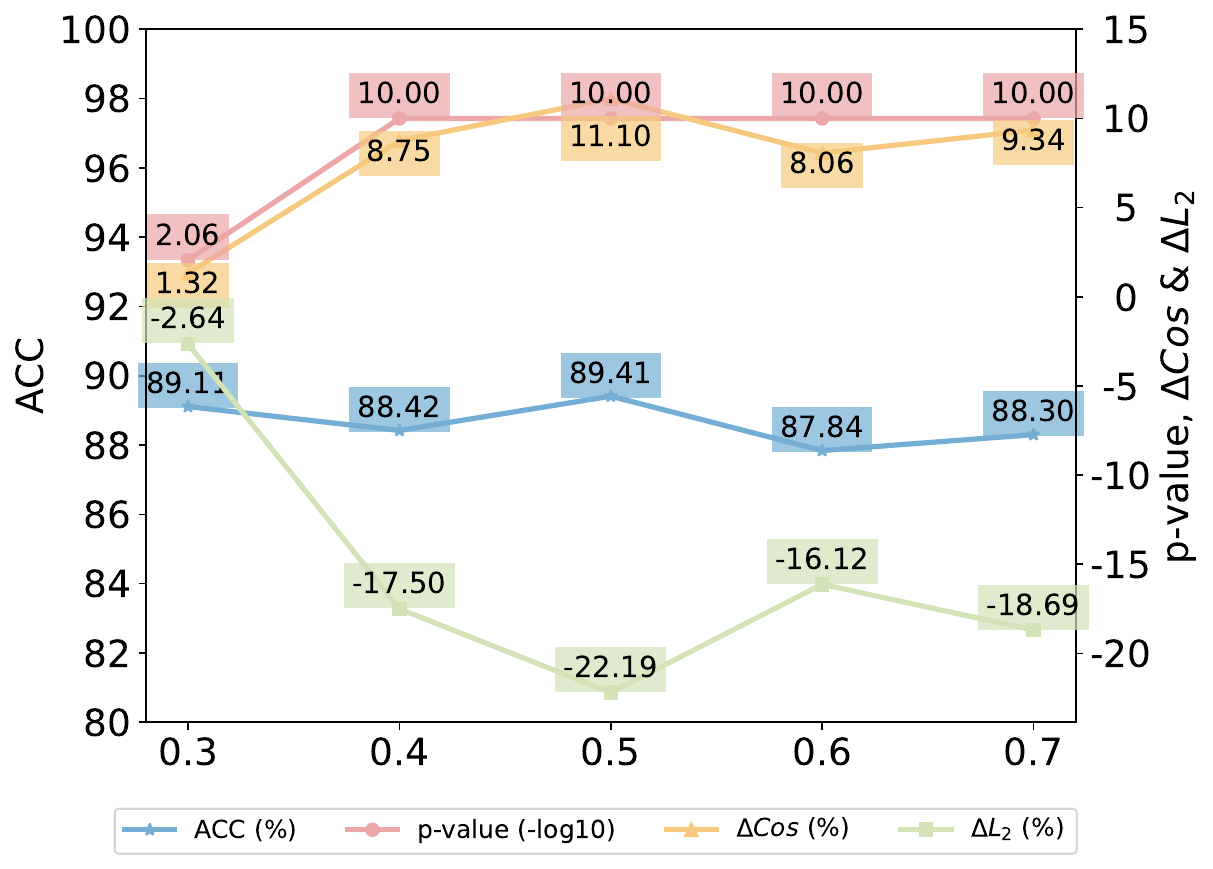}
    \label{fig: sst area cse}
  }
  \subfigure[Detect-Sampling] {
    \includegraphics[width=0.23\linewidth]{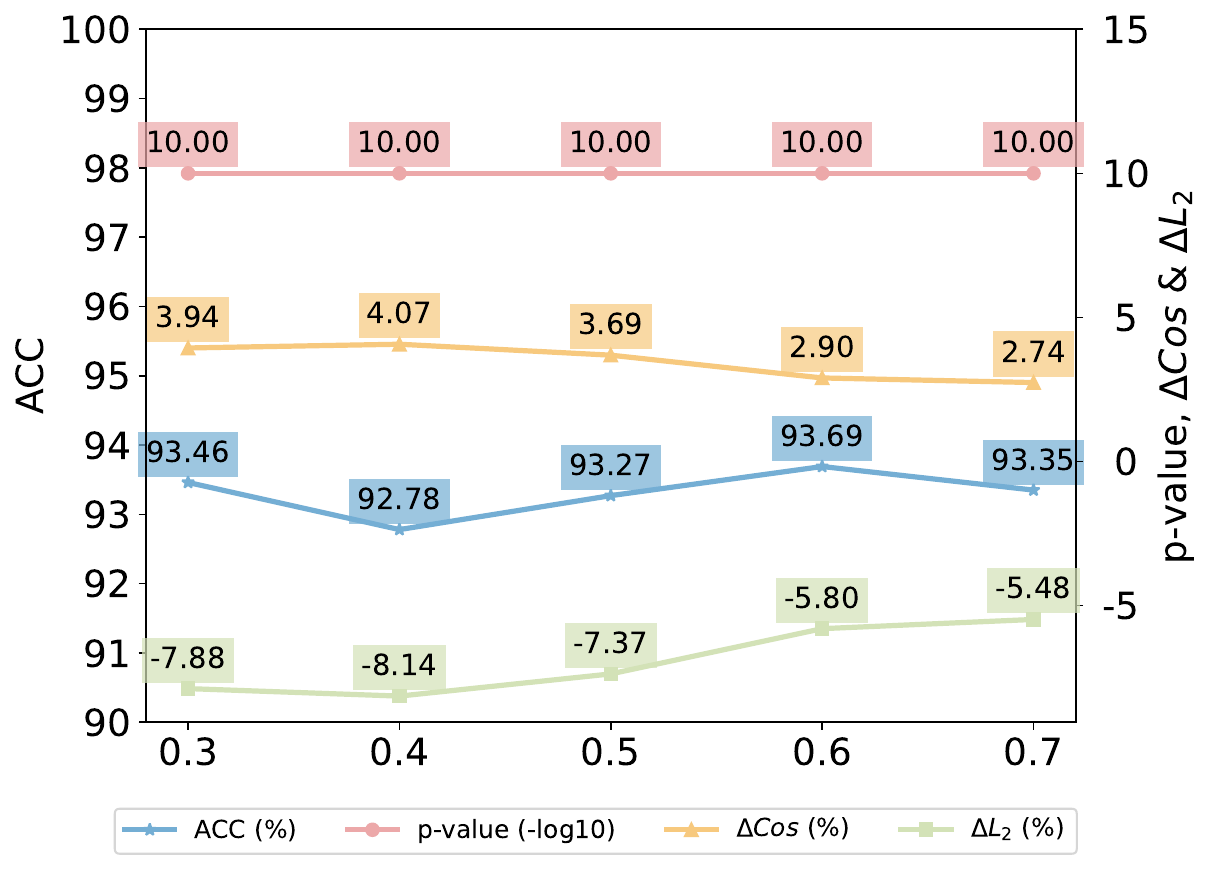}
    \label{fig: sst area sample}
  }
  \subfigure[Dimensionality-Reduction] {
    \includegraphics[width=0.23\linewidth]{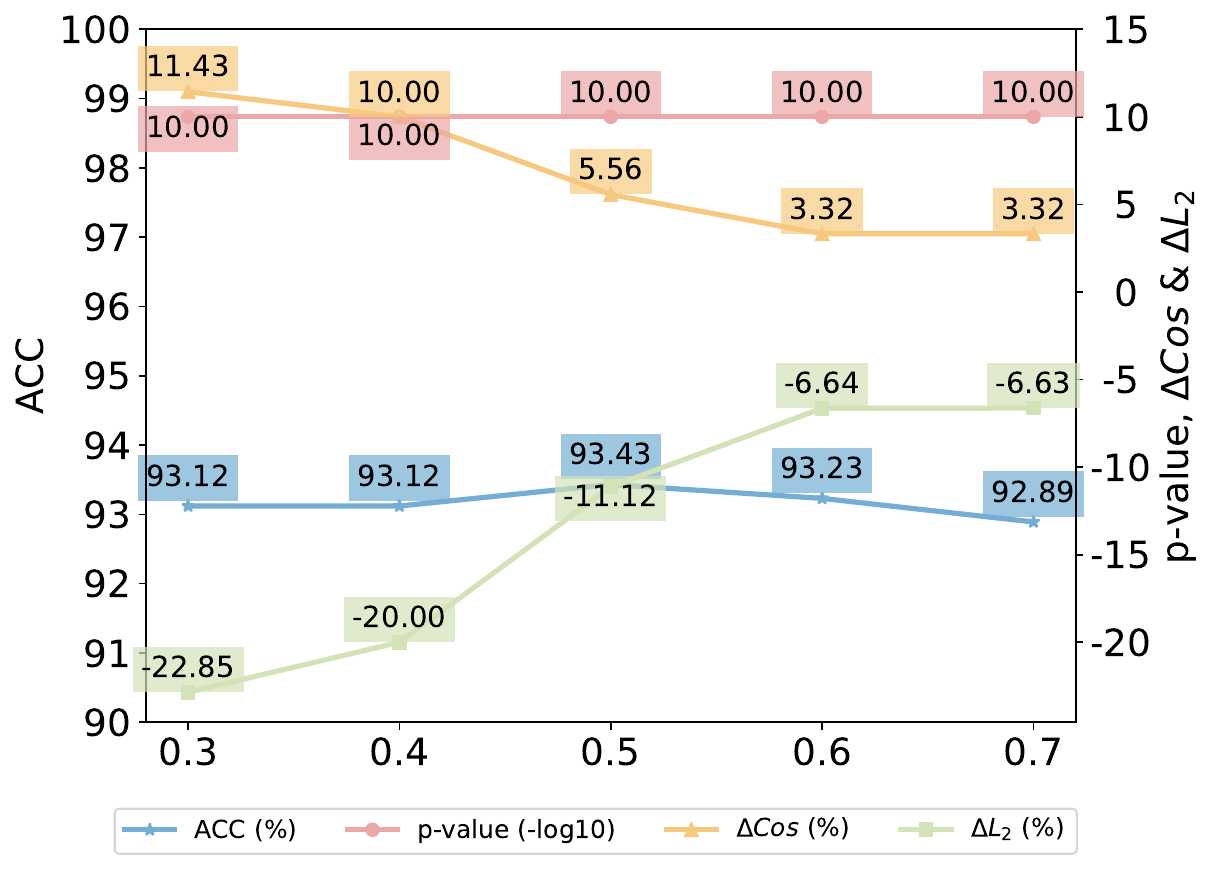}
    \label{fig: sst area dim}
  }
\caption{The impact of the watermark proportion $\alpha$ in four scenarios for the SST2 dataset.}
\label{fig: sst area}
\end{figure*}

\begin{figure*}[t]
\centering
  \subfigure[No Attack] {
    \includegraphics[width=0.23\linewidth]{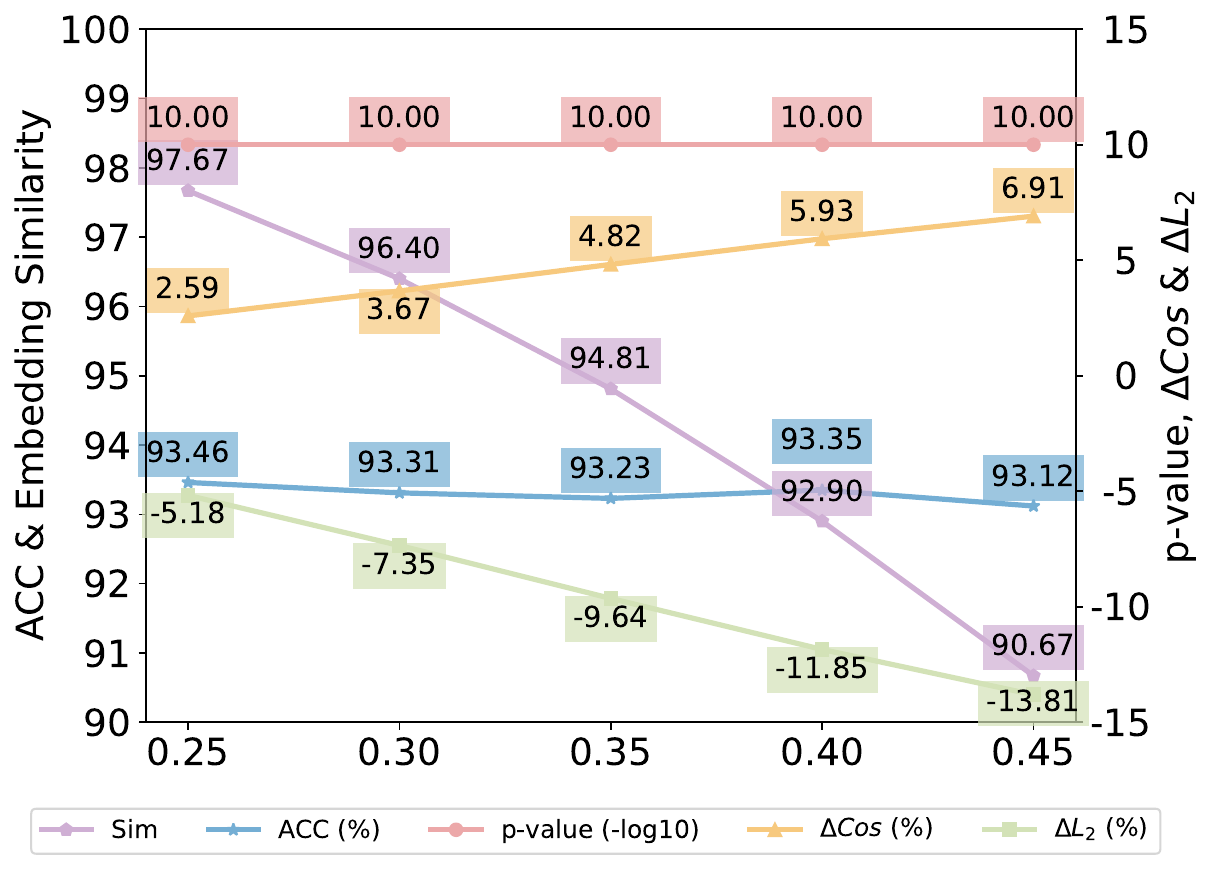}
    \label{fig: sst strength noattack}
  }
  \subfigure[CSE] {
    \includegraphics[width=0.23\linewidth]{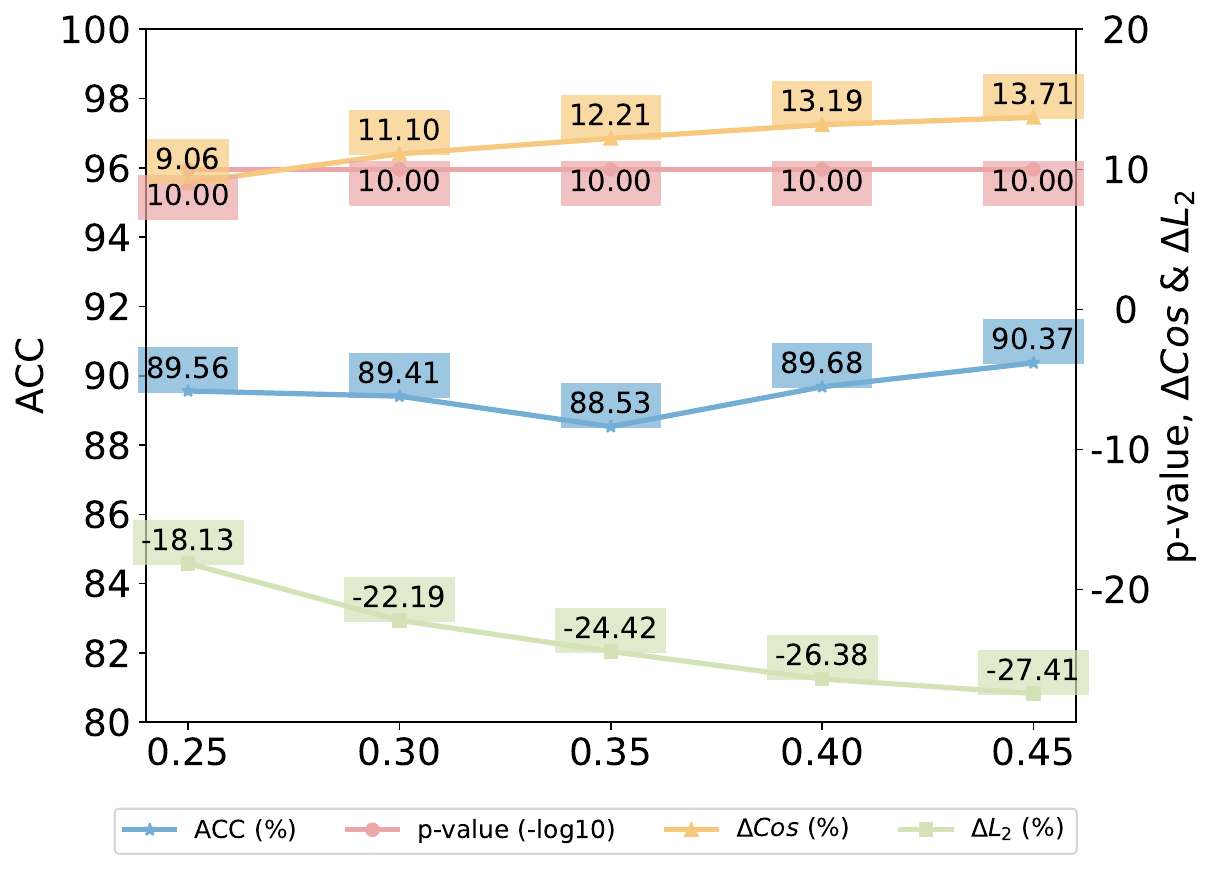}
    \label{fig: sst strength cse}
  }
  \subfigure[Detect-Sampling] {
    \includegraphics[width=0.23\linewidth]{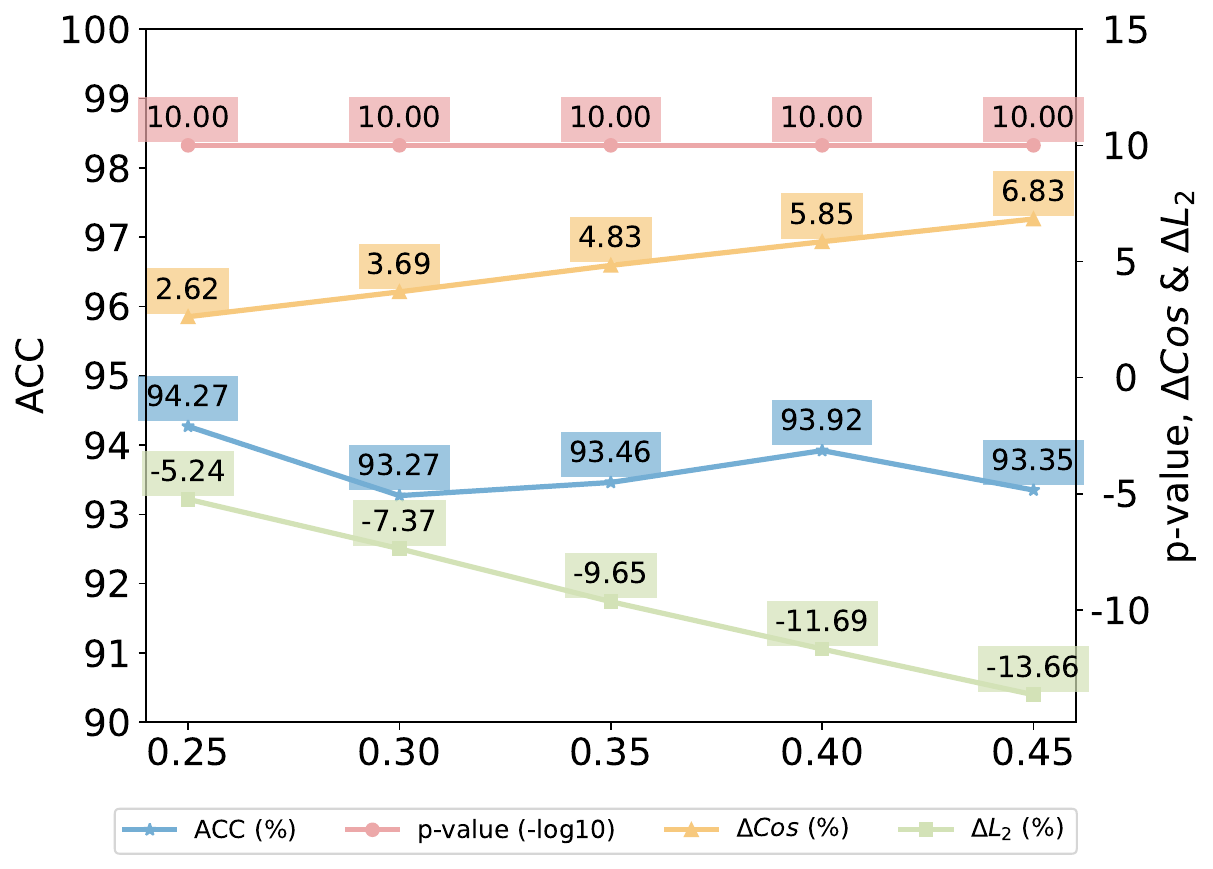}
    \label{fig: sst strength sample}
  }
  \subfigure[Dimensionality-Reduction] {
    \includegraphics[width=0.23\linewidth]{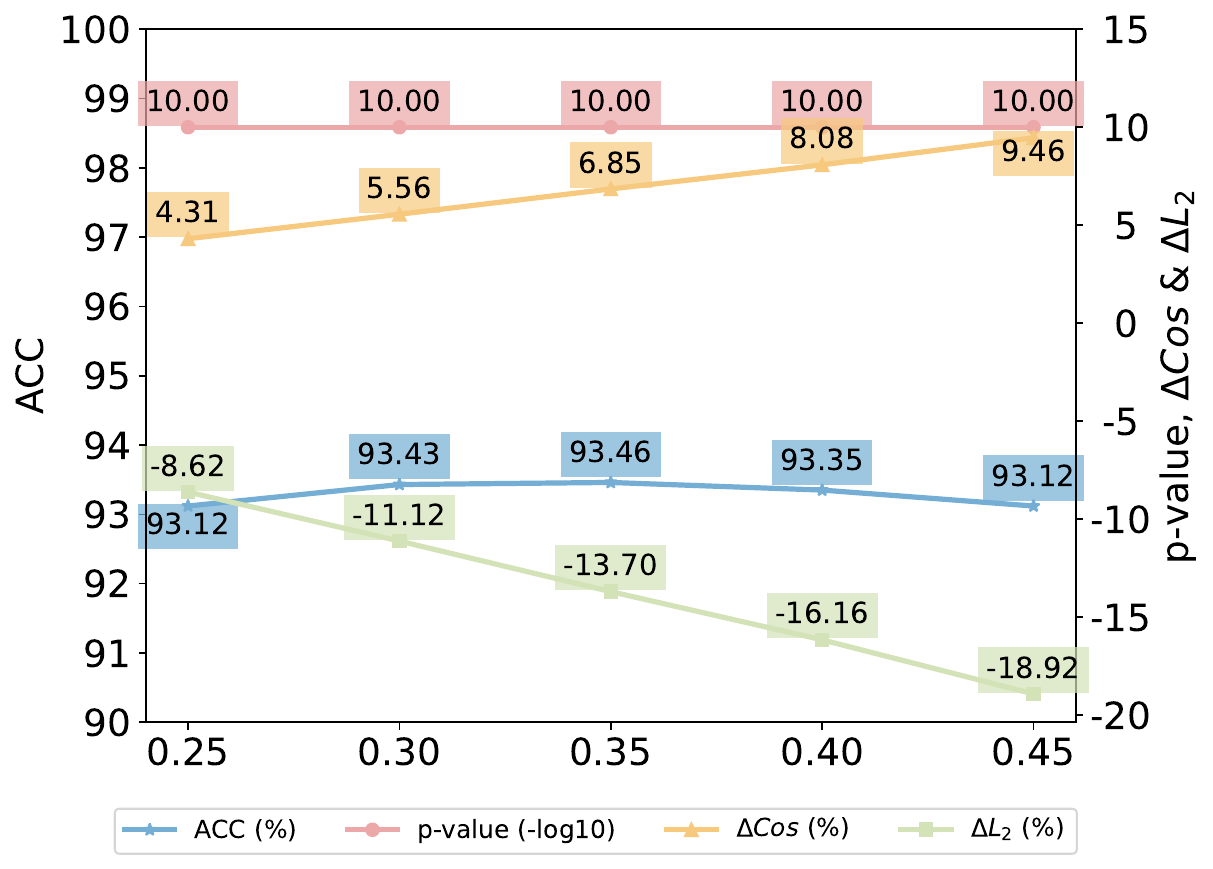}
    \label{fig: sst strength dim}
  }
\caption{The impact of the watermark strength $\delta$ in four scenarios for the SST2 dataset.}
\label{fig: sst strength}
\end{figure*}

\subsubsection{Watermark Attack Details}
To comprehensively evaluate the watermarking method, we construct the \textbf{CSE} and \textbf{Dimensionality-Reduction} (Dim) attack in the stealing stage and the \textbf{Detect-Sampling} attack in the verification stage. Figure \ref{fig: wm attack} shows the overall framework of the three attacks. The detailed construction process and hyperparameters are provided in Appendix \ref{ap: attack detail}.
\paragraph{CSE} CSE attacks \cite{shetty-etal-2024-warden} first clusters the watermarked embeddings, then selects embedding pairs with disparity by comparing corresponding embeddings from the surrogate model, and finally eliminates their top principal components, which can eliminate the influence of the watermark while maintaining service utility. 
\paragraph{Dimensionality-Reduction} 
The attacker can evade watermark detection by altering the dimensionality of the embedding vectors returned by the victim's API. Such manipulation may suppress the potential watermark signal, resulting in embeddings that deviate substantially from the originals. To address this threat, we construct the Dimensionality-Reduction attack. The dimensionality reduction method used is PCA. To calculate watermark verification metrics (e.g., cosine similarity and $L_2$ distance), we align the dimensions of the embedding by calculating the transformation matrix $\mathbf{w}_t$ from the reduced-dimensional embedding $\mathbf{\hat{e}}_o^{*}$ to the watermark embedding $\mathbf{e}_p$ on the non-verification watermark data: $\mathbf{e}_p=\mathbf{\hat{e}}_o^{*} \mathbf{w}_t$.
\paragraph{Detect-Sampling} The service provider needs to query the suspicious model in the watermark verification stage, where abnormal behavior may alert the adversary and result in access denial. To evaluate the stealthiness of watermark verification, we propose the Detect-Sampling attack. We first train the binary detector to measure the difference in text. For high-confidence verification queries, the Detect-Sampling attack returns the randomly sampled embedding of the same dimension to evade watermark detection. The backbone model of the binary detector is BERT (bert-base-cased), which connects four layers of MLP.

\subsubsection{Implementation Details} 
The watermark mapping model is trained on the STSBenchmark (a semantic textual similarity dataset, different from the watermark detection dataset) \cite{muennighoff2022mteb}, optimized by the AdamW optimizer with the learning rate of $1\times10^{-5}$. The neighbor number of the Local Outlier Factor is set to 50, the PCA dimension $h'$ is set to 6, and the watermark proportion $\alpha$ and strength $\delta$ are set to 0.50 and 0.30. We construct each experiment three times using different random seeds and report the average results with standard deviation. For more implementation details and hyperparameter settings, please refer to Appendix \ref{ap: imp}.

\begin{table*}[t]
\caption{Experimental results of verifiability analysis for SemMark.}
\label{tab: verif result}
\centering
\resizebox{\linewidth}{!}{
\begin{tabular}{llcccccccccccc}
\toprule

\multirow{2}{*}{\textbf{Method}} & \multirow{2}{*}{\textbf{Attack}} & \multicolumn{3}{c}{\textbf{SST2}} & \multicolumn{3}{c}{\textbf{MIND}} & \multicolumn{3}{c}{\textbf{AGNews}} & \multicolumn{3}{c}{\textbf{Enron Spam}} \\
\cmidrule(lr){3-5}  \cmidrule(lr){6-8}     \cmidrule(lr){9-11}  \cmidrule(lr){12-14}
& & \textbf{p-value} & \textbf{$\Delta$ Cos} & \textbf{$\Delta$ $\boldsymbol{L_2}$} & \textbf{p-value} & \textbf{$\Delta$ Cos} & \textbf{$\Delta$ $\boldsymbol{L_2}$} & \textbf{p-value} & \textbf{$\Delta$ Cos} & \textbf{$\Delta$ $\boldsymbol{L_2}$} & \textbf{p-value} & \textbf{$\Delta$ Cos} & \textbf{$\Delta$ $\boldsymbol{L_2}$} \\
\midrule

\multirow{5}{*}{\textbf{SemMark}}
& \textbf{Original} & $>0.71$ & -0.06 &  0.12 & $>0.46$ & 0.08 & -0.16 & $>0.46$ & -0.04 & 0.07 & $>0.17$ & 0.01 & -0.02 \\
& \textbf{No Attack} & $<10^{-10}$ & 3.67 & -7.35 & $<10^{-10}$ & 15.51 & -31.02 & $<10^{-10}$ & 7.49 & -14.98 & $<10^{-10}$ & 5.95 & -11.90 \\
& \textbf{CSE} & $<10^{-10}$ & 11.10 & -22.19 & $<10^{-6}$ & 0.79 & -1.57 & $<10^{-10}$ & 15.97 & -31.93 & $<10^{-10}$ & 16.90 & -33.80 \\
& \textbf{Detect} & $<10^{-10}$ & 3.69 & -7.37 & $<10^{-10}$ & 13.23 & -26.46 & $<10^{-10}$ & 7.36 & -14.72 & $<10^{-10}$ & 0.13 & -0.26 \\
& \textbf{Dim} & $<10^{-10}$ & 5.56 & -11.12 & $<10^{-10}$ & 16.83 & -33.66 & $<10^{-10}$ & 10.31 & -20.61 & $<10^{-10}$ & 9.01 & -18.02 \\
\midrule

\multirow{1}{*}{\textbf{w/o} $\mathcal{M}$}
& \textbf{Original} & $<10^{-2}$ & -0.37 & 0.74 & $<10^{-10}$ & 1.25 & -2.51 & $<10^{-4}$ & 0.63 & -1.26 & $<10^{-10}$ & -2.01 & 4.01 \\

\bottomrule
\end{tabular}
}
\end{table*}

\subsection{Main Results}
Table \ref{tab: main result} shows the experimental results of all methods on four datasets and four scenarios, where ``Original'' means that the EaaS provider does not inject watermarks into the original embedding, and the attacker utilizes the original embedding to copy the model.  Table \ref{tab: sim result} shows the cosine similarity between the watermarked embedding and the original embedding. 
Our method effectively defends against potential watermark attacks while maintaining the watermark embedding closely related to the original embedding, achieving excellent performance in downstream tasks. We have the following observations and analyses:

\textbf{Semantic-based paradigm makes the watermark process more natural and imperceptible.} 
Trigger-based methods (EmbMarker, WARDEN, and EspeW) require constructing texts containing numerous trigger words during verification, which can be easily identified by adversaries and are vulnerable to Detect-Sampling attacks. In contrast, the semantics-based paradigm exploits the inherent semantic properties of embeddings to partition text, ensuring that the verification process remains natural and covert, thereby enhancing the stealthiness of watermarking methods.

\textbf{Diverse semantic-aware watermarks are difficult to eliminate by watermark attacks.} Trigger-based methods typically pre-construct a fixed set of watermark embeddings, which limits their diversity and scalability and leaves them vulnerable to CSE attack. Linear transformation-based methods (WET) rely heavily on the transformation matrix and are susceptible to perturbations induced by dimensionality changes. Our Method defends against all watermark attacks by employing the watermark mapping model to inject semantic-aware watermarks with adaptive weights, thereby producing more diverse watermark embeddings that are difficult to eliminate through clustering analysis and dimensionality change.

\textbf{Adaptive watermark weight mechanism effectively maintains the closeness between watermark embeddings and original embeddings.} As shown in Table \ref{tab: sim result}, EmbMarker achieves the highest cosine similarity between the watermarked embeddings and their original counterparts by injecting a single, identical watermark. Injecting diverse watermarks (WET and SemMark) potentially compromises closeness to the original embedding, for example, WET achieves $<$30\% similarity across all four datasets. Our method significantly mitigates this phenomenon by cleverly injecting diverse watermarks ($>$\textbf{94}\%). The adaptive watermark weighting mechanism assigns more weight to outliers, significantly improving the similarity with the original embedding. Therefore, our watermarking scheme is both diverse and harmless.

\subsection{Impact of Watermark Proportion}
We analyze the impact of the watermark proportion on the task, similarity, and detection performance. The experimental results of the SST2 dataset are shown in Figure \ref{fig: sst area}, with additional datasets and analyses provided in Appendix \ref{ap: area}. Our method consistently achieves excellent task, similarity, and detection performance under all settings, effectively defending against potential watermark attacks (p-value < 0.05). Furthermore, benefit from the semantic-aware watermarks generated by the watermark mapping model, increasing the watermark proportion exerts only a negligible impact on the similarity to the original embedding, thereby ensuring the excellent harmlessness of our watermarking method.

\begin{table}[t]
\caption{Experimental results of time complexity.}
\label{tab: time result}
\centering
\resizebox{\linewidth}{!}{
\begin{tabular}{llcccc}
\toprule
\textbf{Model} & \textbf{Dataset} & \textbf{Encoding} & \textbf{Watermark} & \textbf{Ratio} & \textbf{$\text{Time}_{\text{per}}$} \\
\midrule
\multirow{4}{*}{\textbf{SBERT}} & \textbf{SST2} & 525.31 & 27.04 & 4.90 & 0.0004 \\
& \textbf{MIND} & 763.75 & 46.37 & 5.72 & 0.0005 \\
& \textbf{AGNews} & 970.59 & 62.25 & 6.02 & 0.0005 \\
& \textbf{Enron Spam} & 248.87 & 15.72 & 5.94 & 0.0005 \\
\midrule
\multirow{4}{*}{\textbf{Jina-v3}} & \textbf{SST2} & 2901.97 & 25.39 & 0.88 & 0.0004 \\
& \textbf{MIND} & 4279.30 & 45.11 & 1.04 & 0.0005 \\
& \textbf{AGNews} & 5230.16 & 56.22 & 1.06 & 0.0005 \\
& \textbf{Enron Spam} & 1386.98 & 14.46 & 1.03 & 0.0005 \\
\midrule
\multirow{4}{*}{\textbf{Qwen-8B}} & \textbf{SST2} & 3140.66 & 29.77 & 0.94 & 0.0004 \\
& \textbf{MIND} & 4650.77 & 62.62 & 1.33  & 0.0006 \\
& \textbf{AGNews} & 10136.59 & 72.86 & 0.71 & 0.0006 \\
& \textbf{Enron Spam} & 1458.48 & 17.11 & 1.16 & 0.0005 \\
\bottomrule
\end{tabular}
}
\end{table}

\subsection{Impact of Watermark Strength}
Figure \ref{fig: sst strength} illustrates the impact of watermark strength on the task, similarity, and detection performance on the SST2 dataset. Appendix \ref{ap: strength} presents additional datasets and analyses.
As the watermark strength increases, the $\Delta$ Cos and $\Delta$ $L_2$ metrics improve significantly. 
For example, when the watermark strength increases from 0.25 to 0.45 in the ``No attack'' scenario, the $\Delta$ Cos and $\Delta$ $L_2$ metrics improve by nearly two times.
This phenomenon demonstrates that watermark strength is positively correlated with detection performance.
However, excessive watermark strength inevitably degrades the similarity with the original embedding. Unlike existing methods that adopt uniform watermark weights, our adaptive weight mechanism assigns higher watermark weights to outlier data and lower weights to dense data. This strategy not only diversifies the watermark direction but also substantially mitigates the negative impact of watermark strength on the original embedding.

\subsection{Verifiability Analysis for SemMark}
In Table \ref{tab: verif result}, we analyze the verifiability of SemMark, where ``w/o $\mathcal{M}$'' denotes injecting a fixed watermark embedding rather than generating one via the watermark mapping model $\mathcal{M}$, and ``Original'' means that the EaaS provider does not inject watermarks into the original embedding. As shown in Table \ref{tab: verif result}, even when using a single watermark embedding, the embeddings for the watermarked and non-watermarked regions differ significantly in the original scenario (p-value $< 10^{-2}$), rendering the watermarking method unverifiable. This occurs because the semantic paradigm partitions the watermark region in the semantic space, resulting in distance bias between different regions, which causes discrepancies with single watermark embedding. In contrast, our method considers semantic consistency and similarity in training the watermark mapping model $\mathcal{M}$, constraining the distance between the watermark and the embedding while generating diverse semantic-aware watermarks. Thus, by combining the semantic paradigm with watermark injection through $\mathcal{M}$, SemMark achieves excellent verifiability, diversity, and harmlessness, while remaining stealthy in the verification stage.

\subsection{Time Complexity Analysis}
We analyze the time complexity of our method in Table \ref{tab: time result}, where ``Encoding'' and ``Watermark'' denote the encoding time and watermark injection time of our mothod, ``Ratio'' means the ratio of watermark injection time relative to the total time, and ``$\text{Time}_{\text{per}}$'' indicates the average time to inject the watermark into a sample (in seconds). Considering the response time and network latency of calling the OpenAI API, we adopt open-source encoding models with different embedding dimensions, including SBERT \cite{reimers-gurevych-2019-sentence}, Jina-v3 (jina-embeddings-v3) \cite{sturua2024jina}, and Qwen-8B (Qwen3-Embedding-8B) \cite{qwen3embedding}. As shown in Table \ref{tab: time result}, watermark injection in our method accounts for a very small fraction of the total time, particularly in the Jina-v3 and Qwen-8B models with embedding dimensions of 1024 and 4096 (average \textbf{0.0005} seconds per sample, about \textbf{1}\% total time). This efficiency benefits from the lightweight design of the watermark mapping model and only needs to inject watermarks in the watermark region, enabling fast and effective generation of semantic-aware watermarks.

\section{Conclusion}
In this paper, we comprehensively analyze the existing EaaS watermarking methods and propose a novel semantic paradigm. We partition the embedding space based on the natural semantic properties of the embeddings and inject watermarks generated by the watermark mapping model into specific regions with the adaptive weight mechanism.  To comprehensively evaluate the EaaS watermarking method, we propose the Detect-Sampling and  Dimensionality-Reduction attack to evaluate the stealth and resistance to dimensionality changes of watermarking methods. We construct experiments on four popular NLP datasets and three watermark removal attacks, and our watermarking method demonstrates excellent verifiability, diversity, stealthiness, and harmlessness.

\section{Acknowledgments}
This work is supported by the Postdoctoral Fellowship Program of CPSF under Grant Number GZC20251076, and the National Natural Science Foundation of China (No.U2336202).

\bibliographystyle{ACM-Reference-Format}
\bibliography{sample-base}

\cleardoublepage

\appendix

\section{Dataset}
\label{ap: dataset}
SST2 (Stanford Sentiment Treebank) \cite{socher-etal-2013-recursive} is a comprehensive dataset in natural language processing, originally constructed from movie reviews. This dataset includes both positive and negative sentiments and is widely used in sentiment classification tasks.

MIND \cite{wu-etal-2020-mind} is a large-scale news recommendation dataset derived from user behavior logs of Microsoft News. This dataset contains rich textual content including title, abstract, body, category and entities, and is widely used in the field of news recommendation and recommendation systems. MIND covers 18 news categories, including sports, finance, lifestyle, health, weather, and entertainment. We use the MIND dataset for news classification tasks.

AGNews \cite{10.5555/2969239.2969312} is a large-scale text classification dataset derived from AG’s corpus of news articles. Each data item includes the title and explanation, and is widely used in the fields of text classification and text representation learning. AGNews covers 4 categories, including World, Sports, Business, and Technology.

Enron Spam \cite{metsis2006spam} is a classic email dataset derived from user messages of the Enron Corpus. This dataset includes both ham and spam emails and is widely used in the fields of spam identification and text classification tasks.

Table \ref{tab: dataset} shows the detailed statistics of these datasets. For each dataset, we adhere to the official data splits. Following \citet{peng-etal-2023-copying}, we use the validation set instead of the test set to verify the performance of downstream tasks in the SST2 dataset since its test set has no labels.

\section{Baseline Settings}
We use the official settings of baselines, the details are as follows:
\begin{itemize}
    \item \textbf{EmbMarker} \cite{peng-etal-2023-copying}: The maximum number of trigger words is set to 4. The trigger set size is set to 20, which is selected based on the frequency interval [0.5\%, 1\%] in the WikiText dataset \cite{merity2017pointer}. The WikiText dataset contains 1,801,350 entries, which are used to select moderate-frequency words as watermark triggers.
    \item \textbf{WARDEN} \cite{shetty-etal-2024-warden}: WARDEN follows the default configuration and settings of EmbMarker, and the number of watermarks is set to 4.
    \item \textbf{EspeW} \cite{wang-etal-2025-robust}: EspeW follows the default configuration and settings of EmbMarker, and the number of watermarks is set to 4, the watermark proportion is set to 50\%.
    \item \textbf{WET} \cite{shetty-etal-2025-wet}: The number of correlations is set to 25, and the number of watermarked dimensions is set to 1536.
\end{itemize}

\section{Implementation Details}
\label{ap: imp}
Considering the sparsity of high-dimensional space, we first reduce the dimensionality of the original embedding using PCA, and then utilize LSH to partition the embedding space into watermark and non-watermark regions. The PCA dimension $h'$ is set to 6, and the number of random normal vectors is set to 6.
We use the STSBenchmark to train the watermark mapping model, which is a semantic textual similarity dataset \cite{muennighoff2022mteb}. The watermark mapping model is optimized by the AdamW optimizer with the learning rate of $1\times10^{-5}$. The scaling weight $\tau$ in semantic consistency loss is set to 1.5, the margin hyperparameter $\eta$ in semantic similarity loss is set to 0.5, the weight $\gamma$ for semantic similarity loss is set to 0.5, and the watermark mapping model is trained for 100 epochs.
In the watermark injection stage, we use the test set of the dataset as the surrogate dataset $\mathcal{D}_s$ to estimate the local outlier factor for embedding,  the number of local neighbors $k$ of the Local Outlier Factor is set to 50, the watermark proportion $\alpha$ is set to 0.50,  the maximum watermark strength $\delta$ and weight margin $\epsilon$ are set to 0.30 and 0.05. 
In the watermark verification stage, we use the test set of the dataset to construct the watermark verification set, and the size $m$ is set to 500.
We construct each experiment three times using different random seeds and report the average results with standard deviation. All models and datasets are accessible via HuggingFace. All experiments are conducted on NVIDIA A100 80GB GPUs and implemented using the Transformers library and PyTorch.

\begin{table}[t]
\caption{The statistics of datasets.}
\centering
\begin{tabular}{c|cccc}
\toprule
\textbf{Dataset} & \# \textbf{Train} & \# \textbf{Test} & \textbf{Avg. Len.} & \# \textbf{Class} \\
\midrule
\textbf{SST2} & 67,349 & 872 & 54.17 & 2 \\
\textbf{MIND} & 97,791 & 32,592 & 66.14 & 18 \\
\textbf{AG News} & 120,000 & 7600 & 236.41 & 4 \\
\textbf{Enron Spam} & 31,716 & 2000 & 34.57 & 2 \\
\bottomrule
\end{tabular}
\label{tab: dataset}
\end{table}

\section{Watermark Removal Attack Details}
\label{ap: attack detail}
\paragraph{CSE} 
CSE attack \cite{shetty-etal-2024-warden} uses the surrogate model and 
and the model that deploys watermarking technology to encode data. The attack first clusters the watermarked embeddings and then selects embedding pairs with disparity by comparing them with the corresponding embeddings from the surrogate model. These samples with distinctive distance changes are considered suspicious watermarked samples. Finally, their top principal components are eliminated to remove the watermark signal. The CSE attack can eliminate the influence of the watermark while maintaining service utility. Following \citet{shetty-etal-2024-warden}, the surrogate model is Sentence-BERT (paraphrase-MiniLM-L6-v2) \cite{reimers-gurevych-2019-sentence}.  The clustering algorithm is Kmeans \cite{10.5555/1283383.1283494}, the number of clusters is set to 20, and the number of elimination principal components is set to 50.

\paragraph{Dimensionality-Reduction} 
Considering that the attacker can modify the dimension of the embedding returned by the victim API to evade watermark detection, we construct the Dimensionality-Reduction attack to eliminate the potential watermark signal by reducing the dimension of the embedding through PCA to make it different from the original embedding. Specifically, we utilize PCA to reduce the dimension of the embedding returned by the victim API, and then use these embeddings to train the attacker's model (e.g., the victim API is OpenAI’s text-embedding-002, and the dimension of the returned embedding is reduced from 1536 to 1024). For the service provider, the inconsistent dimension makes it difficult to calculate the similarity of the embedding to verify the watermark. To align the dimension, we calculate the transformation matrix $\mathbf{w}_t$ from the reduced-dimensional embedding $\mathbf{\hat{e}}_o^{*}$ to the watermark embedding $\mathbf{e}_p$ on the non-verification watermark data (e.g., training set): $\mathbf{e}_p=\mathbf{\hat{e}}_o^{*} \mathbf{w}_t$, and then obtain $\mathbf{w}_t=\mathbf{\hat{e}}_o^{*+}\mathbf{e}_p$, where $\mathbf{\hat{e}}_o^{*+}$ is the Moore–Penrose pseudoinverse of $\mathbf{\hat{e}}_o^{*}$. In our experiments, the dimension of the Dimensionality-Reduction attack is set to 1024.
 
\begin{table*}[ht]
\caption{The performance of the detector in the Detect-Sampling attack.}
\label{tab: detect-sampling performance}
\centering
\resizebox{\linewidth}{!}{
\begin{tabular}{lcccccccccccc}
\toprule
\multirow{2}{*}{\textbf{Method}} & \multicolumn{3}{c}{\textbf{SST2}} & \multicolumn{3}{c}{\textbf{MIND}} & \multicolumn{3}{c}{\textbf{AGNews}} & \multicolumn{3}{c}{\textbf{Enron Spam}} \\
\cmidrule(lr){2-4}  \cmidrule(lr){5-7}     \cmidrule(lr){8-10}  \cmidrule(lr){11-13}
& \textbf{Precision} $\downarrow$ & \textbf{Recall} $\downarrow$ & \textbf{F1} $\downarrow$ & \textbf{Precision} $\downarrow$ & \textbf{Recall} $\downarrow$ & \textbf{F1} $\downarrow$ & \textbf{Precision} $\downarrow$ & \textbf{Recall} $\downarrow$ & \textbf{F1} $\downarrow$ & \textbf{Precision} $\downarrow$ & \textbf{Recall} $\downarrow$ & \textbf{F1} $\downarrow$  \\
\midrule
\textbf{EmbMarker} & 78.40 & 98.00 & 87.11 & 89.09 & 98.00 & 93.33 & 98.99 & 98.00 & 98.49 & 59.04 & 98.00 & 73.68 \\
\textbf{WARDEN} & 78.33 & 99.38 & 87.60 & 89.33 & 99.38 & 94.08 & 98.15 & 99.38 & 98.76 & 60.69 & 99.38 & 75.36 \\
\textbf{SemMark} & 6.13 & 1.37 & 2.23 & 59.39 & 11.70 & 19.55 & 45.45 & 2.00 & 3.83 & 50.49 & 61.60 & 55.50 \\

\bottomrule
\end{tabular}
}
\end{table*}

\begin{table*}[ht]
\caption{The impact of token length on the Detect-Sampling detector's performance.}
\label{tab: detect-sampling token length}
\centering
\resizebox{\linewidth}{!}{
\begin{tabular}{lcccccccccccc}
\toprule
\multirow{2}{*}{\textbf{Method}} & \multicolumn{3}{c}{\textbf{SST2}} & \multicolumn{3}{c}{\textbf{MIND}} & \multicolumn{3}{c}{\textbf{AGNews}} & \multicolumn{3}{c}{\textbf{Enron Spam}} \\
\cmidrule(lr){2-4}  \cmidrule(lr){5-7}     \cmidrule(lr){8-10}  \cmidrule(lr){11-13}
& \textbf{Precision} $\downarrow$ & \textbf{Recall} $\downarrow$ & \textbf{F1} $\downarrow$ & \textbf{Precision} $\downarrow$ & \textbf{Recall} $\downarrow$ & \textbf{F1} $\downarrow$ & \textbf{Precision} $\downarrow$ & \textbf{Recall} $\downarrow$ & \textbf{F1} $\downarrow$ & \textbf{Precision} $\downarrow$ & \textbf{Recall} $\downarrow$ & \textbf{F1} $\downarrow$ \\
\midrule
\rowcolor[gray]{0.9} \multicolumn{13}{c}{(0, 10]} \\
\textbf{EmbMarker} & 72.33 & 98.80 & 83.52 & 86.82 & 98.80 & 92.42 & 54.29 & 98.80 & 70.07 & 61.52 & 98.80 & 75.83 \\
\textbf{WARDEN} & 69.96 & 99.20 & 82.05 & 84.50 & 99.20 & 91.26 & 53.68 & 99.20 & 69.66 & 59.33 & 99.20 & 74.25 \\
\textbf{SemMark} & 49.73 & 75.00 & 59.81 & 51.01 & 86.00 & 64.04 & 51.31 & 90.40 & 65.46 & 50.42 & 83.20 & 62.79 \\
\midrule
\rowcolor[gray]{0.9} \multicolumn{13}{c}{(10, 20]} \\
\textbf{EmbMarker} & 91.91 & 100.00 & 95.79 & 91.91 & 100.00 & 95.79 & 64.85 & 100.00 & 78.68 & 61.50 & 100.00 & 76.16 \\
\textbf{WARDEN} & 90.25 & 100.00 & 94.88 & 91.74 & 100.00 & 95.69 & 63.61 & 100.00 & 77.76 & 61.12 & 100.00 & 75.87 \\
\textbf{SemMark} & 46.92 & 36.60 & 41.12 & 49.71 & 34.80 & 40.94 & 49.55 & 54.80 & 52.04 & 51.84 & 64.80 & 57.60 \\
\midrule
\rowcolor[gray]{0.9} \multicolumn{13}{c}{(20, 30]} \\
\textbf{EmbMarker} & 95.24 & 100.00 & 97.56 & 93.11 & 100.00 & 96.43 & 99.01 & 100.00 & 99.50 & 57.41 & 100.00 & 72.94 \\
\textbf{WARDEN} & 96.34 & 100.00 & 98.14 & 90.91 & 100.00 & 95.24 & 99.21 & 100.00 & 99.60 & 57.47 & 100.00 & 72.99 \\
\textbf{SemMark} & 47.78 & 28.00 & 35.31 & 44.23 & 9.20 & 15.23 & 48.38 & 32.80 & 39.09 & 51.71 & 63.40 & 56.96 \\
\midrule
\rowcolor[gray]{0.9} \multicolumn{13}{c}{(30, 40]} \\
\textbf{EmbMarker} & 97.47 & 100.00 & 98.72 & 93.46 & 100.00 & 96.62 & 99.01 & 100.00 & 99.50 & 62.27 & 100.00 & 76.75 \\
\textbf{WARDEN} & 97.85 & 100.00 & 98.91 & 91.58 & 100.00 & 95.60 & 99.01 & 100.00 & 99.50 & 60.98 & 100.00 & 75.76 \\
\textbf{SemMark} & 48.78 & 24.00 & 32.17 & 49.44 & 8.80 & 14.94 & 49.20 & 24.60 & 32.80 & 51.77 & 64.20 & 57.32 \\
\midrule
\rowcolor[gray]{0.9} \multicolumn{13}{c}{(40, 50]} \\
\textbf{EmbMarker} & 96.71 & 100.00 & 98.33 & 91.41 & 100.00 & 95.51 & 97.47 & 100.00 & 98.72 & 60.53 & 100.00 & 75.41 \\
\textbf{WARDEN} & 98.04 & 100.00 & 99.01 & 90.09 & 100.00 & 94.79 & 97.85 & 100.00 & 98.91 & 60.68 & 100.00 & 75.53 \\
\textbf{SemMark} & 49.15 & 23.20 & 31.52 & 50.56 & 9.00 & 15.28 & 54.74 & 20.80 & 30.14 & 51.75 & 65.00 & 57.62 \\

\bottomrule
\end{tabular}
}
\end{table*}

\paragraph{Detect-Sampling}
The service provider needs to call the suspicious model in the watermark verification stage, and abnormal behavior may cause the attacker to defend and deny access. For example, the trigger word-based methods \cite{peng-etal-2023-copying, shetty-etal-2024-warden, wang-etal-2025-robust} use trigger words as backdoors to insert watermarks, and concatenate trigger words to query suspicious models in the watermark verification stage. To evaluate the stealthiness of the watermarking method in the verification stage, we propose the Detect-Sampling attack.  We first train the binary detector to measure the difference in text. For queries with significant differences (Detector threshold is set to 0.5), the Detect-Sampling attack returns randomly sampled embeddings of the same dimension to confuse the service provider and evade watermark detection. The backbone model of the binary detector is BERT (bert-base-cased), which connects four layers of MLP. The training data includes normal and abnormal texts. We sample 10,000 examples as normal text on the MultiNLI dataset \cite{N18-1101}, which is a multi-genre natural language inference corpus derived from ten different genres of written and spoken English. The abnormal text is constructed by concatenating randomly sampled tokens and has the same size as the normal text. The loss function of binary detector is Binary Cross-Entropy (BCE), optimized by AdamW optimizer with the learning rate of $5\times10^{-5}$. The training epoch is 10.

\begin{table*}[ht]
\caption{The impact of verification data size on watermark detection performance.}
\label{tab: all data num result}
\centering
\resizebox{0.9\linewidth}{!}{
\begin{tabular}{lcccccccccccc}
\toprule

\multirow{2}{*}{\textbf{Attack}} & \multicolumn{3}{c}{\textbf{SST2}} & \multicolumn{3}{c}{\textbf{MIND}} & \multicolumn{3}{c}{\textbf{AGNews}} & \multicolumn{3}{c}{\textbf{Enron Spam}} \\
\cmidrule(lr){2-4}  \cmidrule(lr){5-7}     \cmidrule(lr){8-10}  \cmidrule(lr){11-13}
& \textbf{p-value} & \textbf{$\Delta$ Cos} & \textbf{$\Delta$ $\boldsymbol{L_2}$} & \textbf{p-value} & \textbf{$\Delta$ Cos} & \textbf{$\Delta$ $\boldsymbol{L_2}$} & \textbf{p-value} & \textbf{$\Delta$ Cos} & \textbf{$\Delta$ $\boldsymbol{L_2}$} & \textbf{p-value} & \textbf{$\Delta$ Cos} & \textbf{$\Delta$ $\boldsymbol{L_2}$} \\
\midrule
\rowcolor[gray]{0.9} \multicolumn{13}{c}{100} \\
\textbf{Original} & $>0.97$ & -0.04 & 0.09 & $>0.91$ & 0.12 & -0.25 & $>0.11$ & -0.32 & 0.64 & $>0.28$ & 0.01 & -0.01 \\
\textbf{No Attack} & $<10^{-10}$ & 4.13 & -8.25 & $<10^{-10}$ & 16.37 & -32.75 & $<10^{-10}$ & 6.83 & -13.66 & $<10^{-10}$ & 6.30 & -12.59 \\
\textbf{CSE} & $<10^{-8}$ & 11.58 & -23.16 & $<10^{-2}$ & 0.83 & -1.67 & $<10^{-10}$ & 14.79 & -29.59 & $<10^{-10}$ & 19.39 & -38.79 \\
\textbf{Detect} & $<10^{-10}$ & 4.67 & -9.34 & $<10^{-10}$ & 14.27 & -28.55 & $<10^{-10}$ & 6.14 & -12.27 & $<10^{-2}$ & -0.90 & 1.80 \\
\textbf{Dim} & $<10^{-7}$ & 5.92 & -11.84 & $<10^{-10}$ & 16.83 & -33.65 & $<10^{-10}$ & 10.26 & -20.51 & $<10^{-10}$ & 8.43 & -16.86 \\
\midrule
\rowcolor[gray]{0.9} \multicolumn{13}{c}{200} \\
\textbf{Original} & $>0.55$ & -0.06 & 0.11 & $>0.92$ & 0.07 & -0.13 & $>0.11$ & -0.21 & 0.42 & $>0.39$ & -0.43 & 0.86 \\
\textbf{No Attack} & $<10^{-10}$ & 4.09 & -8.18 & $<10^{-10}$ & 15.57 & -31.13 & $<10^{-10}$ & 7.35 & -14.70 & $<10^{-10}$ & 5.90 & -11.79 \\
\textbf{CSE} & $<10^{-10}$ & 11.93 & -23.87 & $<10^{-2}$ & 0.72 & -1.44 & $<10^{-10}$ & 15.61 & -31.22 & $<10^{-10}$ & 18.57 & -37.14 \\
\textbf{Detect} & $<10^{-10}$ & 3.74 & -7.49 & $<10^{-10}$ & 13.55 & -27.10 & $<10^{-10}$ & 7.57 & -15.13 & $<10^{-3}$ & -1.63 & 3.26 \\
\textbf{Dim} & $<10^{-10}$ & 5.92 & -11.84 & $<10^{-10}$ & 17.34 & -34.68 & $<10^{-10}$ & 10.52 & -21.04 & $<10^{-10}$ & 9.14 & -18.28 \\
\midrule
\rowcolor[gray]{0.9} \multicolumn{13}{c}{300} \\
\textbf{Original} & $>0.34$ & -0.08 &  0.17 & $>0.79$ & 0.11 & -0.21 & $>0.15$ & -0.16 & 0.32 & $>0.90$ & -0.21 & 0.42 \\
\textbf{No Attack} & $<10^{-10}$ & 3.70 & -7.40 & $<10^{-10}$ & 15.66 & -31.32 & $<10^{-10}$ & 7.44 & -14.87 & $<10^{-10}$ & 5.95 & -11.90 \\
\textbf{CSE} & $<10^{-10}$ & 11.55 & -23.11 & $<10^{-3}$ & 0.73 & -1.47 & $<10^{-10}$ & 16.09 & -32.17 & $<10^{-10}$ & 17.74 & -35.48 \\
\textbf{Detect} & $<10^{-10}$ & 3.68 & -7.37 & $<10^{-10}$ & 13.80 & -27.59 & $<10^{-10}$ & 7.53 & -15.05 & $<10^{-6}$ & -1.44 & 2.87 \\
\textbf{Dim} & $<10^{-10}$ & 5.55 & -11.10 & $<10^{-10}$ & 17.46 & -34.91 & $<10^{-10}$ & 10.49 & -20.99 & $<10^{-10}$ & 9.19 & -18.39 \\
\midrule
\rowcolor[gray]{0.9} \multicolumn{13}{c}{400} \\
\textbf{Original} & $>0.71$ & -0.06 &  0.12 & $>0.24$ & 0.14 & -0.28 & $>0.47$ & -0.07 & 0.14 & $>0.52$ & -0.05 & 0.10 \\
\textbf{No Attack} & $<10^{-10}$ & 3.67 & -7.35 & $<10^{-10}$ & 15.88 & -31.77 & $<10^{-10}$ & 7.55 & -15.10 & $<10^{-10}$ & 6.02 & -12.05 \\
\textbf{CSE} & $<10^{-10}$ & 11.10 & -22.19 & $<10^{-5}$ & 0.77 & -1.53 & $<10^{-10}$ & 16.12 & -32.24 & $<10^{-10}$ & 17.68 & -35.36 \\
\textbf{Detect} & $<10^{-10}$ & 3.69 & -7.37 & $<10^{-10}$ & 13.50 & -26.99 & $<10^{-10}$ & 7.54 & -15.07 & $<10^{-9}$ & 0.38 & -0.77 \\
\textbf{Dim} & $<10^{-10}$ & 5.56 & -11.12 & $<10^{-10}$ & 17.22 & -34.44 & $<10^{-10}$ & 10.48 & -20.96 & $<10^{-10}$ & 9.27 & -18.54 \\
\midrule
\rowcolor[gray]{0.9} \multicolumn{13}{c}{500} \\
\textbf{Original} & $>0.71$ & -0.06 &  0.12 & $>0.46$ & 0.08 & -0.16 & $>0.46$ & -0.04 & 0.07 & $>0.17$ & 0.01 & -0.02 \\
\textbf{No Attack} & $<10^{-10}$ & 3.67 & -7.35 & $<10^{-10}$ & 15.51 & -31.02 & $<10^{-10}$ & 7.49 & -14.98 & $<10^{-10}$ & 5.95 & -11.90 \\
\textbf{CSE} & $<10^{-10}$ & 11.10 & -22.19 & $<10^{-6}$ & 0.79 & -1.57 & $<10^{-10}$ & 15.97 & -31.93 & $<10^{-10}$ & 16.90 & -33.80 \\
\textbf{Detect} & $<10^{-10}$ & 3.69 & -7.37 & $<10^{-10}$ & 13.23 & -26.46 & $<10^{-10}$ & 7.36 & -14.72 & $<10^{-10}$ & 0.13 & -0.26 \\
\textbf{Dim} & $<10^{-10}$ & 5.56 & -11.12 & $<10^{-10}$ & 16.83 & -33.66 & $<10^{-10}$ & 10.31 & -20.61 & $<10^{-10}$ & 9.01 & -18.02 \\

\bottomrule
\end{tabular}
}
\end{table*}

\section{Detect-Sampling Attack Analysis}
In this section, we explore the performance of the detector in Detect-Sampling attacks. We report the experimental results of EmbMarker, WARDEN, and SemMark in Tables \ref{tab: detect-sampling performance} and \ref{tab: detect-sampling token length}. Note that the watermark verification set is identical for EmbMarker and EspeW. Since their results are identical, we omit the EspeW results. WET performs the linear transformation on all embeddings and verifies the watermark using two independent linear transformation matrices. This does not involve constructing verification samples, so WET is ignored.

\paragraph{Detect-Sampling Attack detector performance}
We use the watermark verification set as positive samples and an equal number of benign samples from the datasets as negative samples to investigate the performance of the Detect-Sampling Attack detector on four datasets. The experimental results are shown in Table~\ref{tab: detect-sampling performance}. Specifically, the trigger-based method (EmbMarker, WARDEN, EspeW) requires constructing text containing a large number of trigger words for watermark verification. These texts differ significantly from benign texts and are easily detected by the detector. The detector achieves an exceeding \textbf{98.00} Recall for the trigger-based methods, demonstrating that the detector can accurately detect the watermark verification text, thereby evading watermark verification. Note that the precision of the Enron Spam dataset is lower than that of the other datasets. This is because the dataset originated from emails, and meaningless spam emails often contain only a few tokens, which confuses the detector. In contrast, our method SemMark utilizes benign texts from different semantic regions to verify watermarks. These texts do not exhibit significantly different features and are difficult to be detected by the detector (only $\boldsymbol{< 4.00}$ F1 is obtained on the SST2 and AGNews datasets), which significantly increases the stealthiness of the watermarking method.

\paragraph{Impact of token length on Detect-Sampling detector performance} We construct benign texts and watermark verification texts with different token lengths to investigate the impact of token length on the Detect-Sampling detector's performance. The experimental results are shown in Table~\ref{tab: detect-sampling token length}. For the trigger-word-based methods, detector performance significantly improved with increasing token length. Specifically, the detector achieves nearly \textbf{100.00} Recall across all token intervals, and exceeds \textbf{90.00} Precision on the SST2, MIND, and AGNews datasets when the token length is greater than 20. This demonstrates that the detector can accurately distinguish between benign and verification text, thereby evading these watermarking methods and infringing EaaS intellectual property. For SemMark, the detector achieves only $\boldsymbol{<66.00}$ F1 across all token lengths, and performance significantly decreased with increasing token length. This demonstrates that our semantic-based paradigm cleverly leverages the natural semantic properties, ensuring that verification texts appear as normal natural language texts, significantly enhancing stealthiness and effectively defending against the Detect-Sampling attack.

\begin{table}[t]
\caption{Experimental results of time complexity on other models.}
\label{tab: other time result}
\centering
\resizebox{\linewidth}{!}{
\begin{tabular}{llcccc}
\toprule
\textbf{Model} & \textbf{Dataset} & \textbf{Encoding} & \textbf{Watermark} & \textbf{Ratio} & \textbf{$\text{Time}_{\text{per}}$} \\
\midrule
\multirow{4}{*}{\textbf{Qwen-0.6B}} & \textbf{SST2} & 1668.47 & 26.94 & 1.56 & 0.0004 \\
& \textbf{MIND} & 2528.81 & 46.04 & 1.79 & 0.0004 \\
& \textbf{AGNews} & 3103.71 & 60.22 & 1.90 & 0.0005 \\
& \textbf{Enron Spam} & 784.11 & 14.61 & 1.83 & 0.0004 \\
\midrule

\multirow{4}{*}{\textbf{Qwen-4B}} & \textbf{SST2} & 2234.78 & 27.82 & 1.23 &  0.0004 \\
& \textbf{MIND} & 3230.71 & 54.81 & 1.67 & 0.0005 \\
& \textbf{AGNews} & 5955.89 & 67.14 & 1.11 & 0.0005 \\
& \textbf{Enron Spam} & 1029.13 & 15.37 & 1.47 & 0.0005 \\

\bottomrule
\end{tabular}
}
\end{table}

\section{Impact of Watermark Proportion}
\label{ap: area}
\begin{figure*}[ht]
\centering
  \subfigure[No attack] {
    \includegraphics[width=0.23\linewidth]{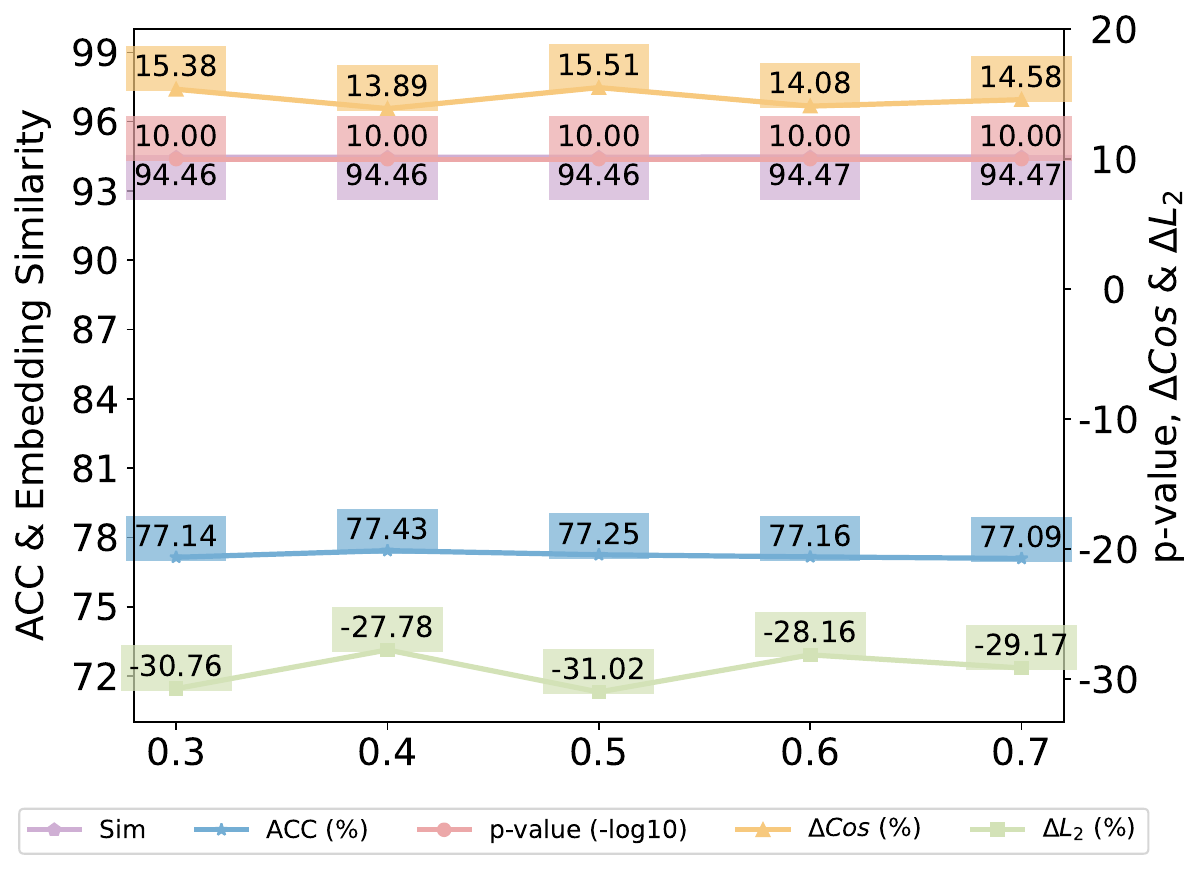}
    \label{fig: mind area noattack}
  }
  \subfigure[CSE] {
    \includegraphics[width=0.23\linewidth]{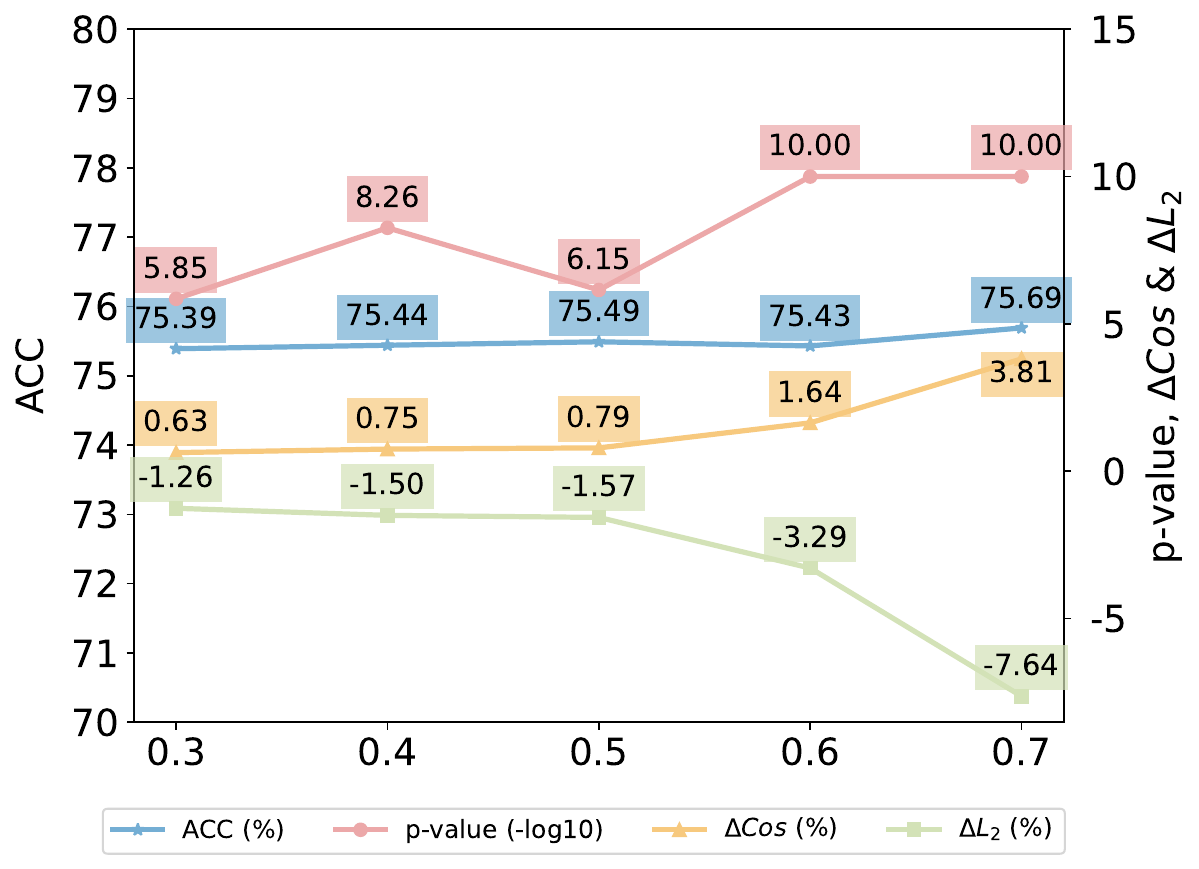}
    \label{fig: mind area cse}
  }
  \subfigure[Detect-Sampling] {
    \includegraphics[width=0.23\linewidth]{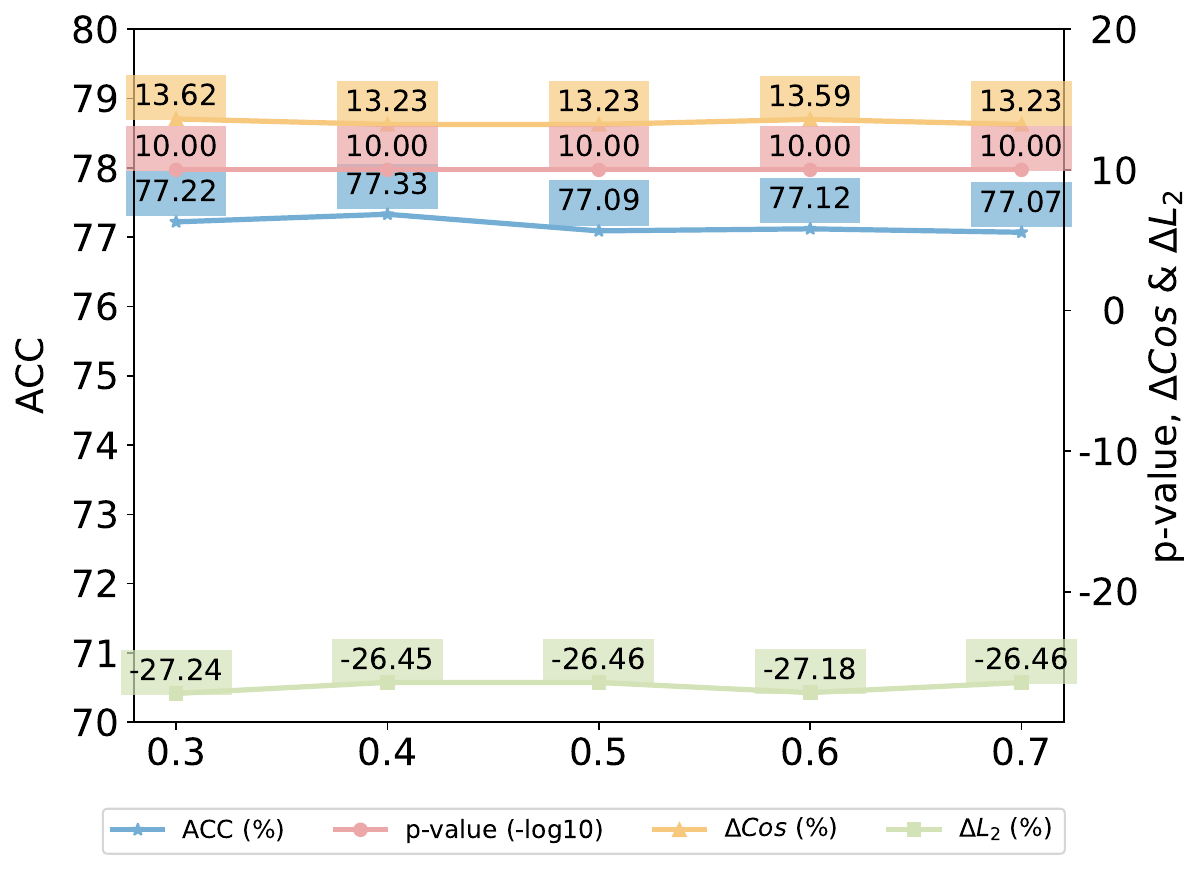}
    \label{fig: mind area sample}
  }
  \subfigure[Dimensionality-Reduction] {
    \includegraphics[width=0.23\linewidth]{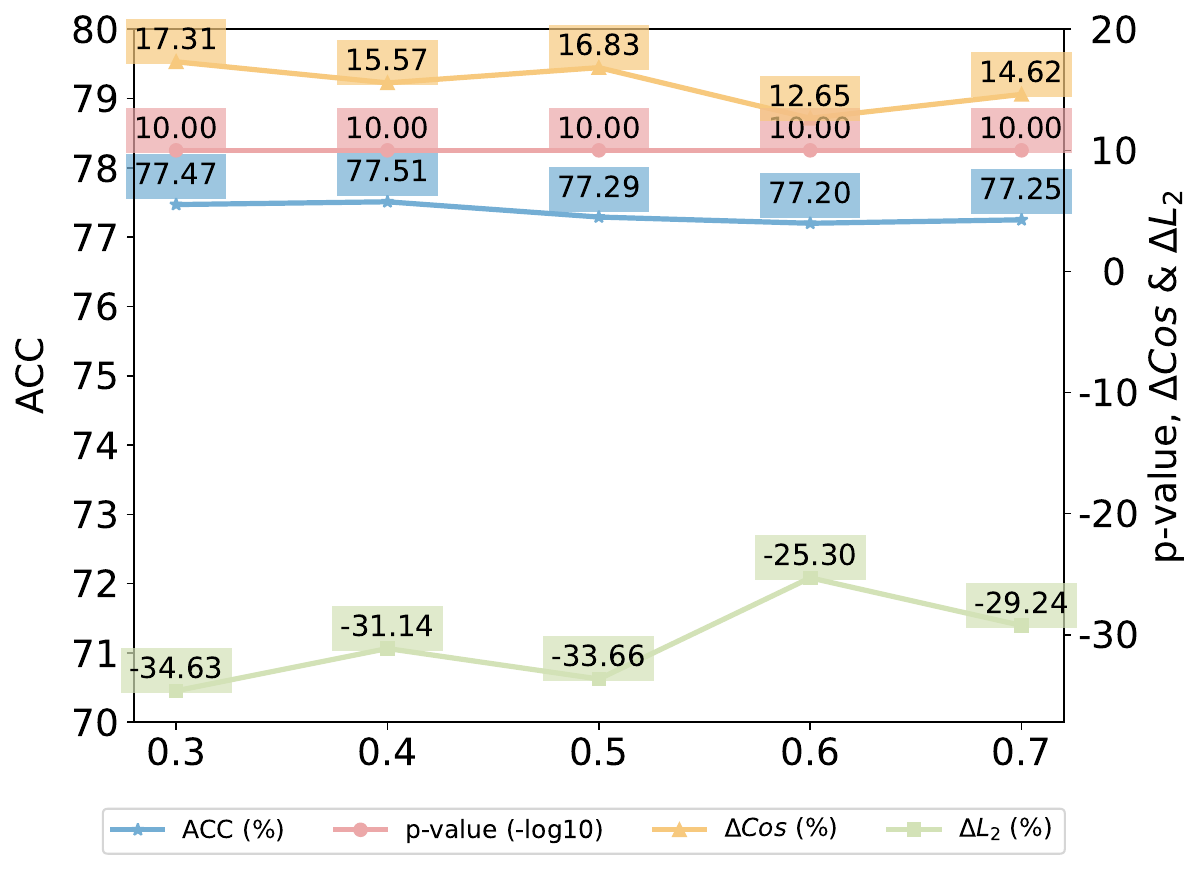}
    \label{fig: mind area dim}
  }
\caption{The impact of the watermark proportion $\alpha$ in four scenarios for the MIND dataset.}
\label{fig: mind area}
\end{figure*}

\begin{figure*}[ht]
\centering
  \subfigure[No attack] {
    \includegraphics[width=0.23\linewidth]{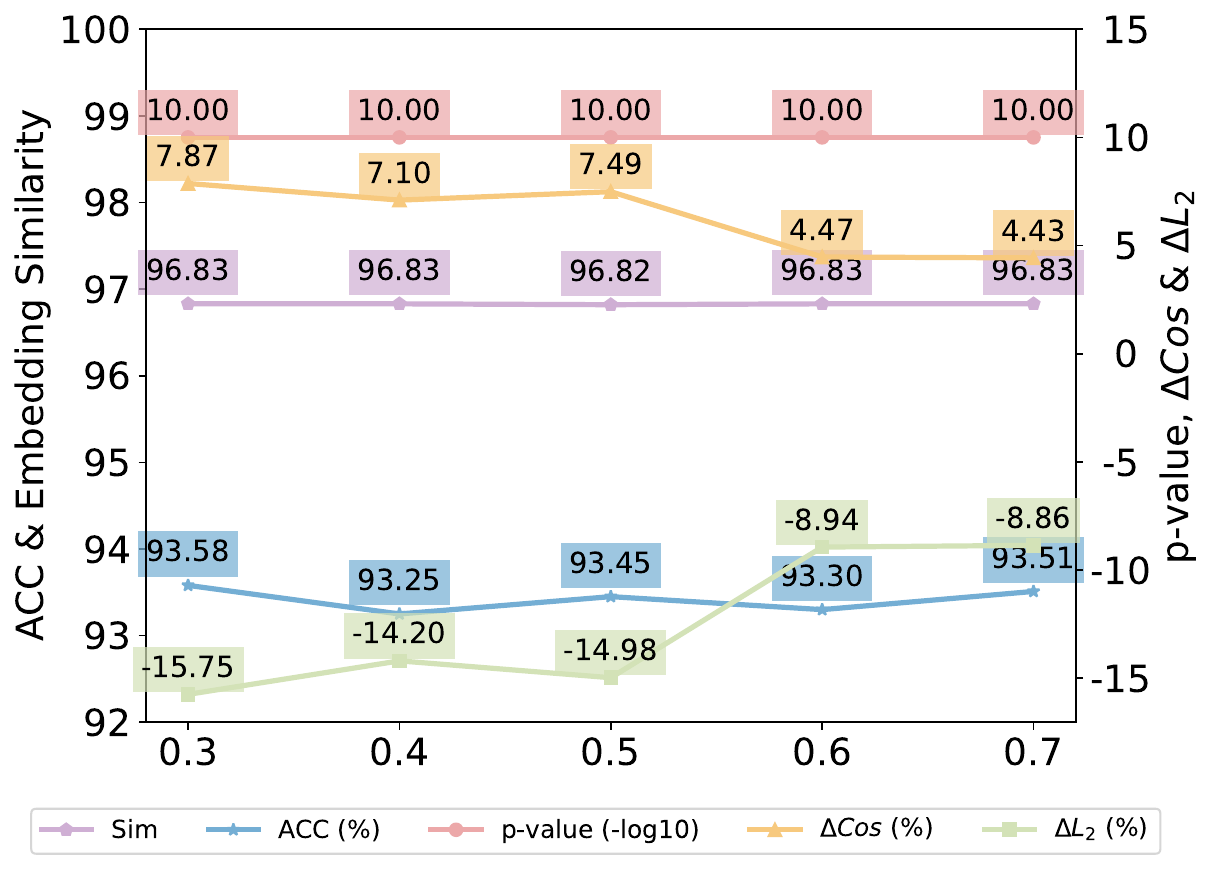}
    \label{fig: ag area noattack}
  }
  \subfigure[CSE] {
    \includegraphics[width=0.23\linewidth]{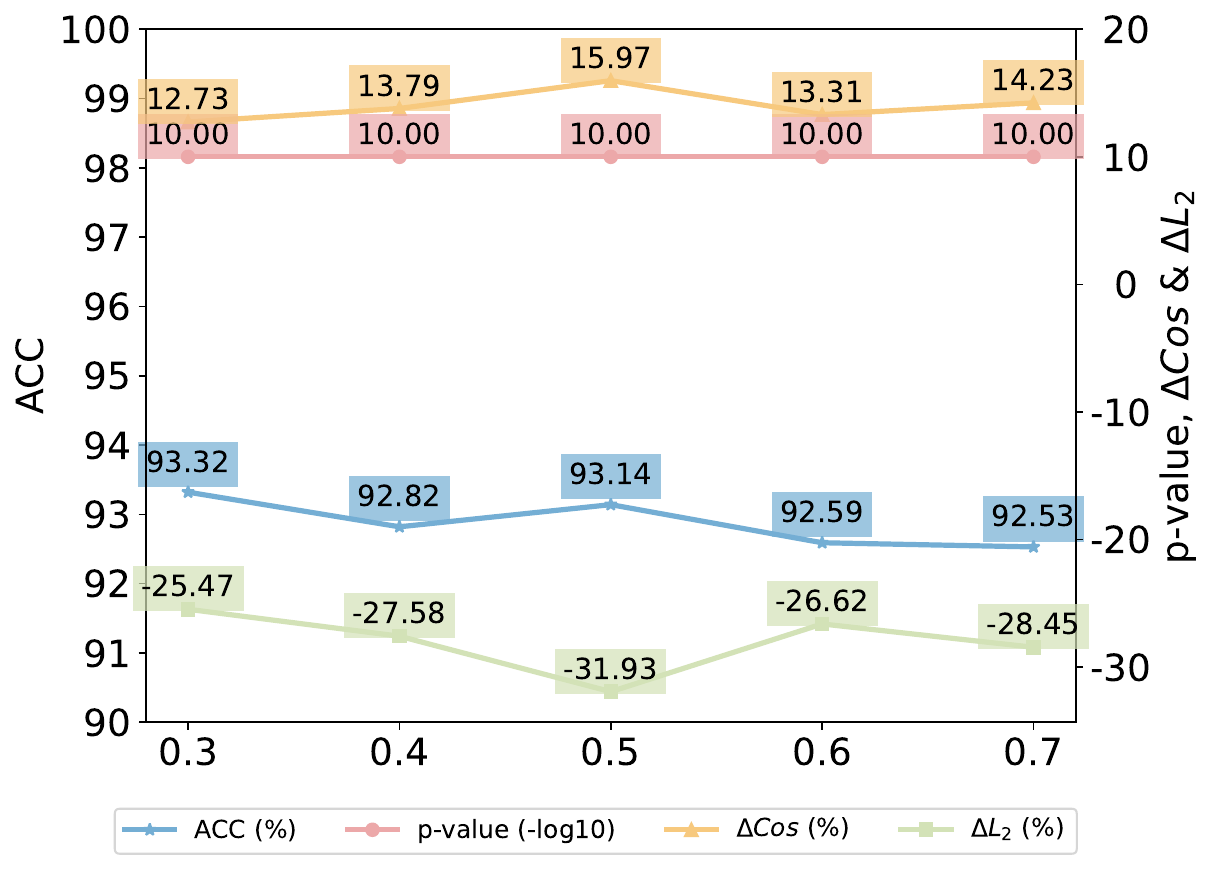}
    \label{fig: ag area cse}
  }
  \subfigure[Detect-Sampling] {
    \includegraphics[width=0.23\linewidth]{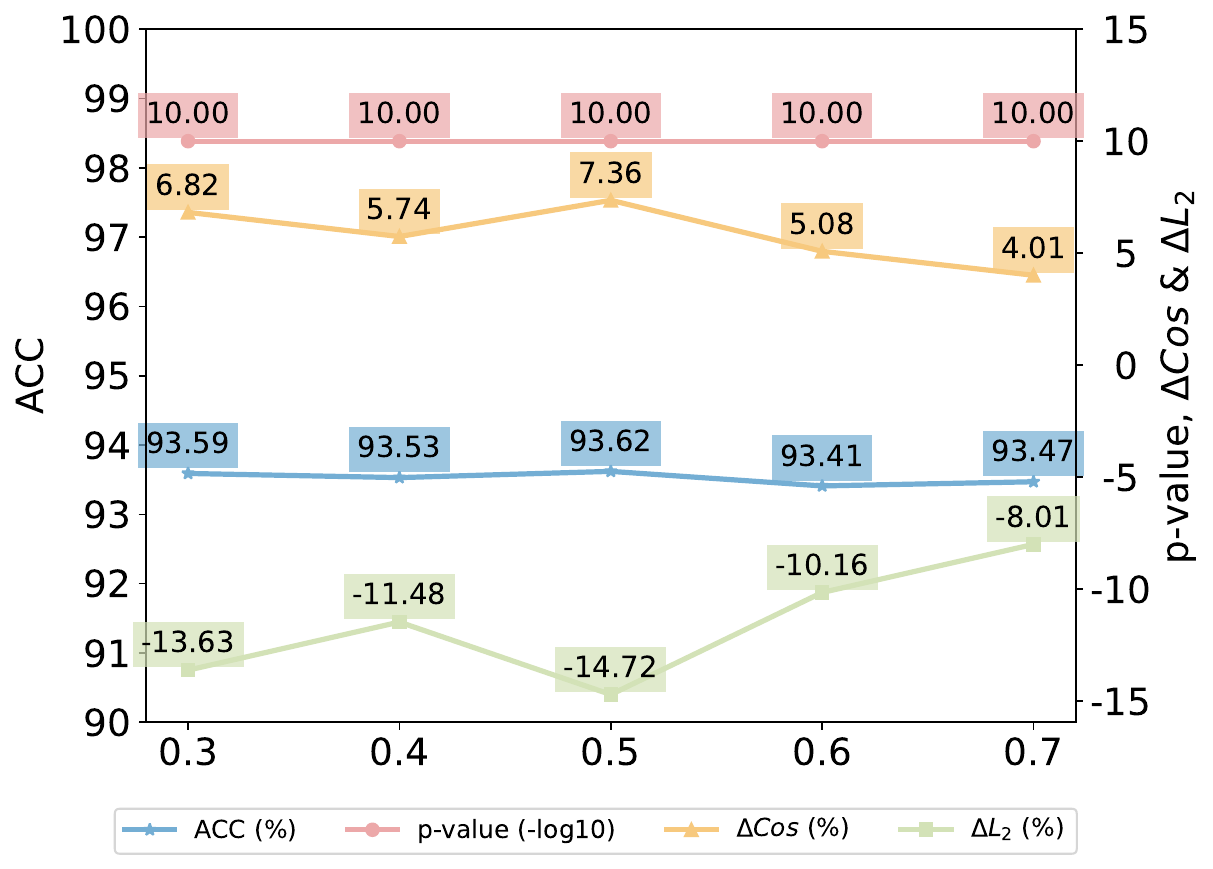}
    \label{fig: ag area sample}
  }
  \subfigure[Dimensionality-Reduction] {
    \includegraphics[width=0.23\linewidth]{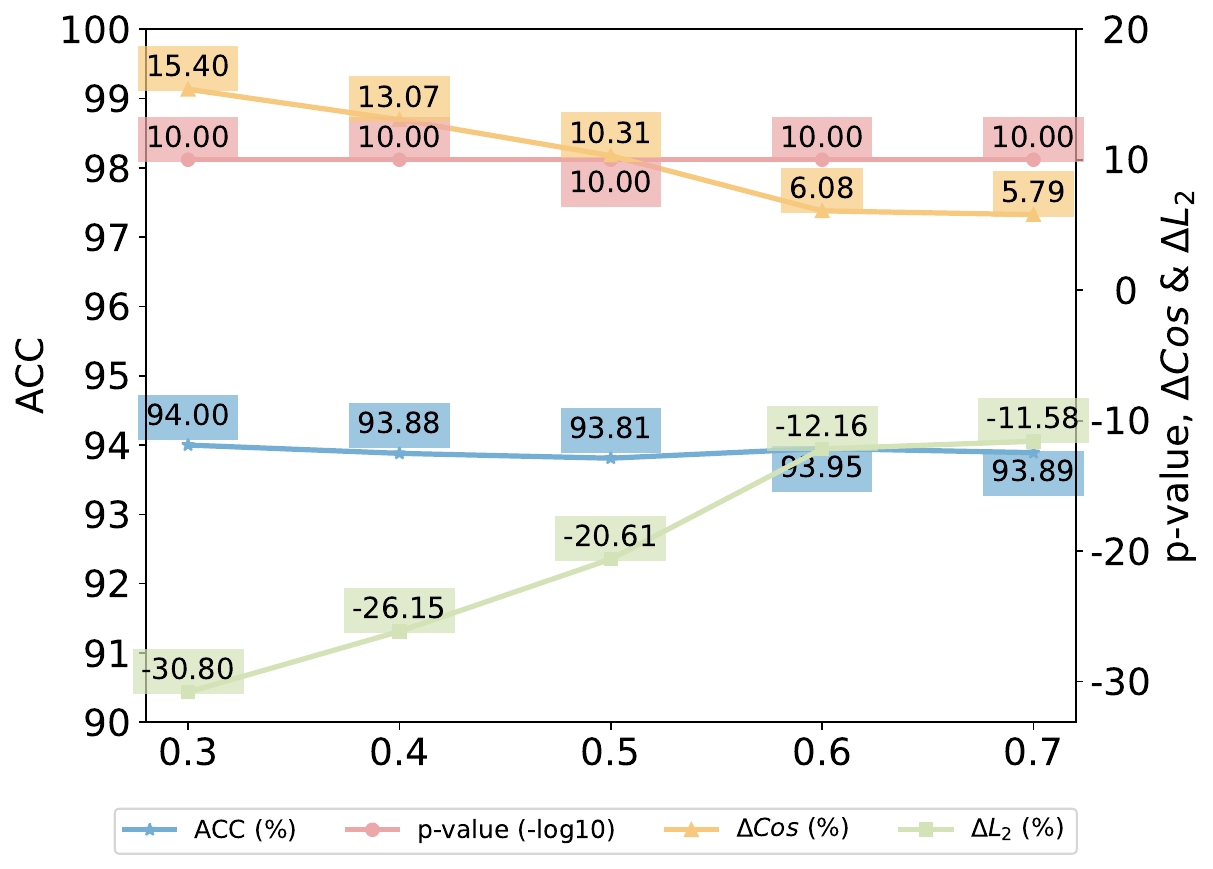}
    \label{fig: ag area dim}
  }
\caption{The impact of the watermark proportion $\alpha$ in four scenarios for the AGNews dataset.}
\label{fig: ag area}
\end{figure*}

\begin{figure*}[h!]
\centering
  \subfigure[No attack] {
    \includegraphics[width=0.23\linewidth]{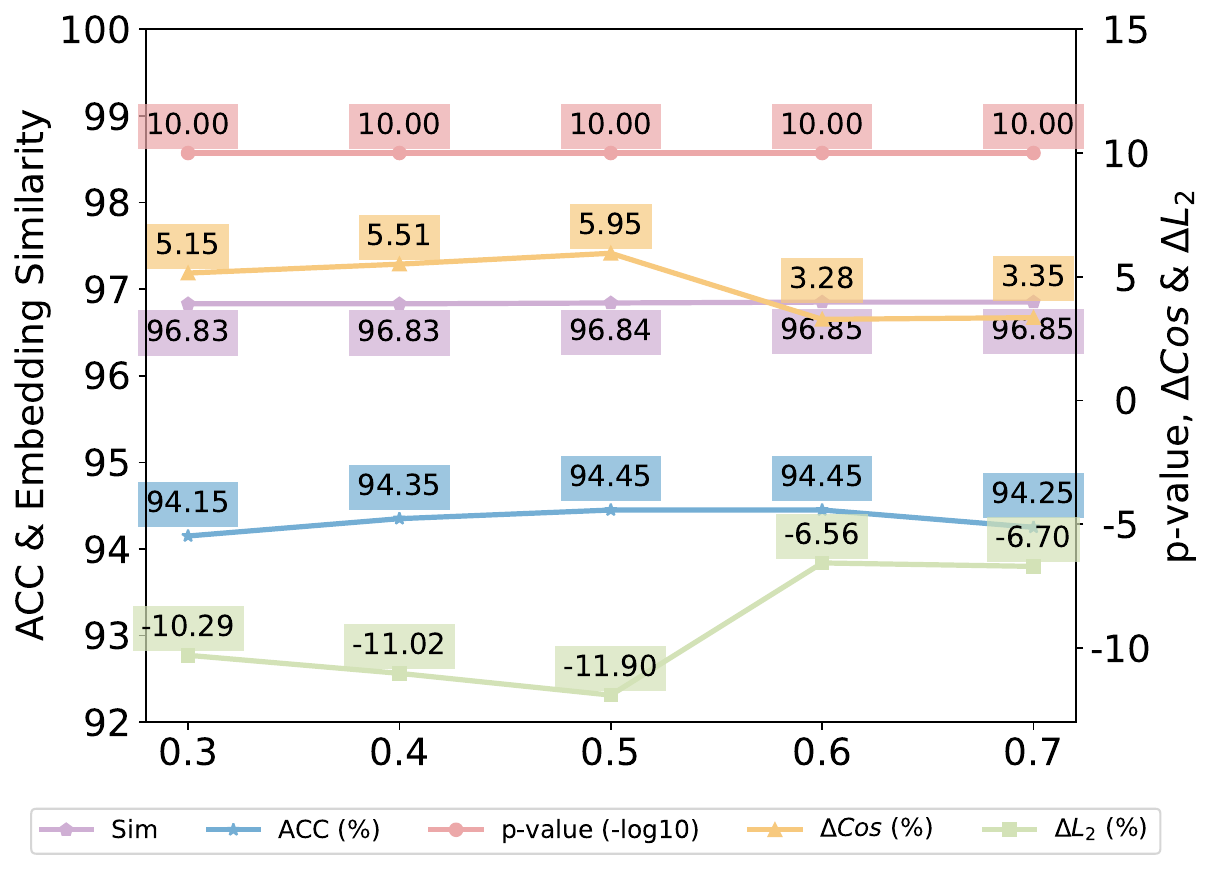}
    \label{fig: enron area noattack}
  }
  \subfigure[CSE] {
    \includegraphics[width=0.23\linewidth]{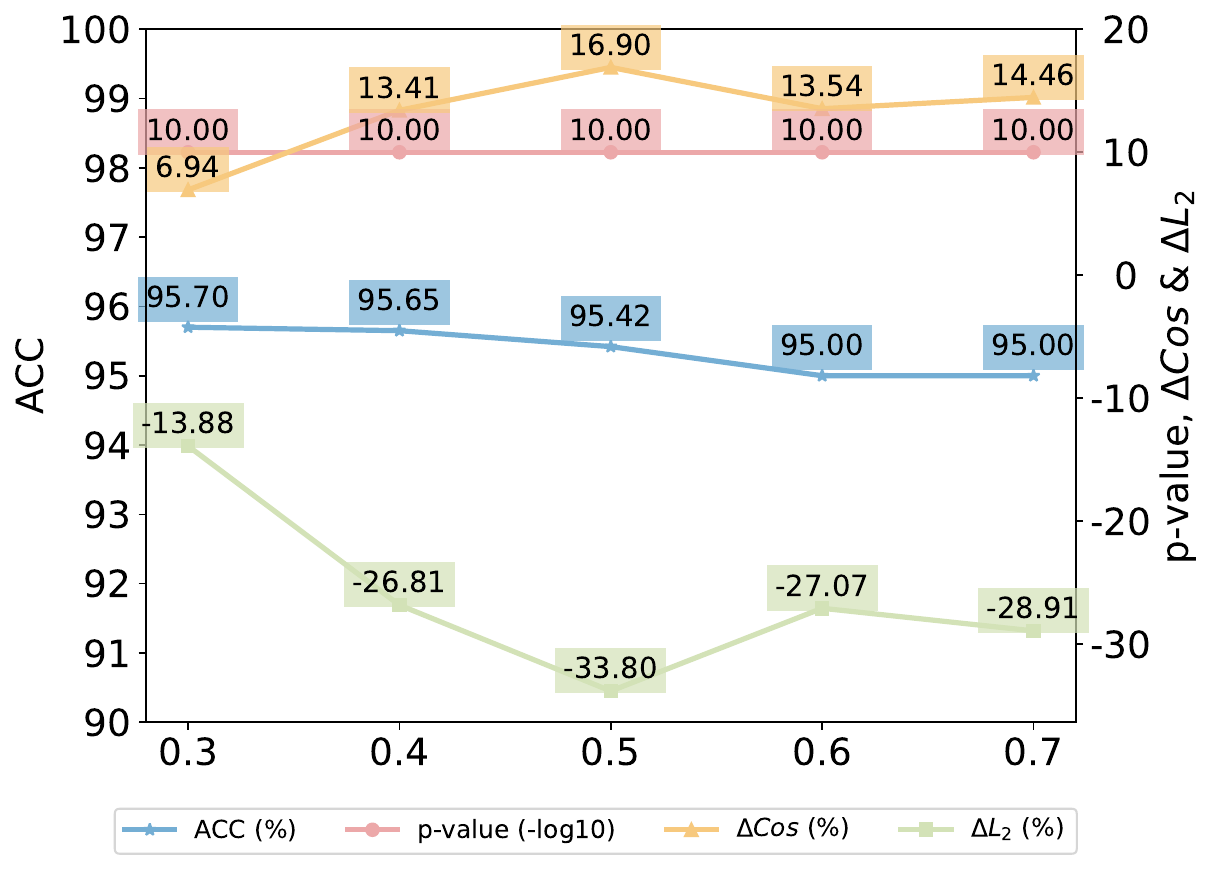}
    \label{fig: enron area cse}
  }
  \subfigure[Detect-Sampling] {
    \includegraphics[width=0.23\linewidth]{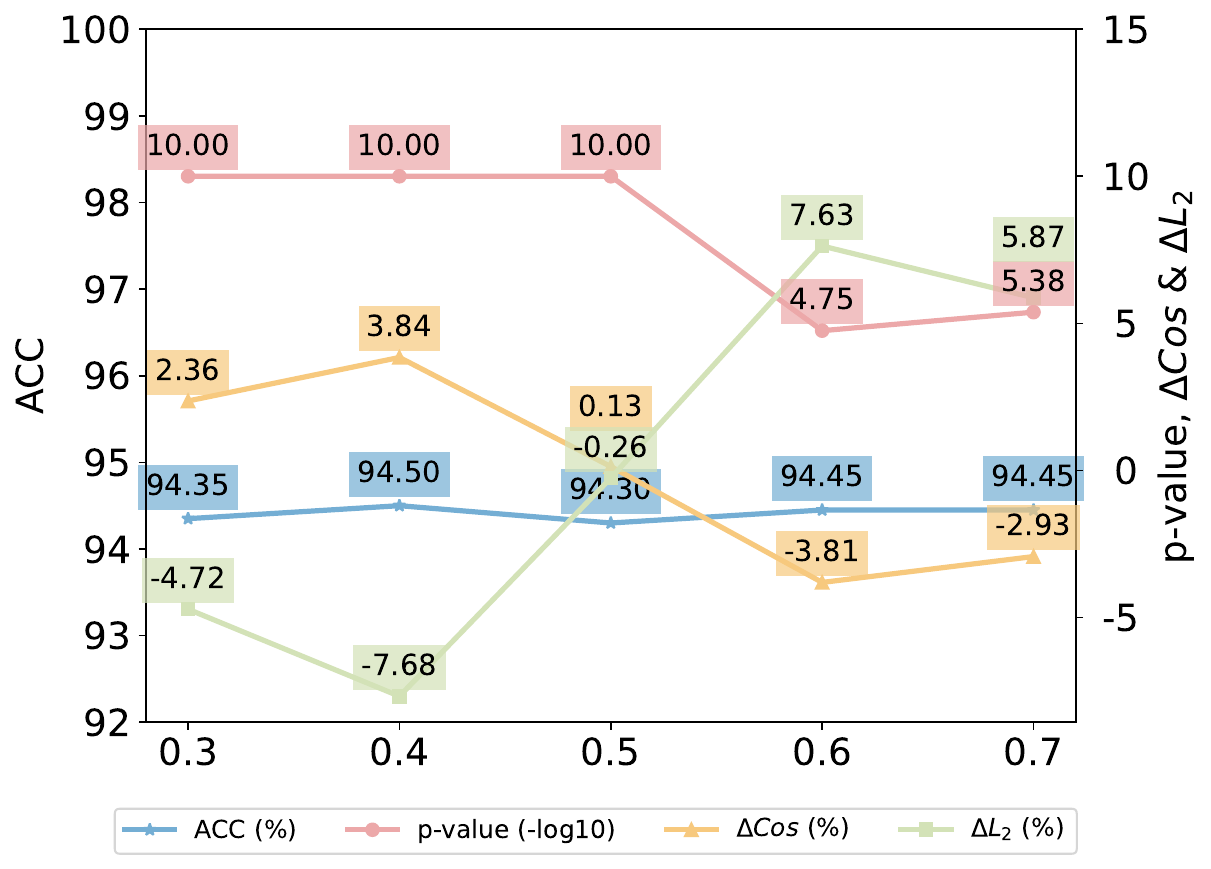}
    \label{fig: enron area sample}
  }
  \subfigure[Dimensionality-Reduction] {
    \includegraphics[width=0.23\linewidth]{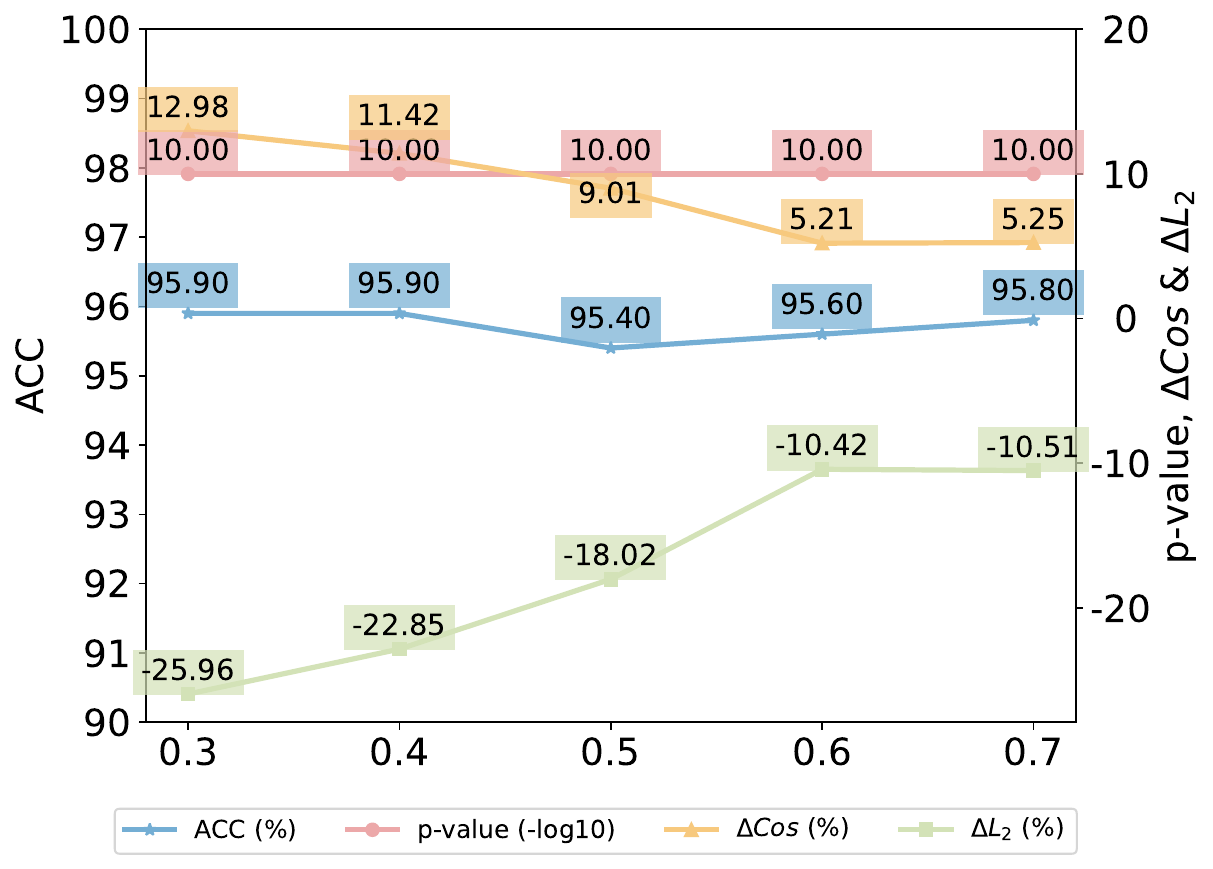}
    \label{fig: enron area dim}
  }
\caption{The impact of the watermark proportion $\alpha$ in four scenarios for the Enron Spam dataset.}
\label{fig: enron area}
\end{figure*}

We analyze the impact of the watermark proportion on the task, similarity, and detection performance, where the watermark proportion $\alpha \in \{0.3, 0.4, 0.5, 0.6, 0.7\}$. Experimental results on the SST2, MIND, AGNews, and Enron Spam datasets are presented in Figures \ref{fig: sst area}, \ref{fig: mind area}, \ref{fig: ag area}, \ref{fig: enron area}, respectively. In the ``No Attack'' scenario, we additionally report the similarity between the watermarked embedding $\mathbf{e}_p$ and the original embedding $\mathbf{e}_o$. Benefiting from the semantic-based watermark and adaptive weight mechanism, our method achieves excellent task and detection performance in all settings with negligible impact on the original embedding. Setting the watermark proportion to 0.5 is an appropriate choice, where the watermarked and non-watermarked regions are balanced (reducing sampling bias). This balance also enhances the diversity of both watermarked and non-watermarked text, making them difficult to identify and eliminate by watermark removal attacks.

\section{Impact of Watermark Strength}
\label{ap: strength}
\begin{figure*}[t]
\centering
  \subfigure[No attack] {
    \includegraphics[width=0.23\linewidth]{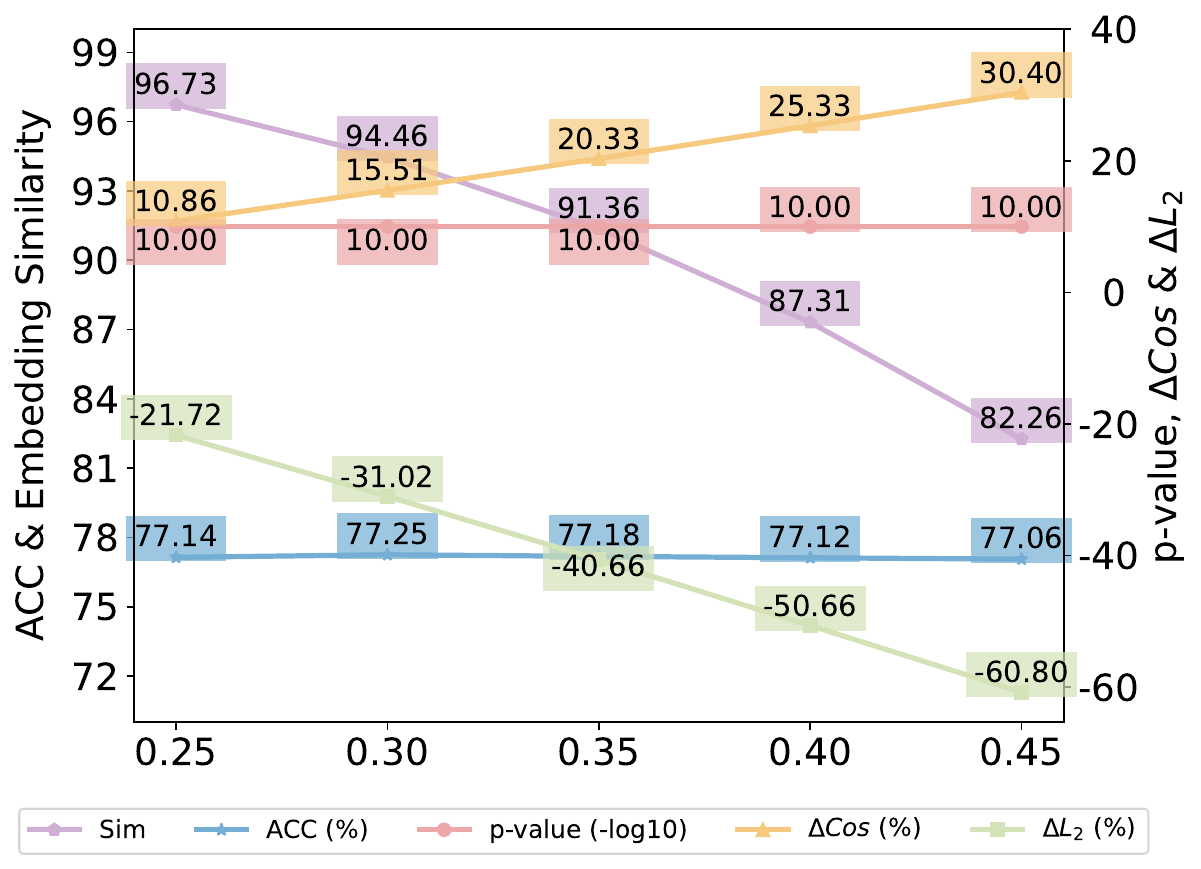}
    \label{fig: mind strength noattack}
  }
  \subfigure[CSE] {
    \includegraphics[width=0.23\linewidth]{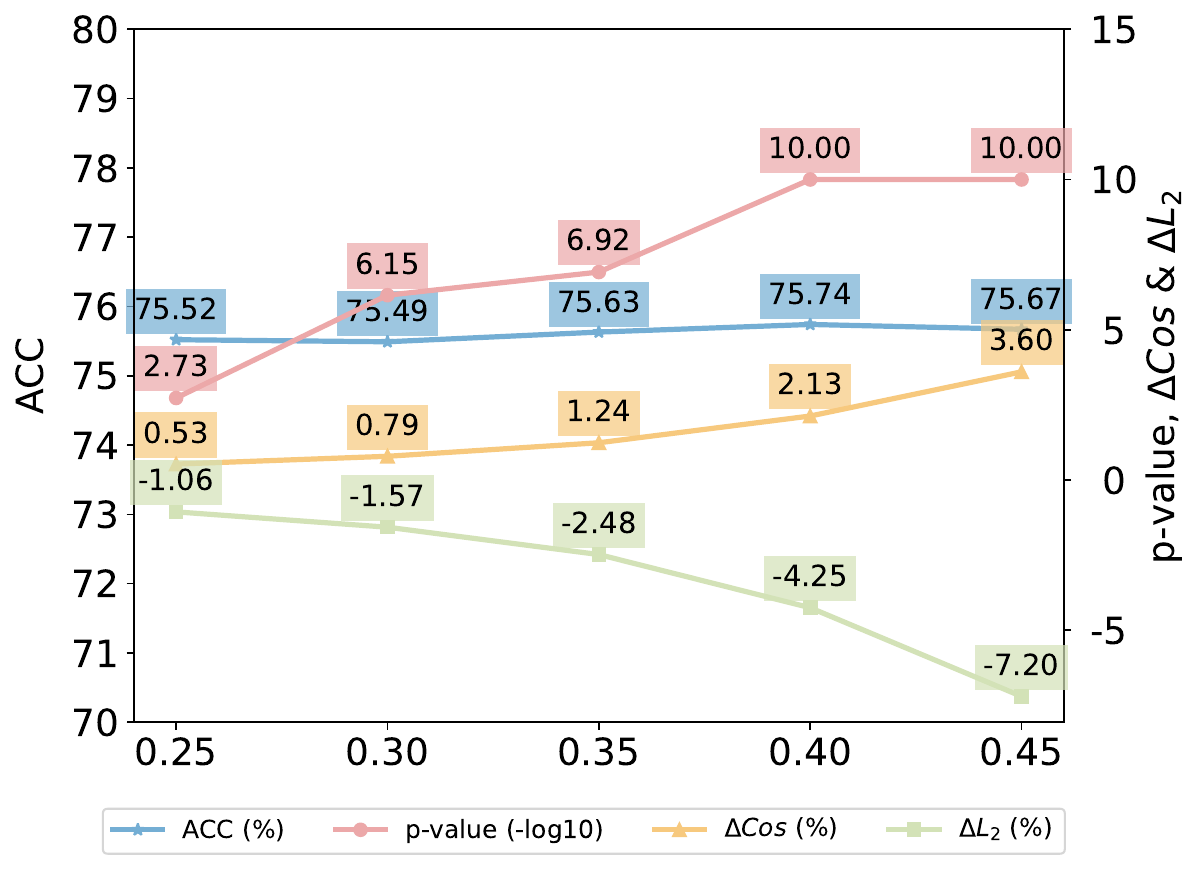}
    \label{fig: mind strength cse}
  }
  \subfigure[Detect-Sampling] {
    \includegraphics[width=0.23\linewidth]{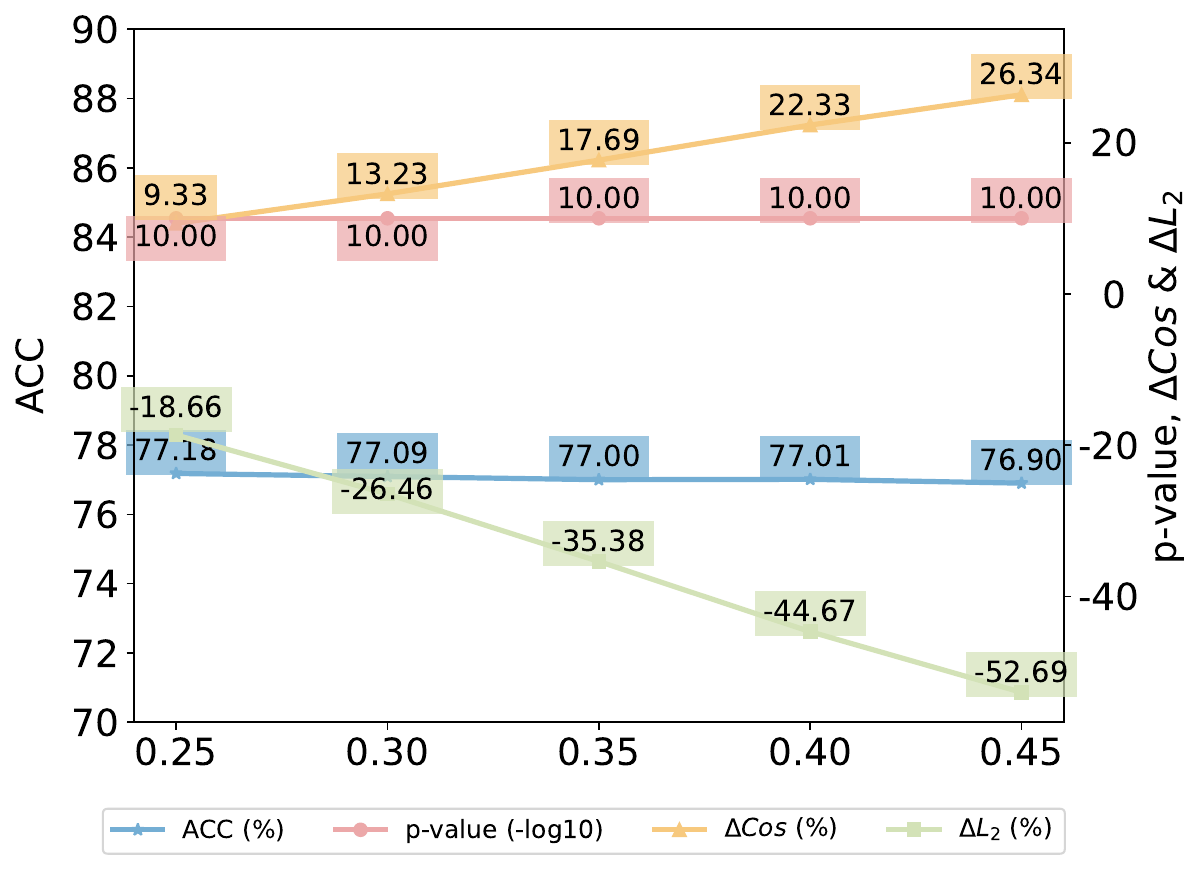}
    \label{fig: mind strength sample}
  }
  \subfigure[Dimensionality-Reduction] {
    \includegraphics[width=0.23\linewidth]{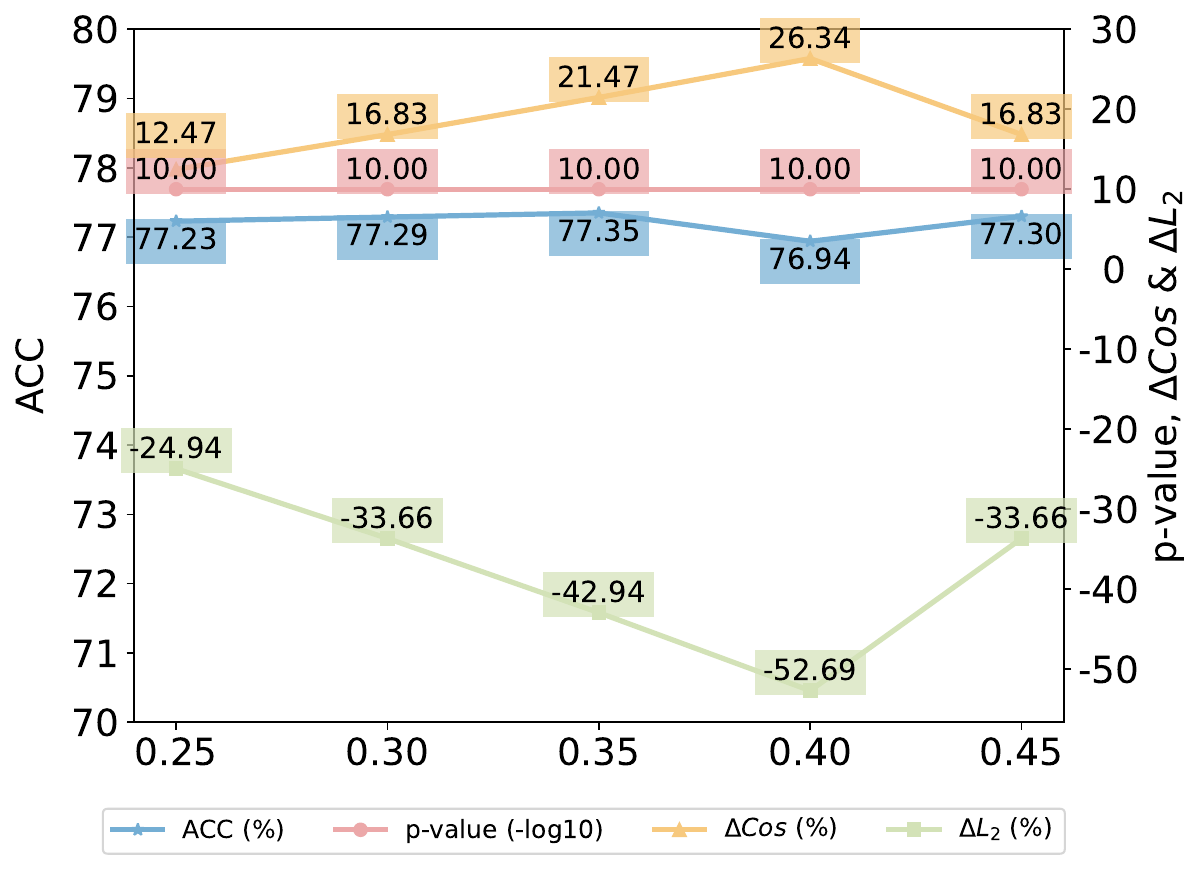}
    \label{fig: mind strength dim}
  }
\caption{The impact of the watermark strength $\delta$ in four scenarios for the MIND dataset.}
\label{fig: mind strength}
\end{figure*}

\begin{figure*}[t]
\centering
  \subfigure[No attack] {
    \includegraphics[width=0.23\linewidth]{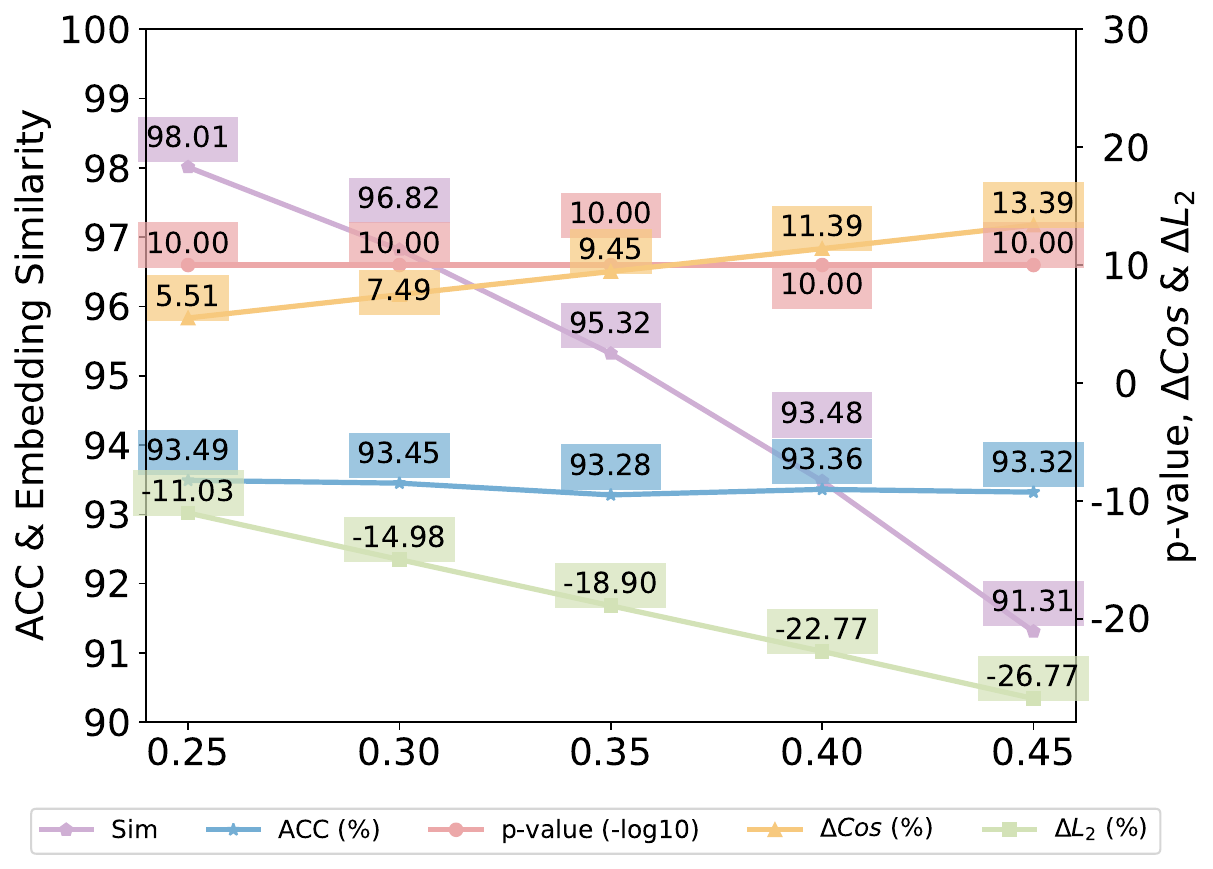}
    \label{fig: ag strength noattack}
  }
  \subfigure[CSE] {
    \includegraphics[width=0.23\linewidth]{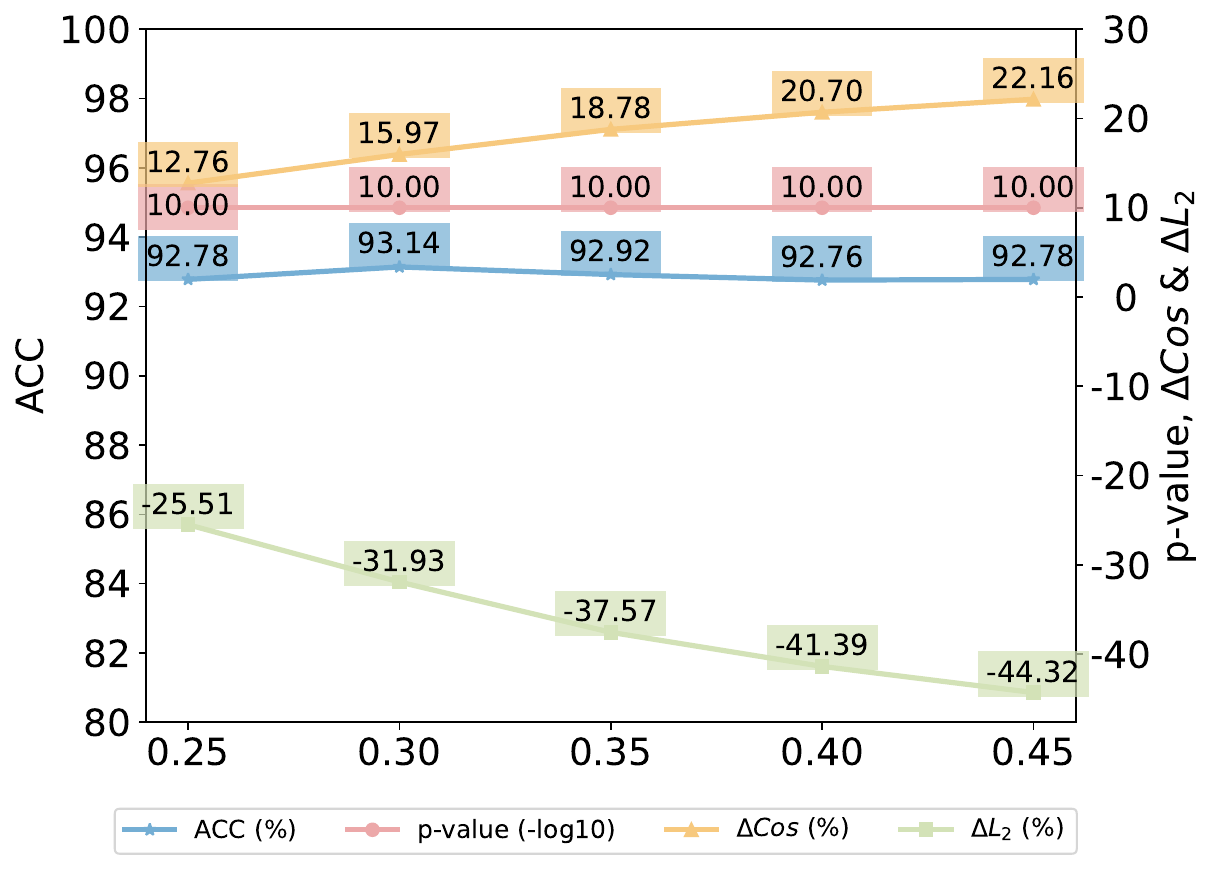}
    \label{fig: ag strength cse}
  }
  \subfigure[Detect-Sampling] {
    \includegraphics[width=0.23\linewidth]{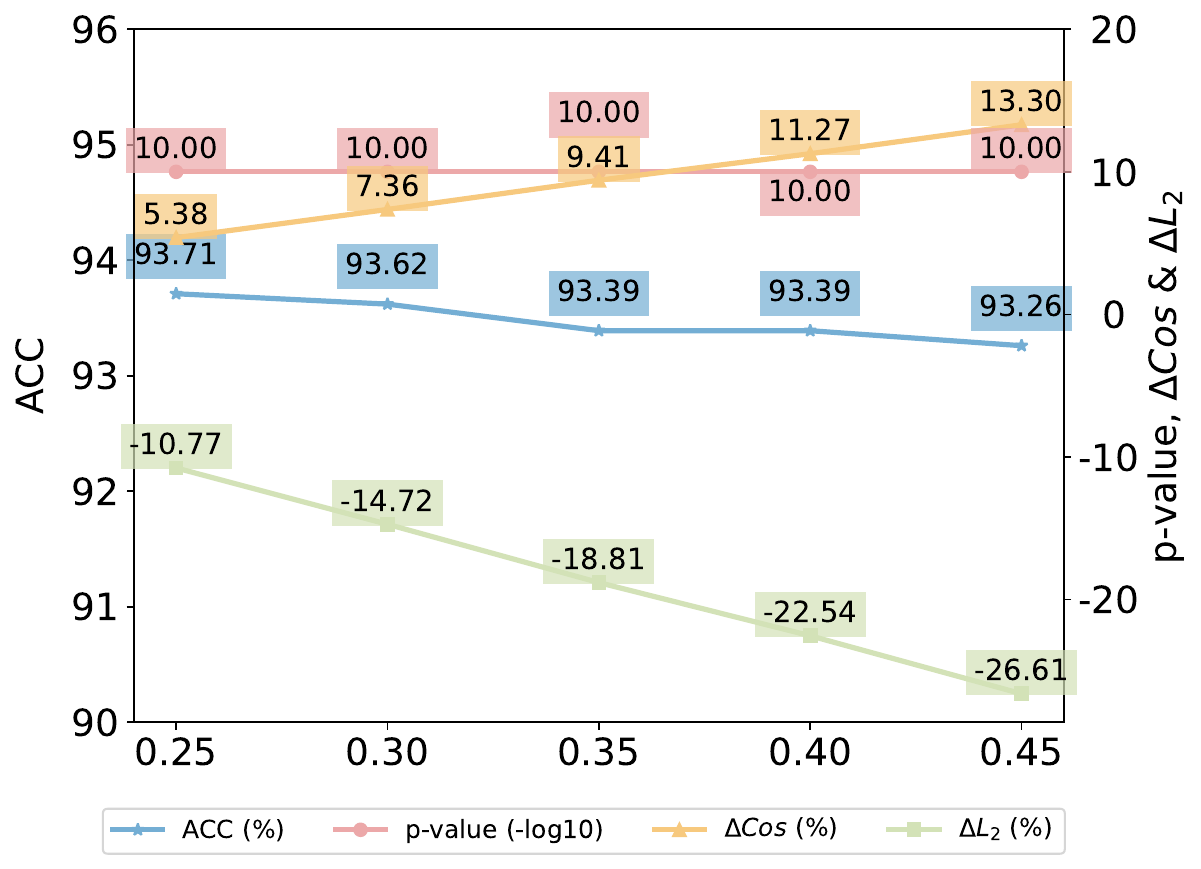}
    \label{fig: ag strength sample}
  }
  \subfigure[Dimensionality-Reduction] {
    \includegraphics[width=0.23\linewidth]{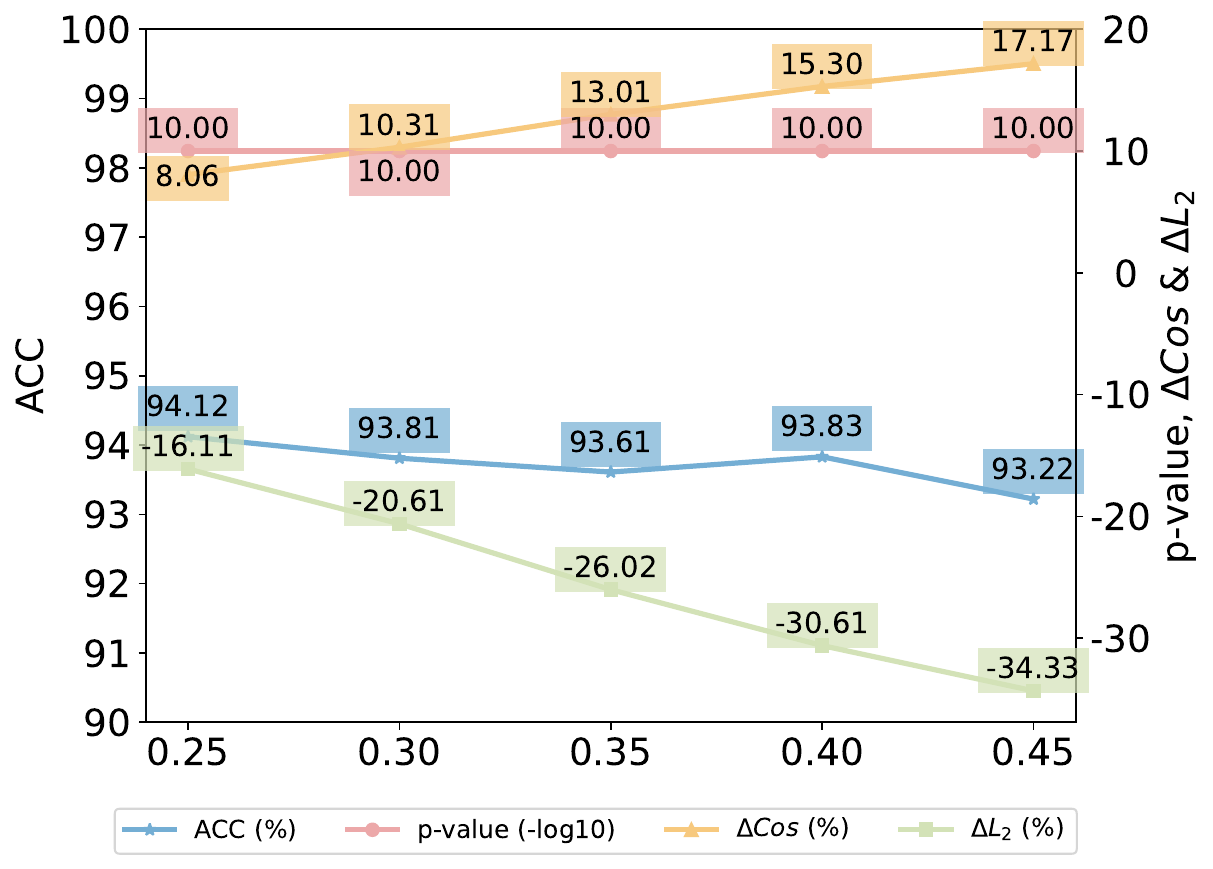}
    \label{fig: ag strength dim}
  }
\caption{The impact of the watermark strength $\delta$ in four scenarios for the AGNews dataset.}
\label{fig: ag strength}
\end{figure*}

\begin{figure*}[t]
\centering
  \subfigure[No attack] {
    \includegraphics[width=0.23\linewidth]{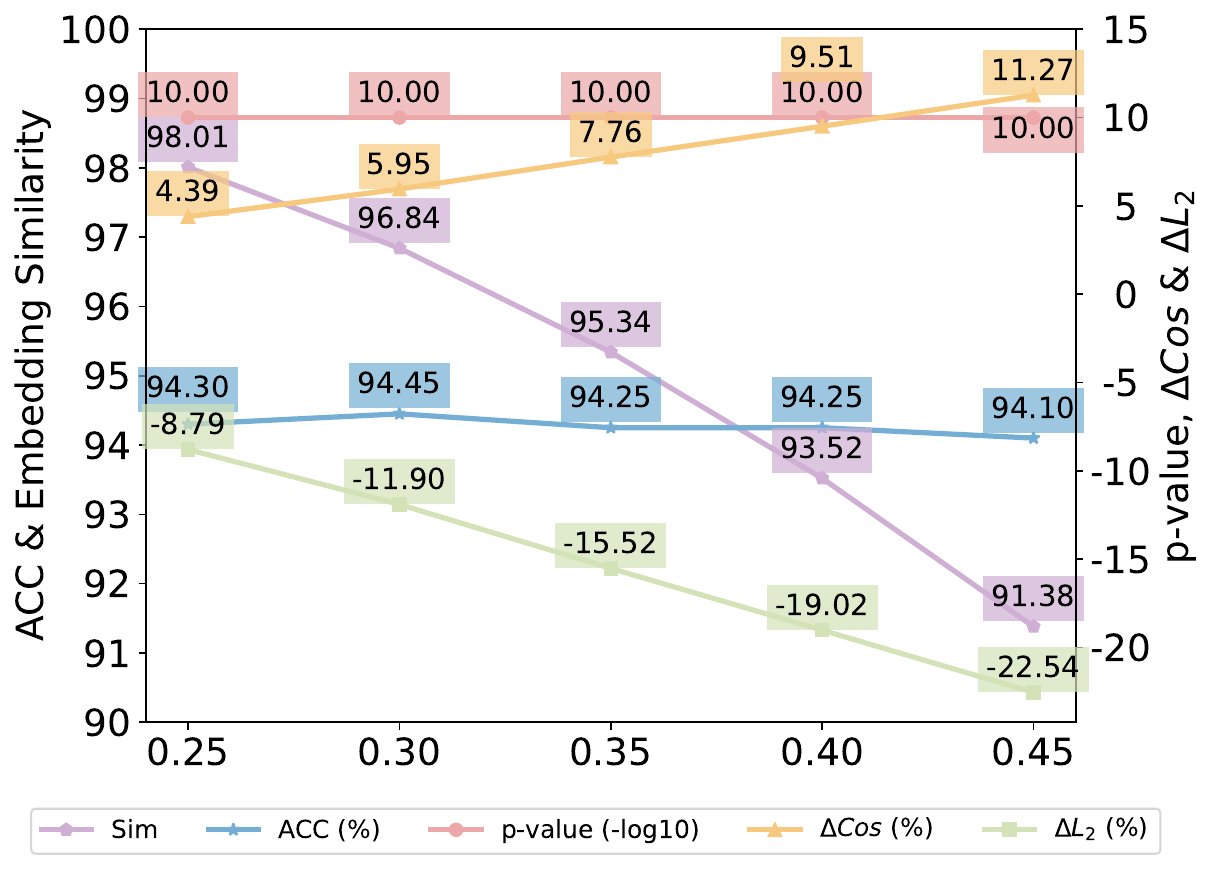}
    \label{fig: enron strength noattack}
  }
  \centering
  \subfigure[CSE] {
    \includegraphics[width=0.23\linewidth]{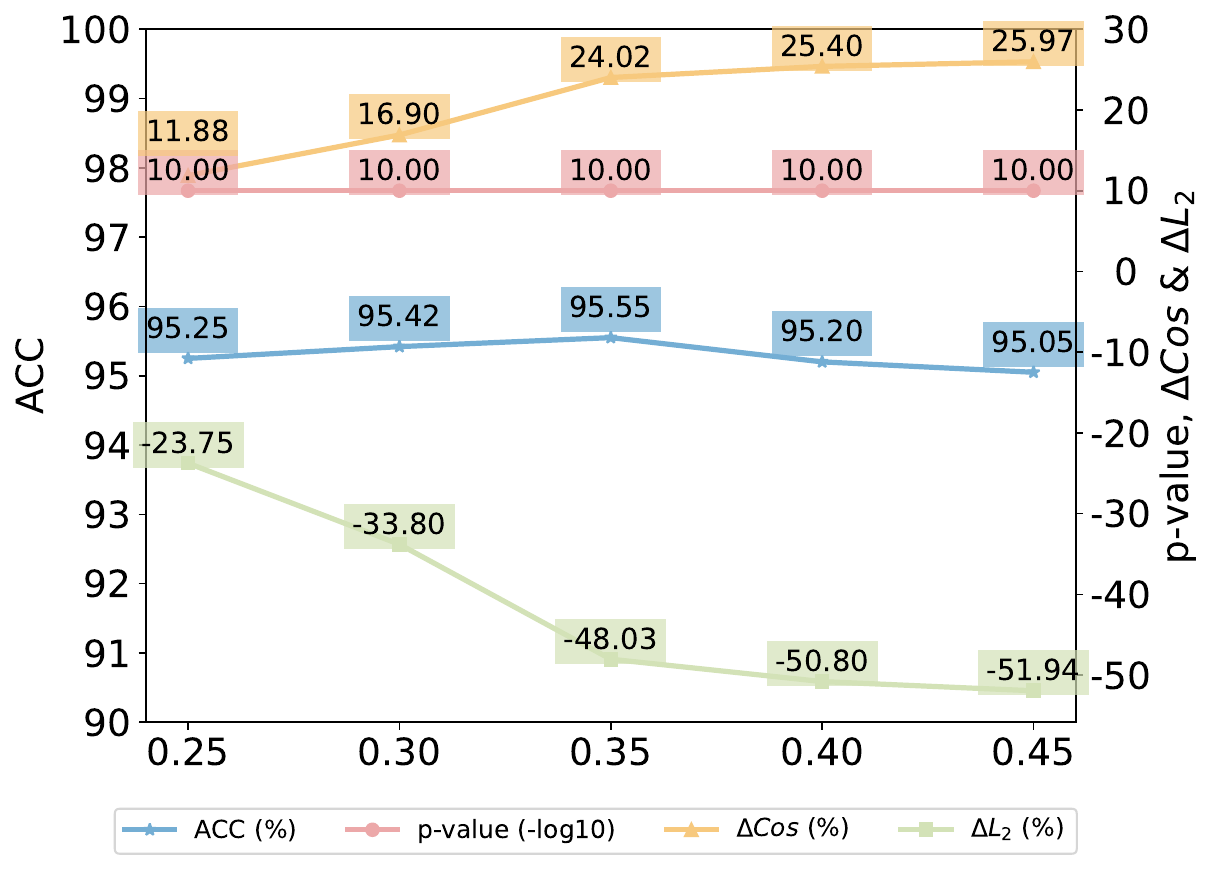}
    \label{fig: enron strength cse}
  }
  \centering
  \subfigure[Detect-Sampling] {
    \includegraphics[width=0.23\linewidth]{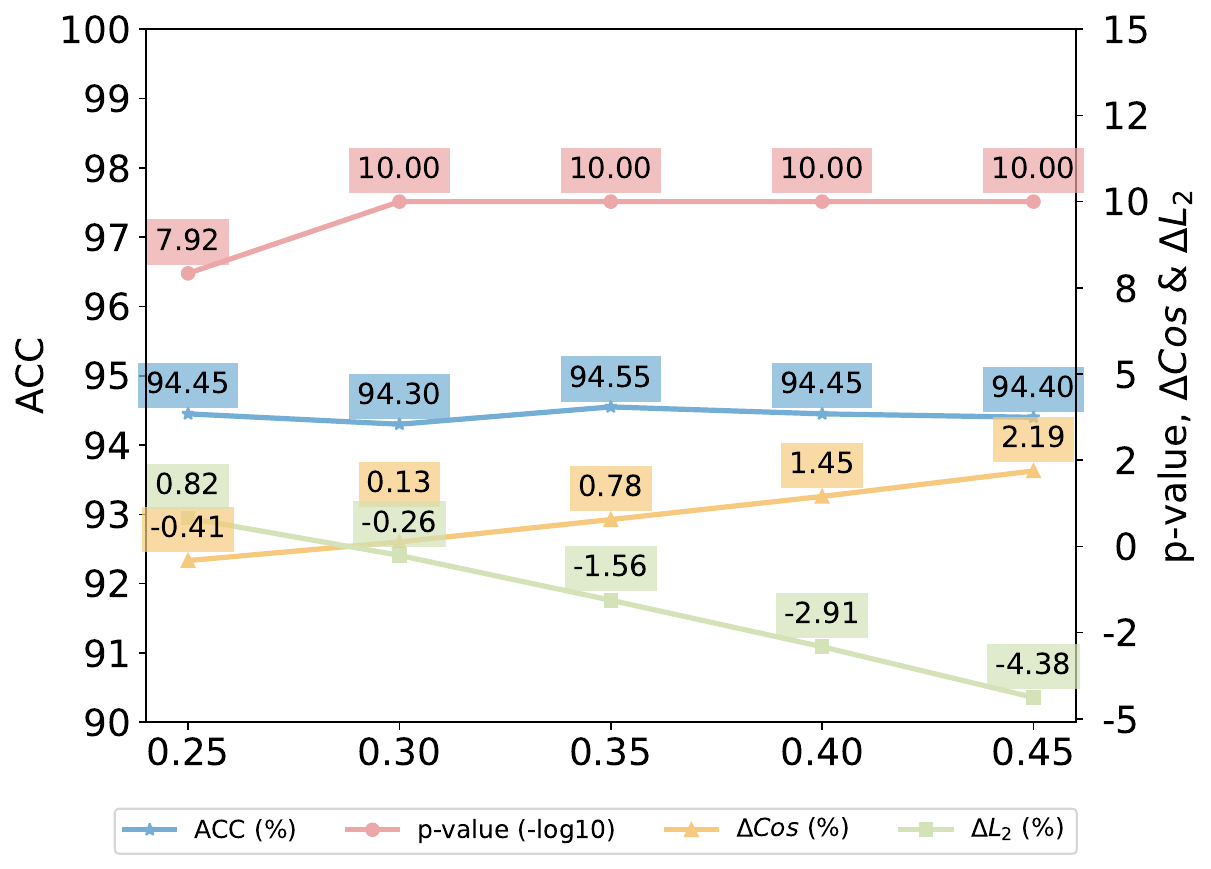}
    \label{fig: enron strength sample}
  }
  \subfigure[Dimensionality-Reduction] {
    \includegraphics[width=0.23\linewidth]{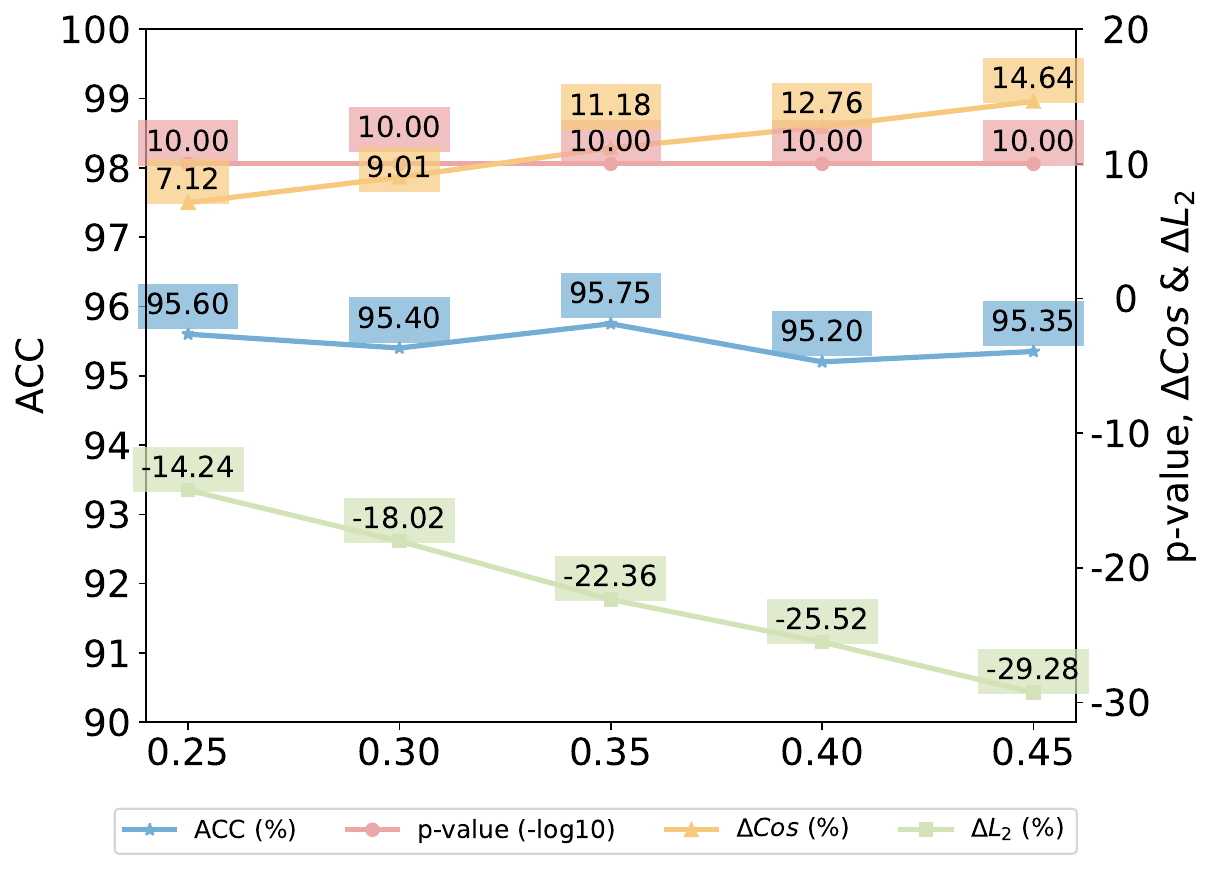}
    \label{fig: enron strength dim}
  }

\caption{The impact of the watermark strength $\delta$ in four scenarios for the Enron Spam dataset.}
\label{fig: enron strength}
\end{figure*}

We investigate the impact of watermark strength on task, similarity, and detection performance, where the watermark strength $\delta \in \{0.25, 0.30, 0.35, 0.40, 0.45\}$. Experimental results on the SST2, MIND, AGNews, and Enron Spam datasets are presented in Figures \ref{fig: sst strength}, \ref{fig: mind strength}, \ref{fig: ag strength}, \ref{fig: enron strength}, respectively. Increasing the watermark strength leads to a significant improvement in both the $\Delta$ Cos and $\Delta L_2$ metrics. For example, when the watermark strength increases from 0.25 to 0.45 in the ``No attack'' scenario of the MIND dataset, the $\Delta$ Cos metric improves by \textbf{19.54}\% (from 10.86 to 30.40), and the $\Delta L_2$ metric improves by \textbf{39.09}\% (from -21.71 to -60.80). This phenomenon demonstrates that watermark strength is positively correlated with detection performance.  However, excessive watermark strength significantly affects the similarity with the original embedding. For example, when the watermark strength increases from 0.25 to 0.45 in the MIND dataset, the similarity between the watermarked embedding and the original embedding decreases by 14.47\% (from $96.73$ to $82.26$). Fortunately, our proposed adaptive weight mechanism substantially mitigates this phenomenon by assigning greater weight to outlier data, thereby reducing its impact on the overall original embedding.

\begin{figure*}[ht]
\centering
  \subfigure[No attack] {
    \includegraphics[width=0.23\linewidth]{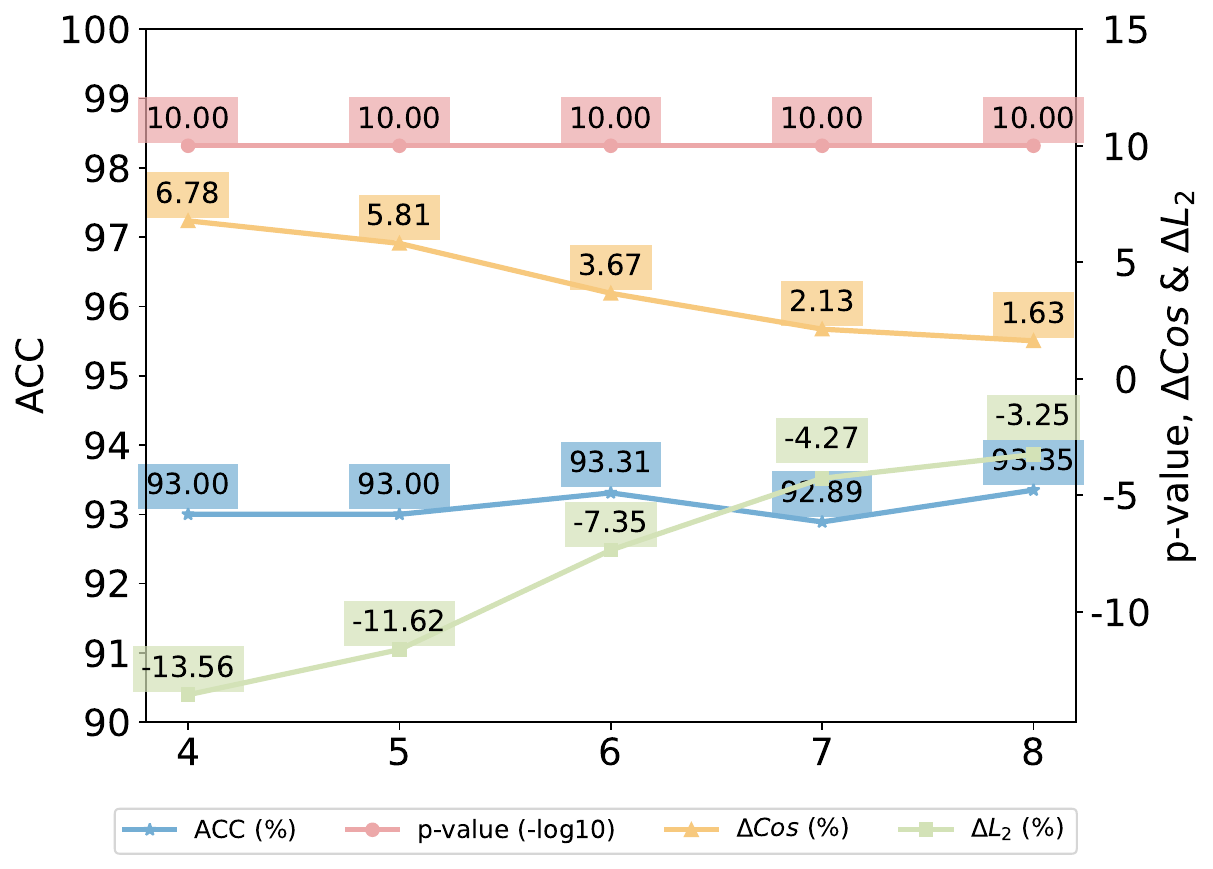}
    \label{fig: sst lsh noattack}
  }
  \subfigure[CSE] {
    \includegraphics[width=0.23\linewidth]{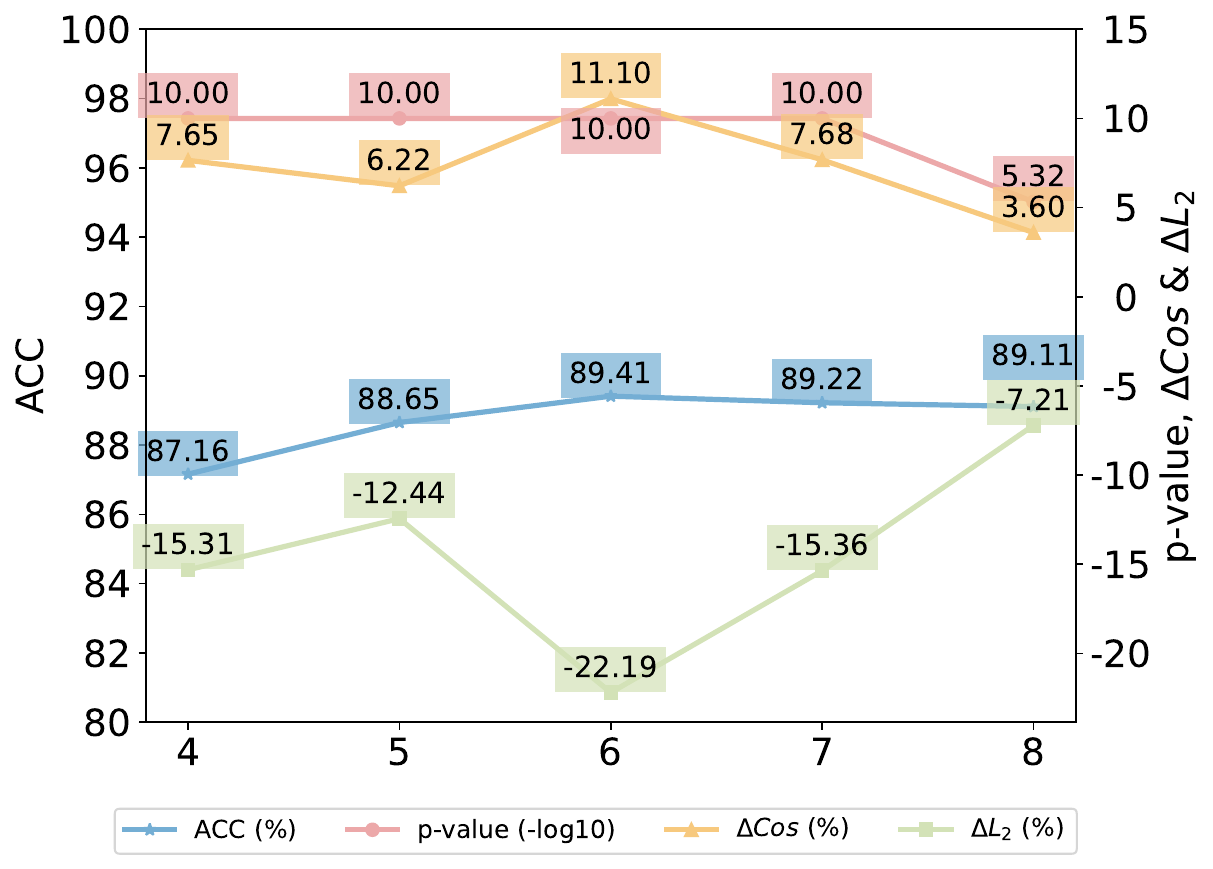}
    \label{fig: sst lsh cse}
  }
  \subfigure[Detect-Sampling] {
    \includegraphics[width=0.23\linewidth]{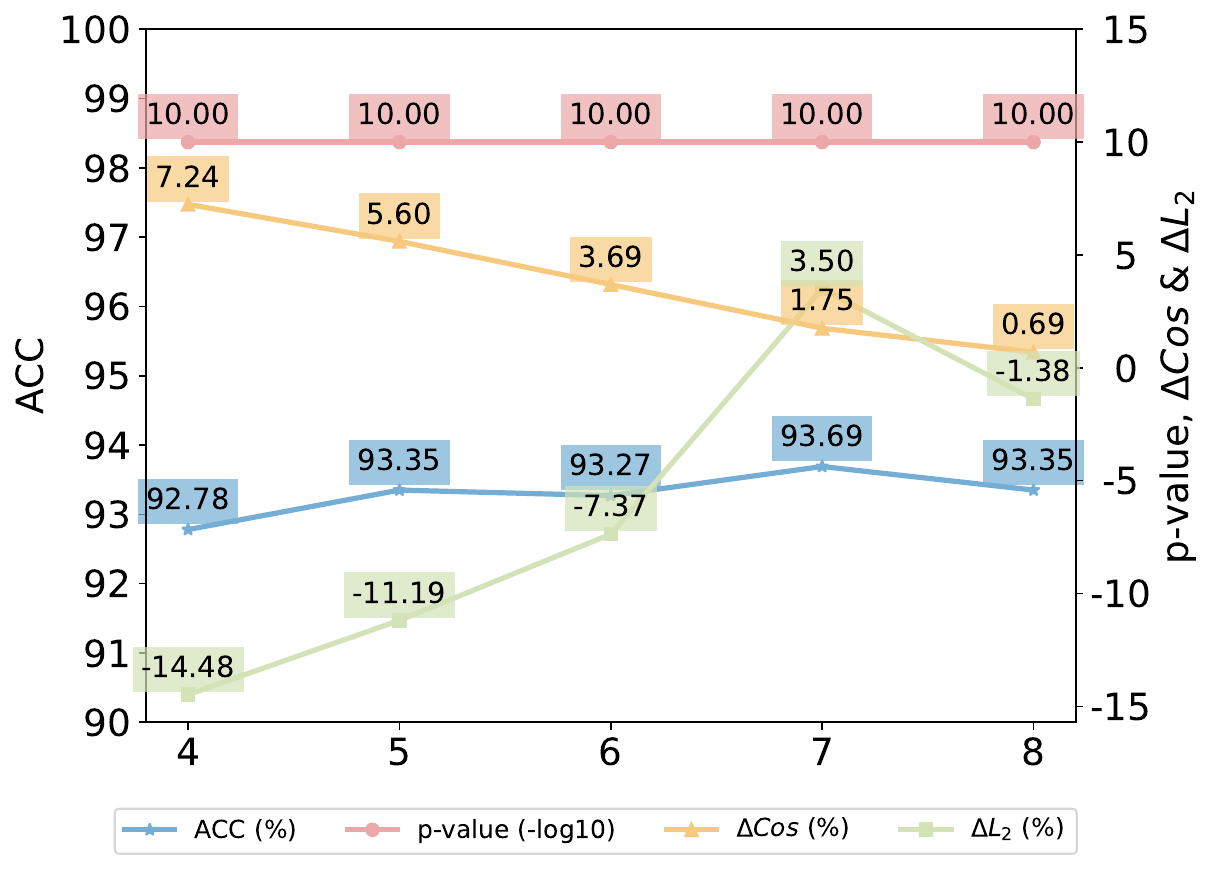}
    \label{fig: sst lsh sample}
  }
  \subfigure[Dimensionality-Reduction] {
    \includegraphics[width=0.23\linewidth]{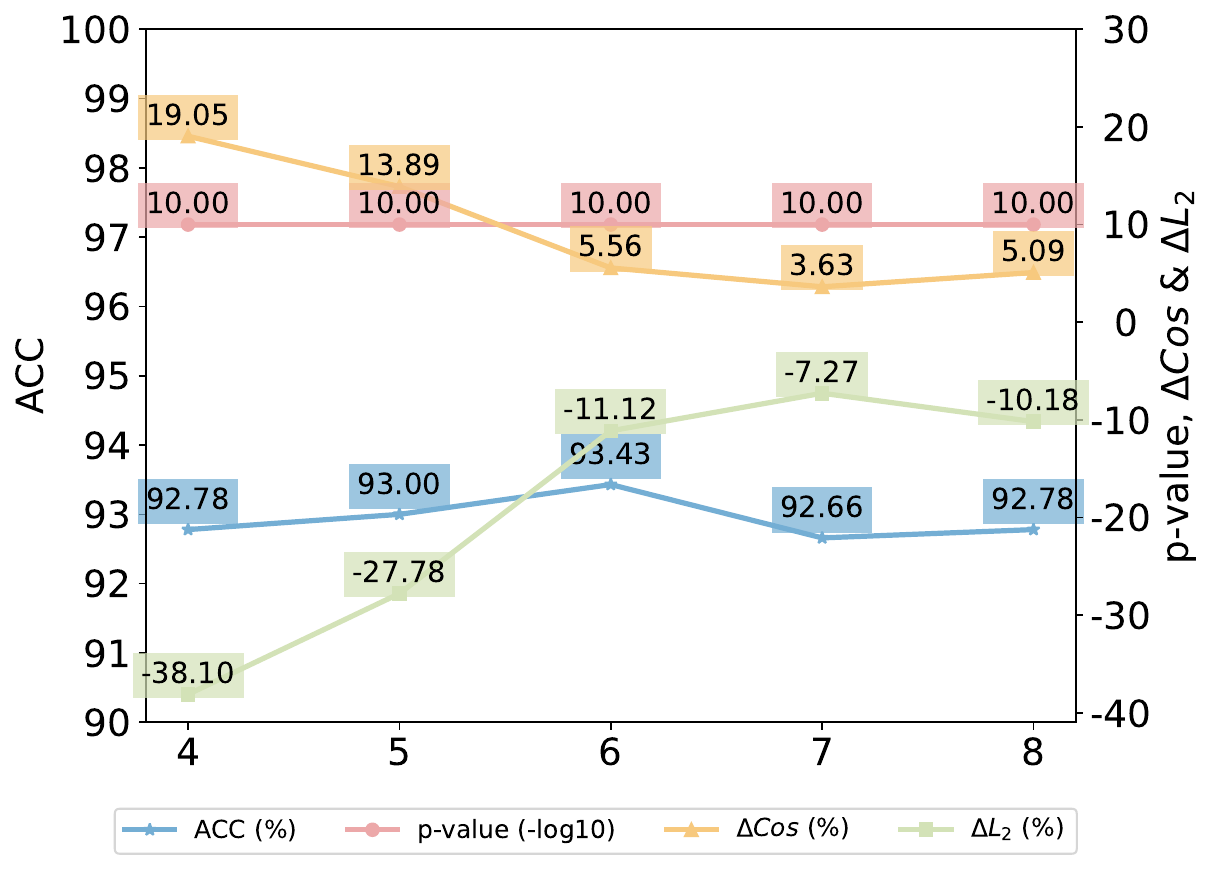}
    \label{fig: sst lsh dim}
  }
\caption{The impact of the number $c$ of random normal vectors in LSH across four scenarios on the SST2 dataset.}
\label{fig: sst lsh}
\end{figure*}

\begin{figure*}[t]
\centering
  \subfigure[No attack] {
    \includegraphics[width=0.23\linewidth]{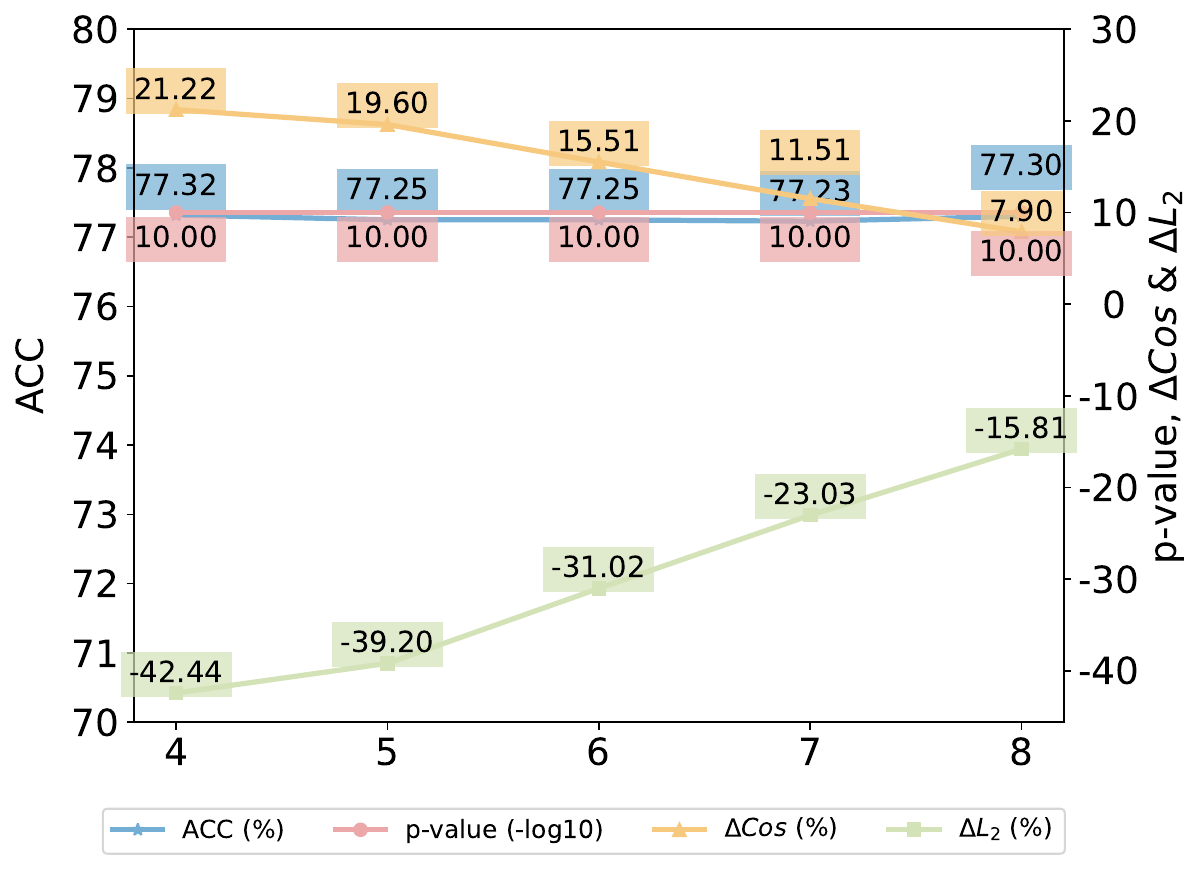}
    \label{fig: mind lsh noattack}
  }
  \subfigure[CSE] {
    \includegraphics[width=0.23\linewidth]{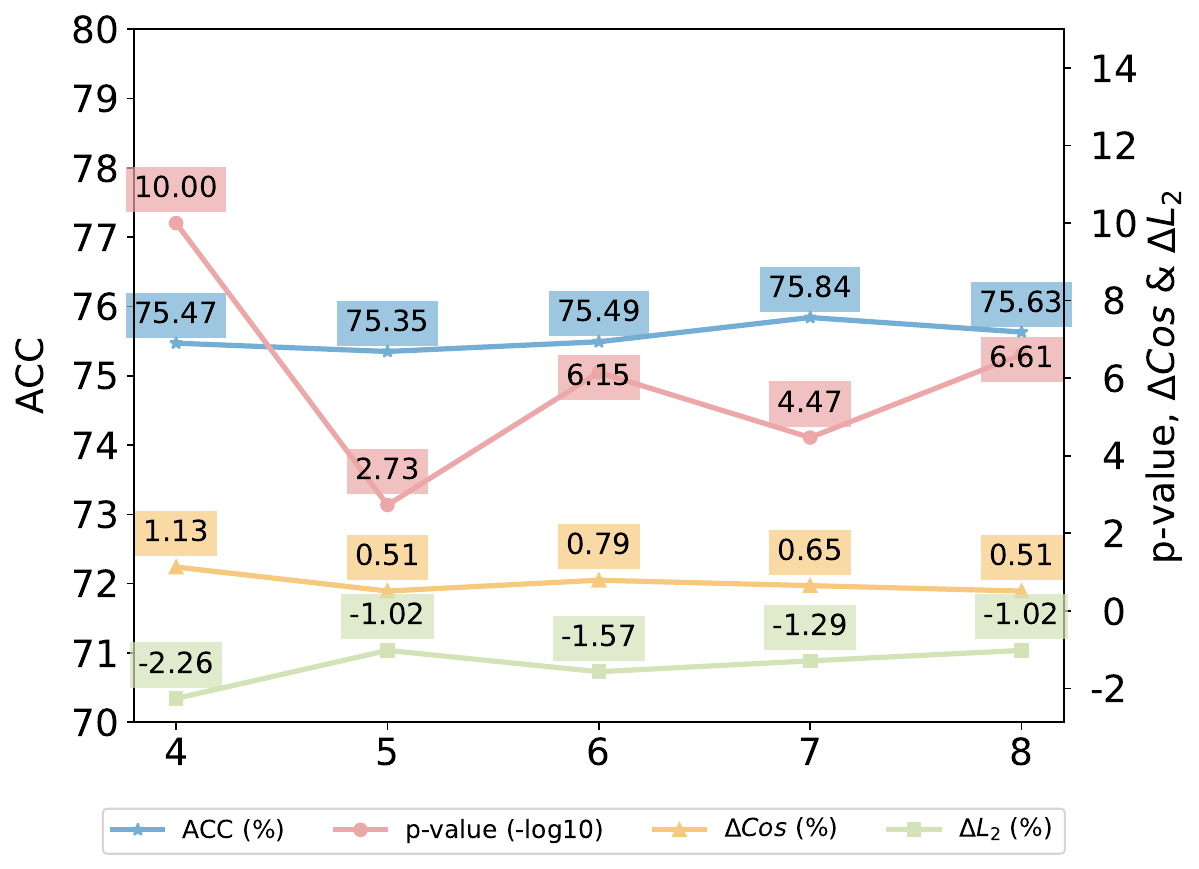}
    \label{fig: mind lsh cse}
  }
  \subfigure[Detect-Sampling] {
    \includegraphics[width=0.23\linewidth]{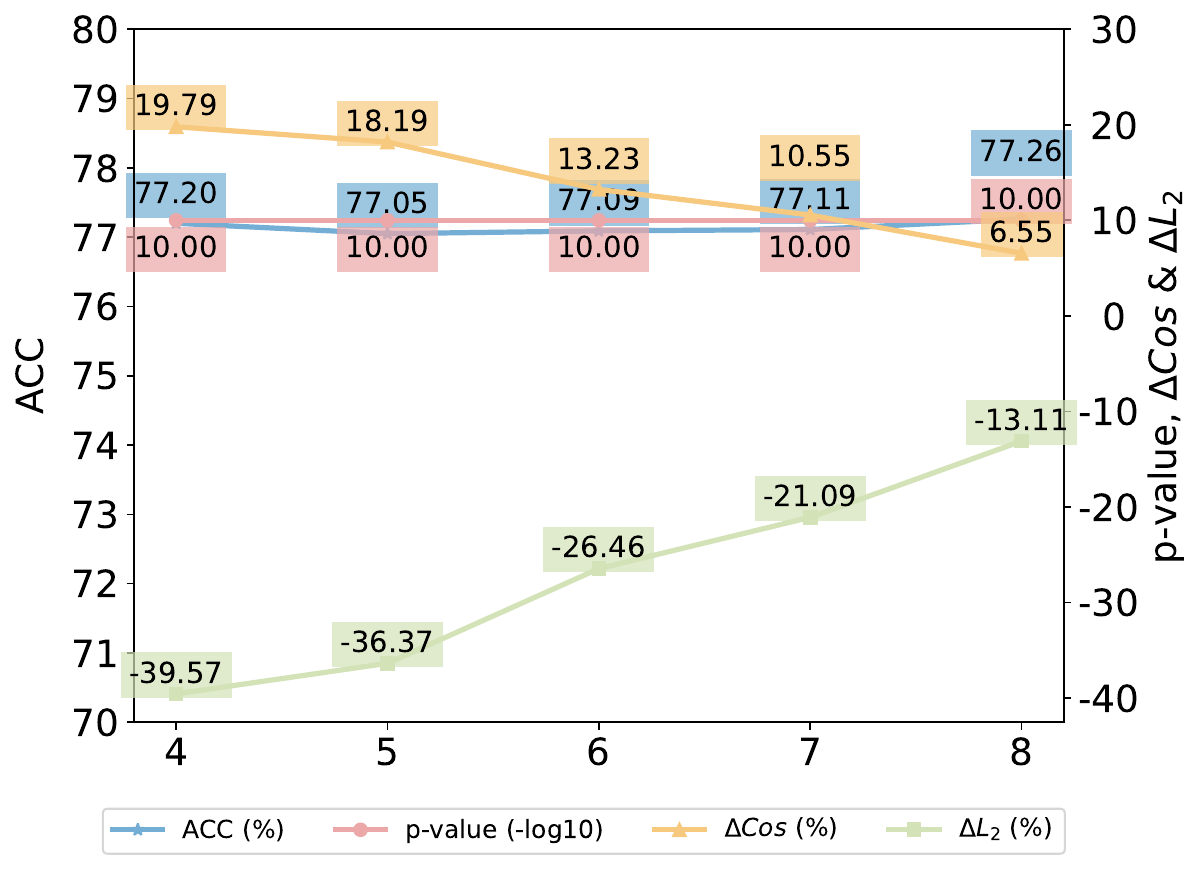}
    \label{fig: mind lsh sample}
  }
  \subfigure[Dimensionality-Reduction] {
    \includegraphics[width=0.23\linewidth]{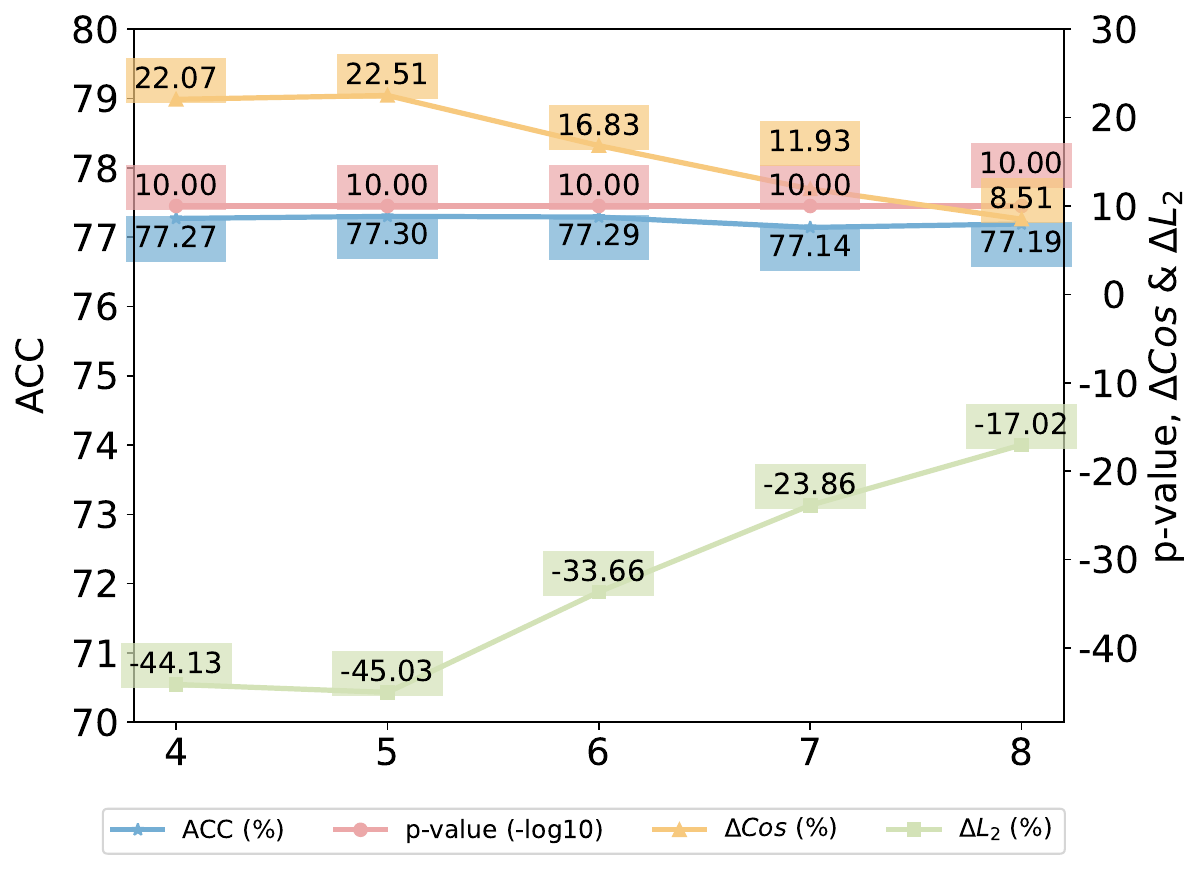}
    \label{fig: mind lsh dim}
  }
\caption{The impact of the number $c$ of random normal vectors in LSH across four scenarios on the MIND dataset.}
\label{fig: mind lsh}
\end{figure*}

\begin{figure*}[t]
\centering
  \subfigure[No attack] {
    \includegraphics[width=0.23\linewidth]{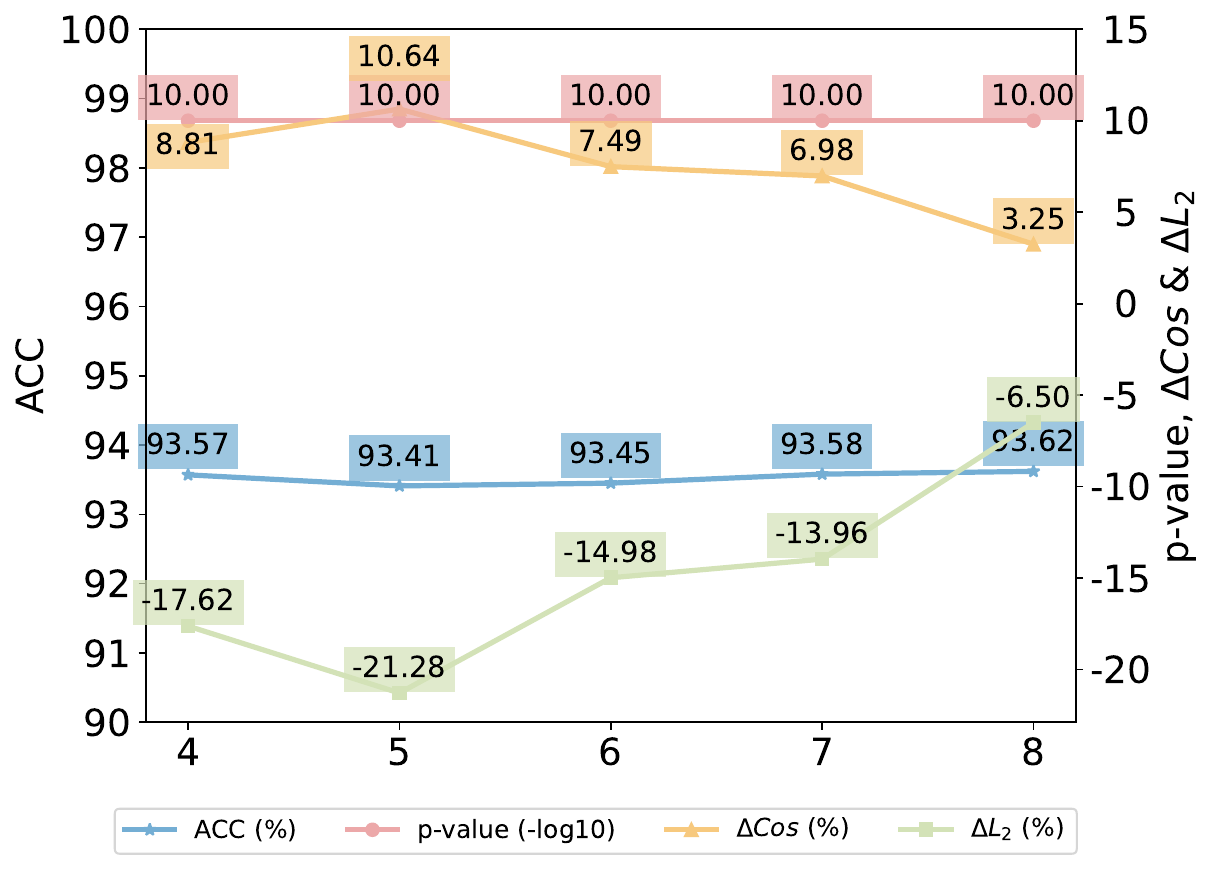}
    \label{fig: ag lsh noattack}
  }
  \subfigure[CSE] {
    \includegraphics[width=0.23\linewidth]{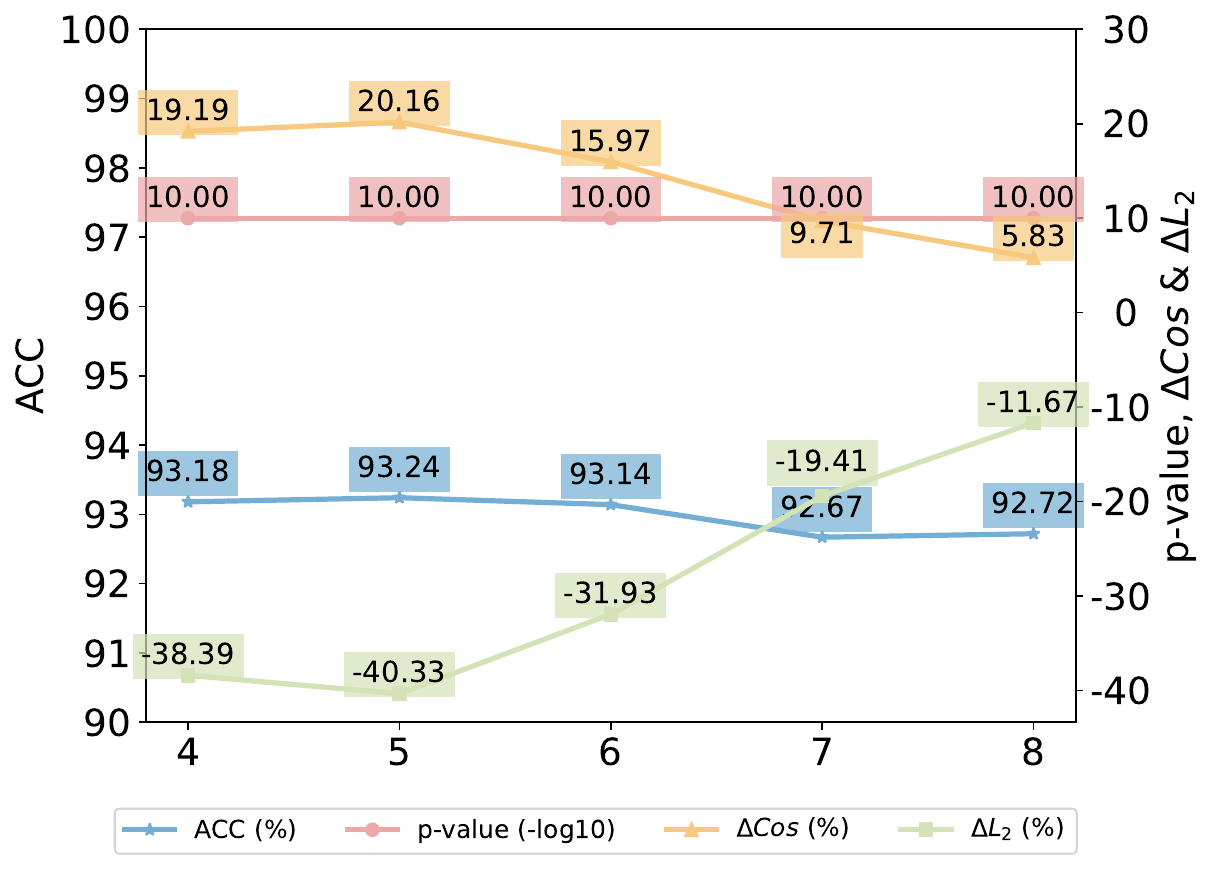}
    \label{fig: ag lsh cse}
  }
  \subfigure[Detect-Sampling] {
    \includegraphics[width=0.23\linewidth]{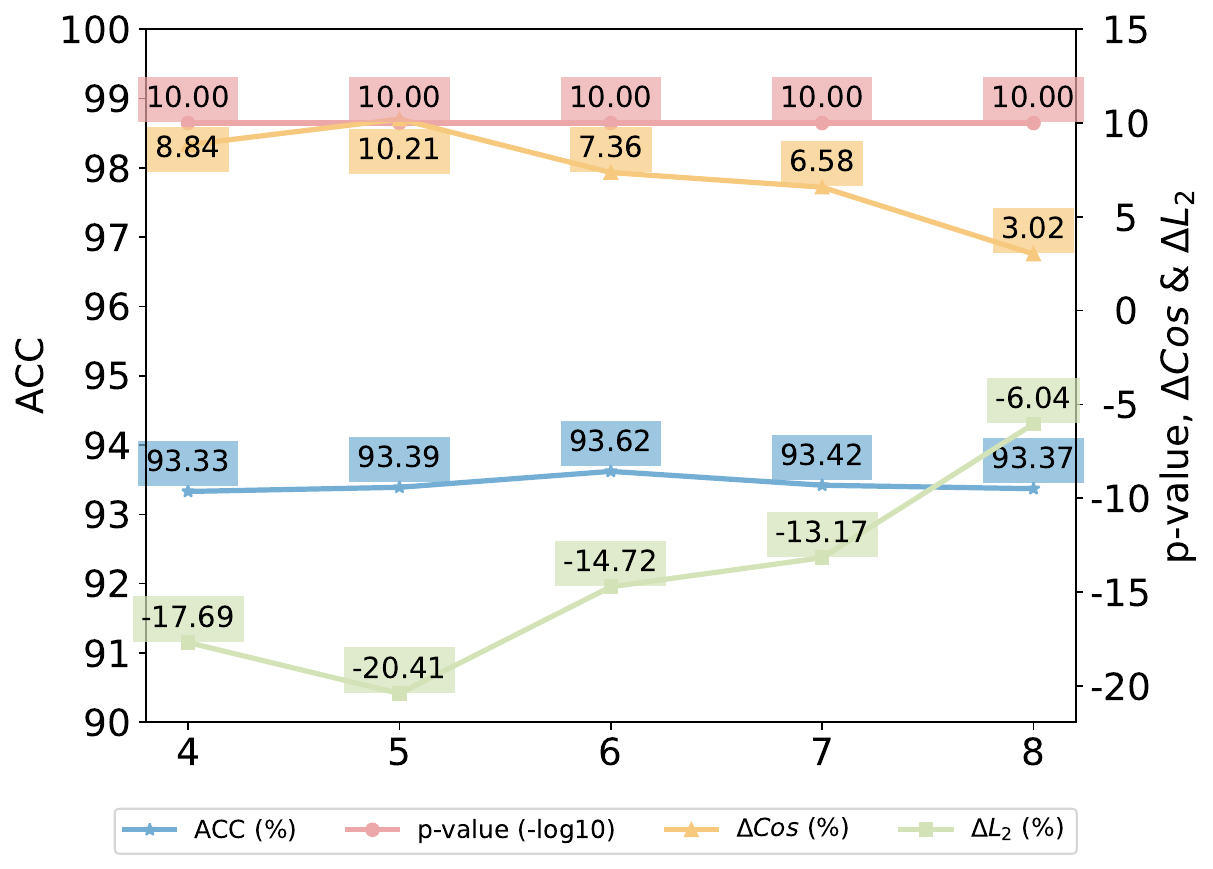}
    \label{fig: ag lsh sample}
  }
  \subfigure[Dimensionality-Reduction] {
    \includegraphics[width=0.23\linewidth]{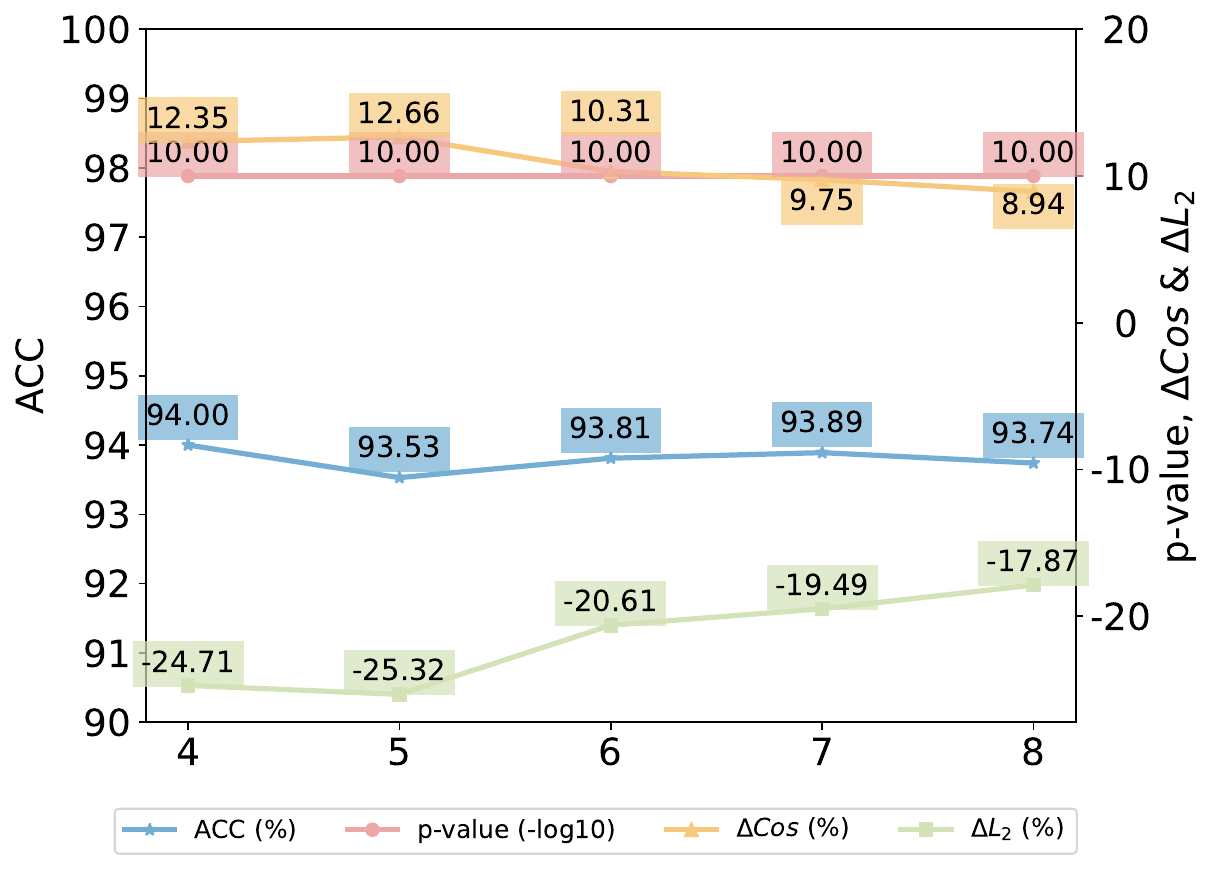}
    \label{fig: ag lsh dim}
  }
\caption{The impact of the number $c$ of random normal vectors in LSH across four scenarios on the AGNews dataset.}
\label{fig: ag lsh}
\end{figure*}

\begin{figure*}[t]
\centering
  \subfigure[No attack] {
    \includegraphics[width=0.23\linewidth]{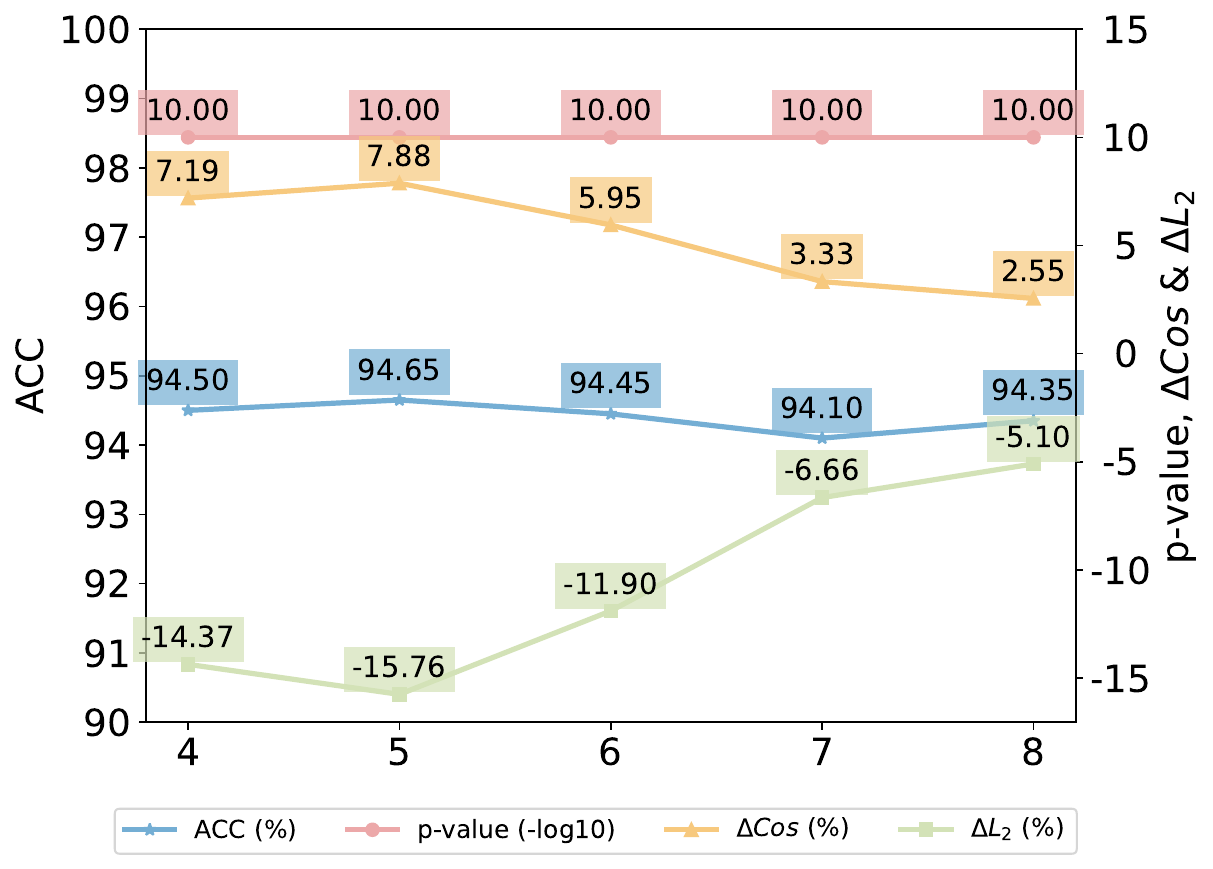}
    \label{fig: enron lsh noattack}
  }
  \subfigure[CSE] {
    \includegraphics[width=0.23\linewidth]{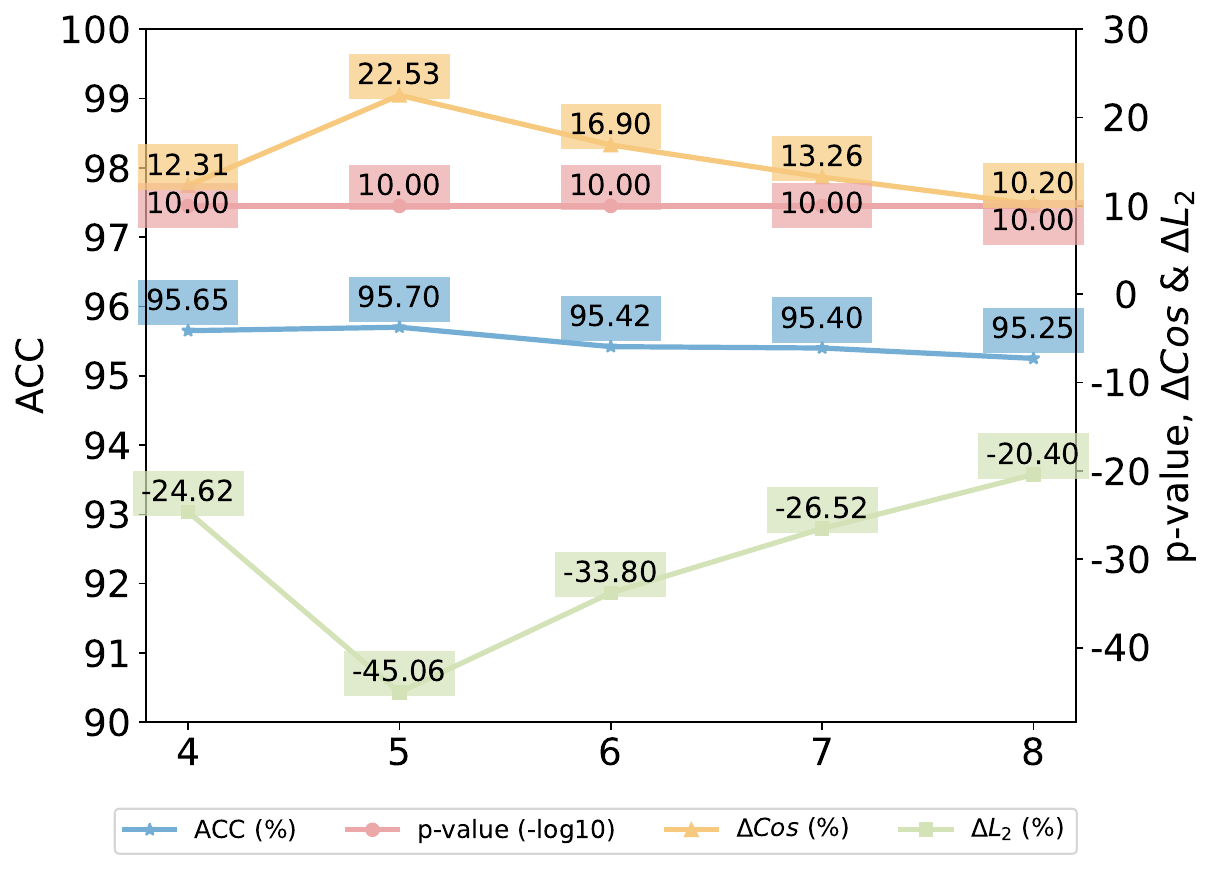}
    \label{fig: enron lsh cse}
  }
  \subfigure[Detect-Sampling] {
    \includegraphics[width=0.23\linewidth]{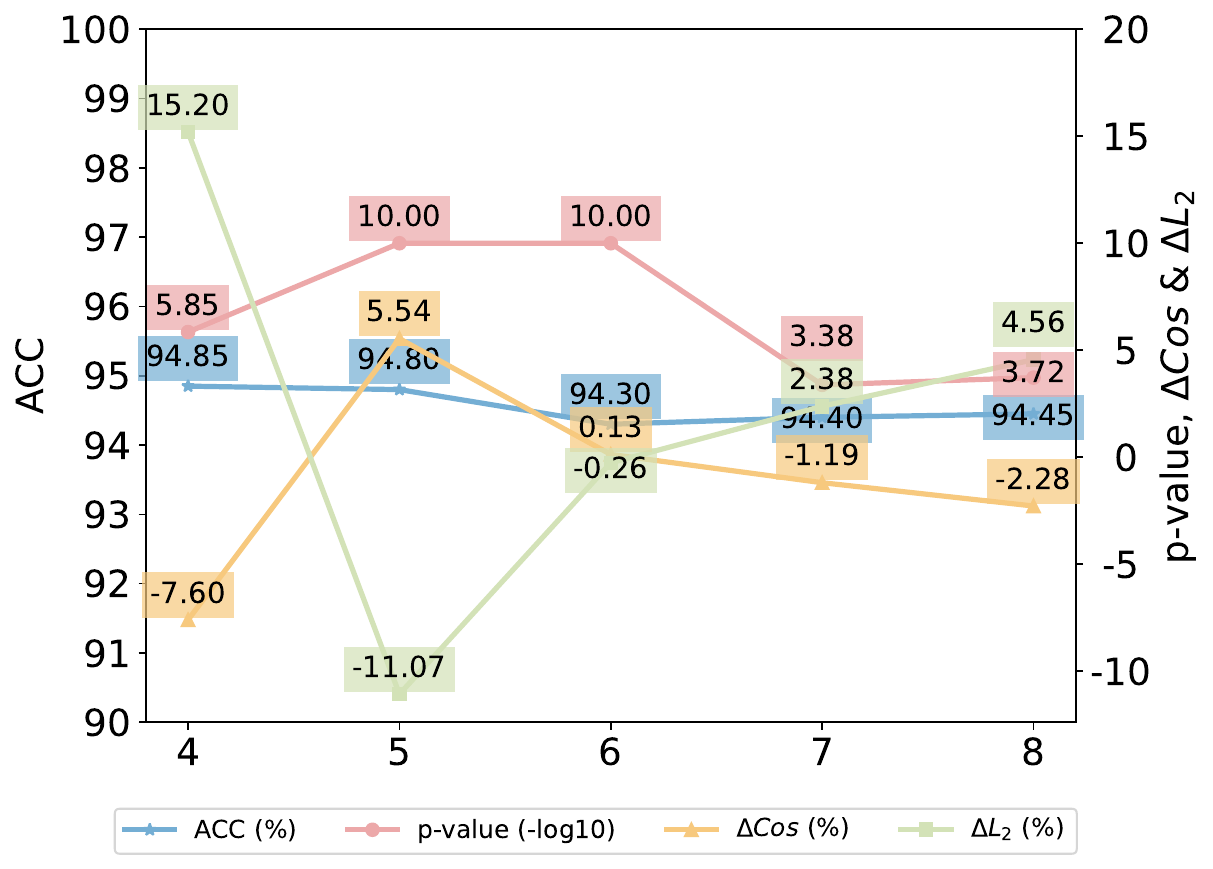}
    \label{fig: enron lsh sample}
  }
  \subfigure[Dimensionality-Reduction] {
    \includegraphics[width=0.23\linewidth]{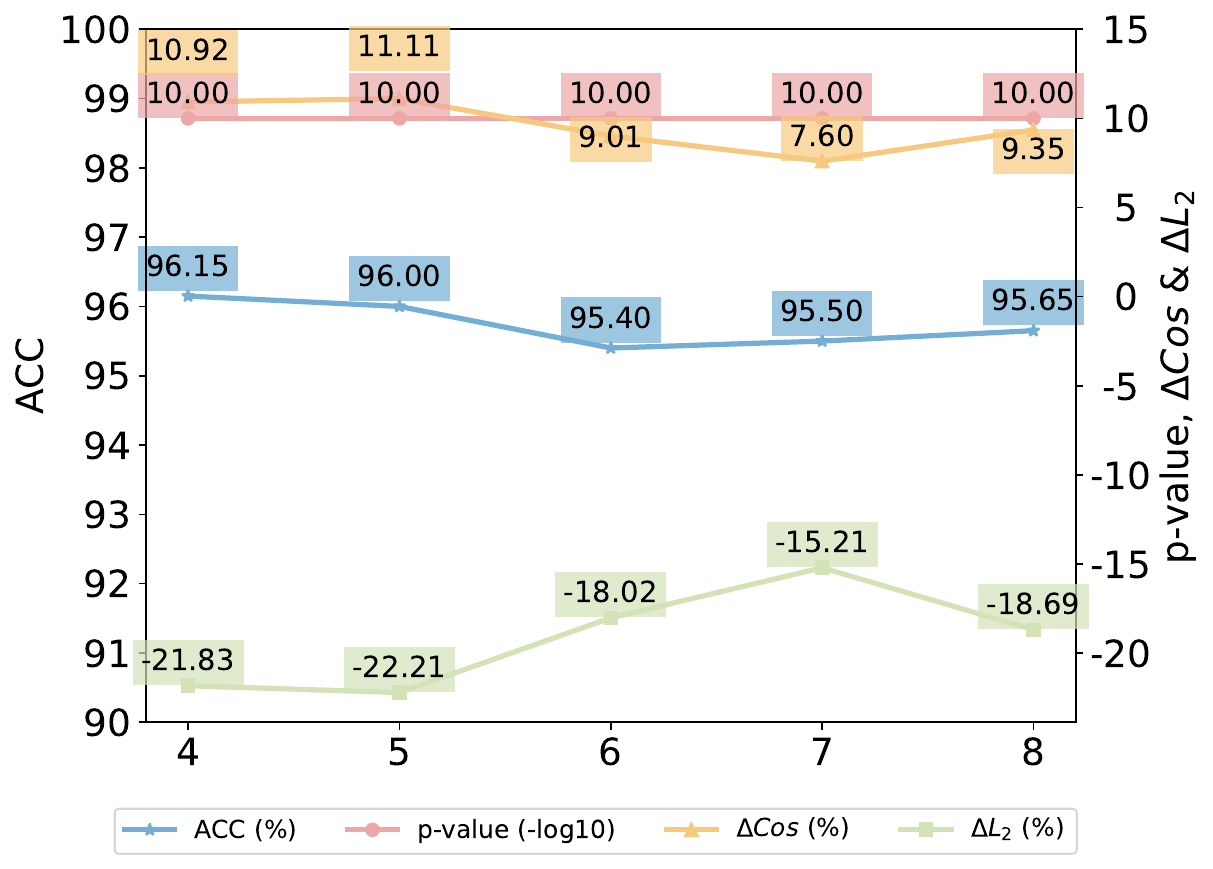}
    \label{fig: enron lsh dim}
  }
\caption{The impact of the number $c$ of random normal vectors in LSH across four scenarios on the Enron Spam dataset.}
\label{fig: enron lsh}
\end{figure*}

\section{Verification Data Analysis}
\label{ap: verif data}
Table \ref{tab: all data num result} explores the impact of verification data size $m$ on detection performance, where $m \in \{100, 200, 300, 400, 500\}$. As shown in Table \ref{tab: all data num result}, our method achieves excellent detection performance across all settings, demonstrating that the semantic-based paradigm can efficiently verify watermarks with only a small amount of data (e.g., $m=100$). In contrast, trigger-based methods concatenate triggers to verify copyright in the verification stage, sacrificing their stealthiness. Linear transformation methods apply linear transformations on all embeddings, sacrificing their harmlessness. Unlike these methods, our approach partitions the embedding space into watermarked and non-watermarked regions based on locality-sensitive hashing and principal component analysis. During watermark verification, only the normal text across different regions needs to be compared. This design restricts watermark injection to the watermarked region, eliminating the need for constructing complex verification text. Furthermore, locality-sensitive hashing and principal component analysis make it difficult for attackers to reverse-engineer our algorithm, thereby further enhancing both the stealthiness and harmlessness.

\section{Similarity Performance Analysis for SemMark}
\begin{figure*}[ht]
\centering
  \subfigure[SST2] {
    \includegraphics[width=0.23\linewidth]{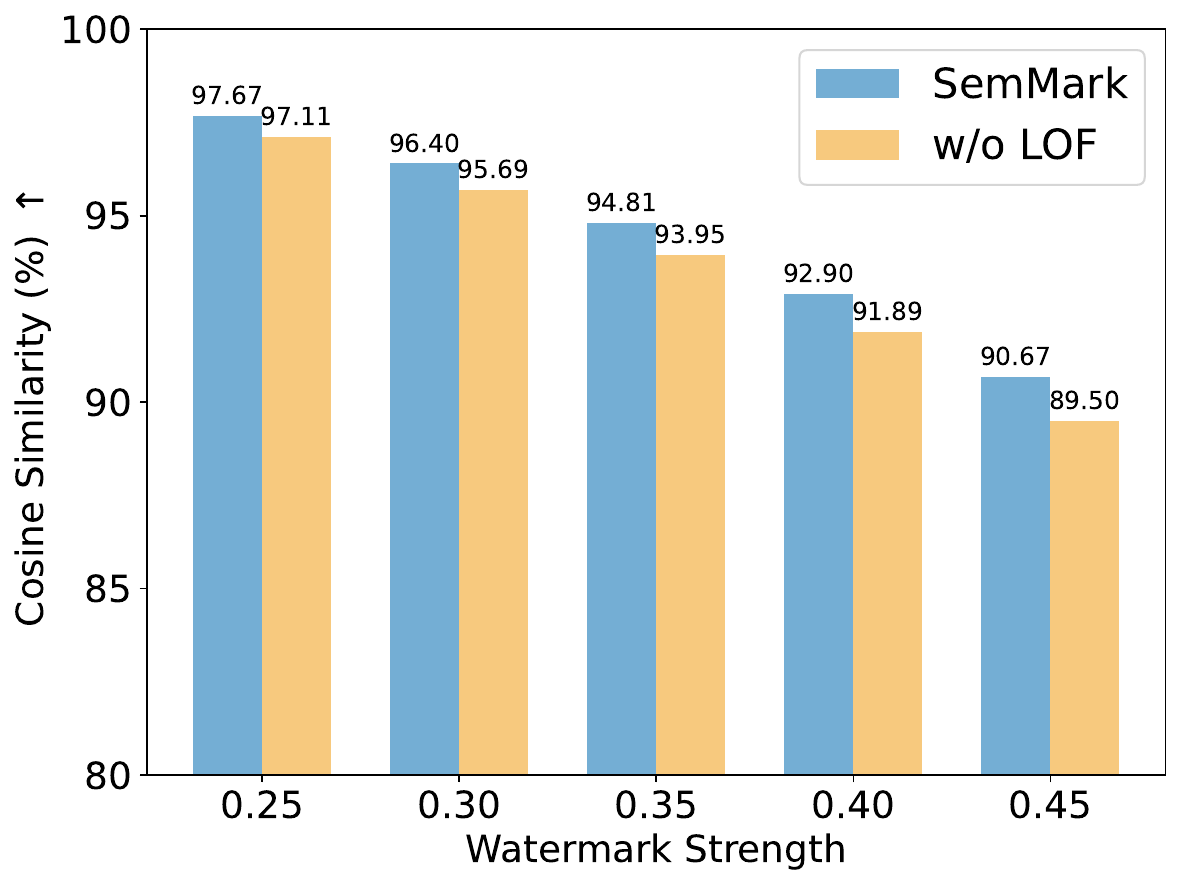}
  }
  \subfigure[MIND] {
    \includegraphics[width=0.23\linewidth]{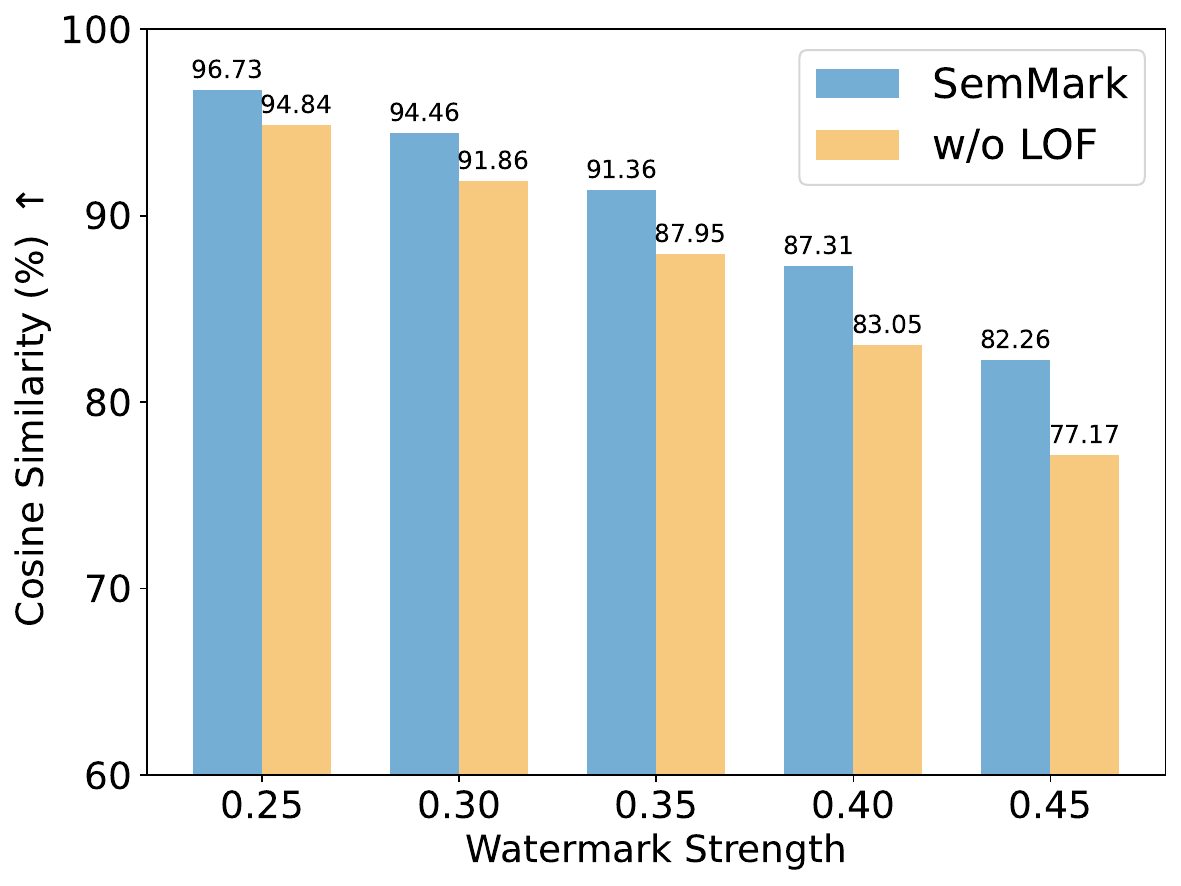}
  }
  \subfigure[AGNews] {
    \includegraphics[width=0.23\linewidth]{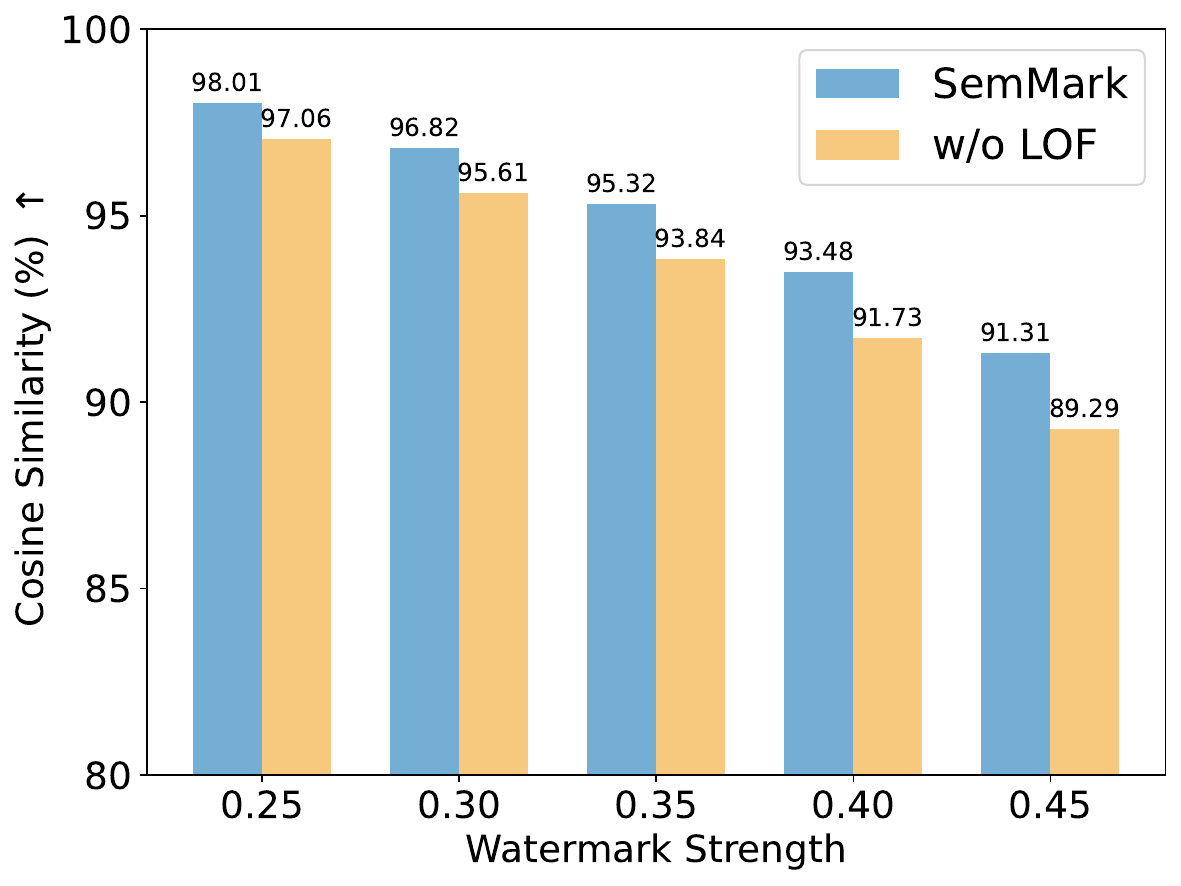}
  }
  \subfigure[Enron Spam] {
    \includegraphics[width=0.23\linewidth]{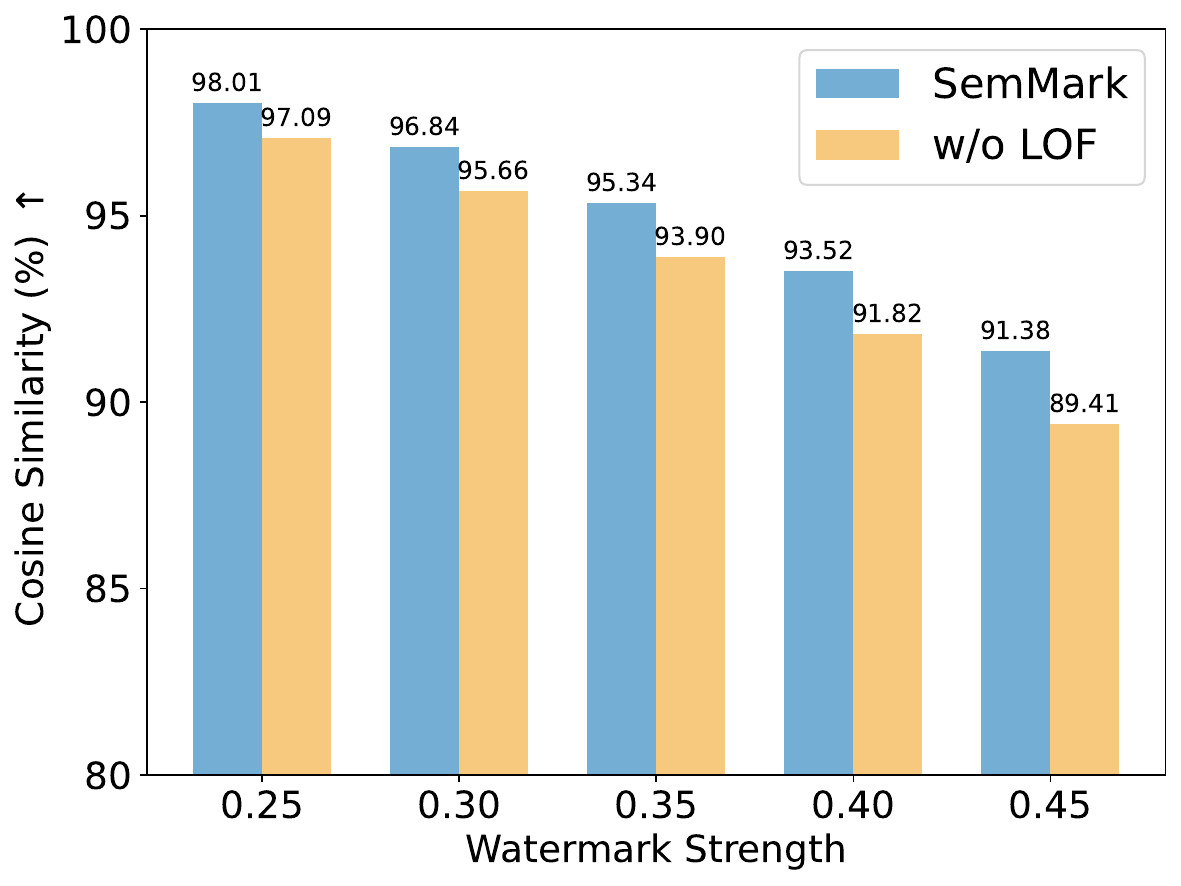}
  }
\caption{Similarity performance analysis of SemMark on the cosine similarity metrics.}
\label{fig: cos result}
\end{figure*}

\begin{figure*}[ht]
\centering
  \subfigure[SST2] {
    \includegraphics[width=0.23\linewidth]{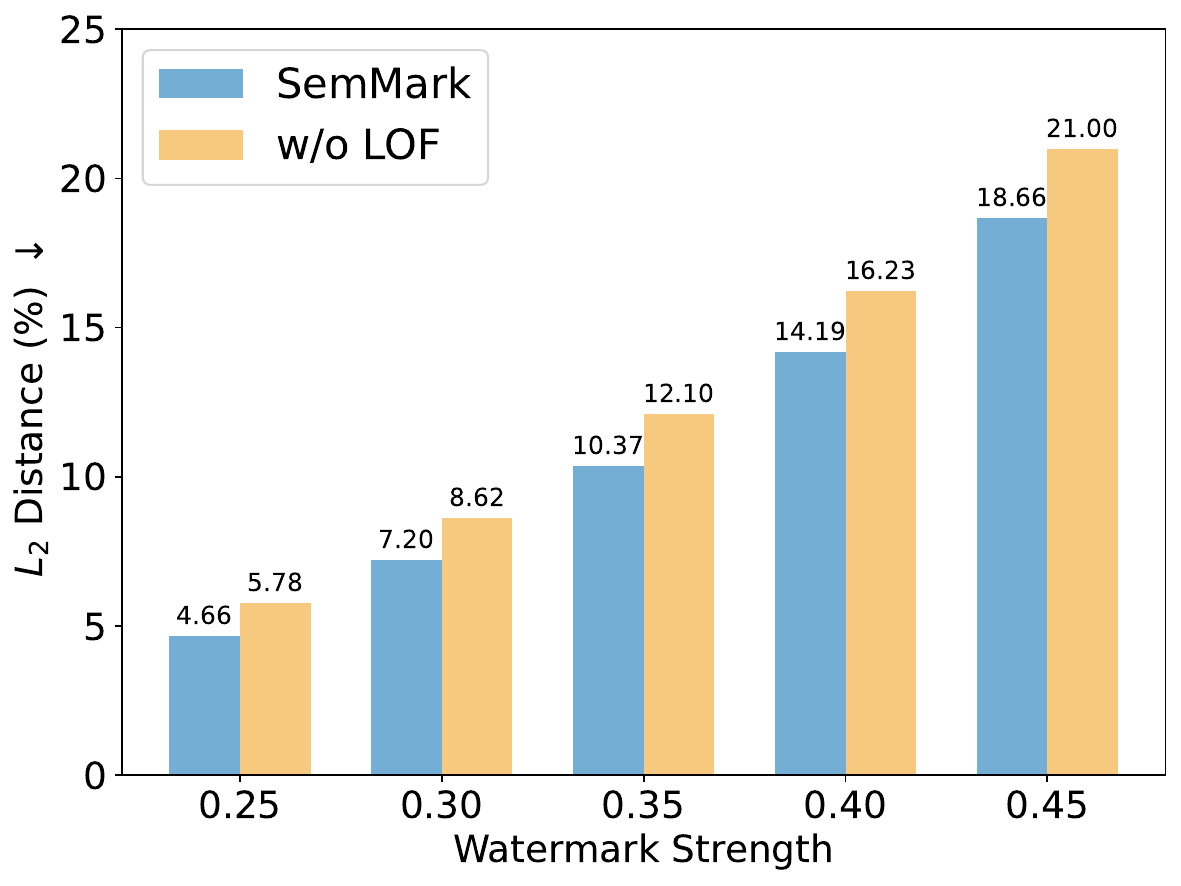}
  }
  \subfigure[MIND] {
    \includegraphics[width=0.23\linewidth]{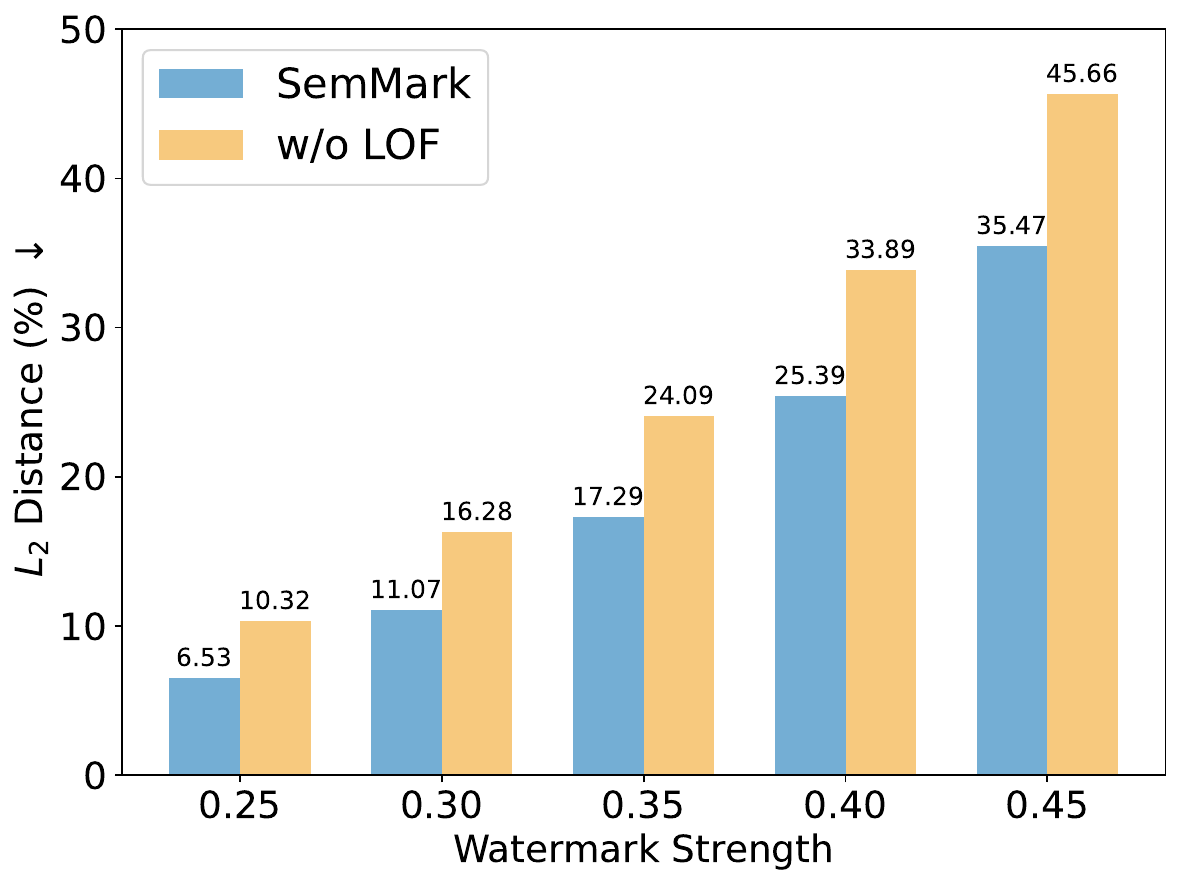}
  }
  \subfigure[AGNews] {
    \includegraphics[width=0.23\linewidth]{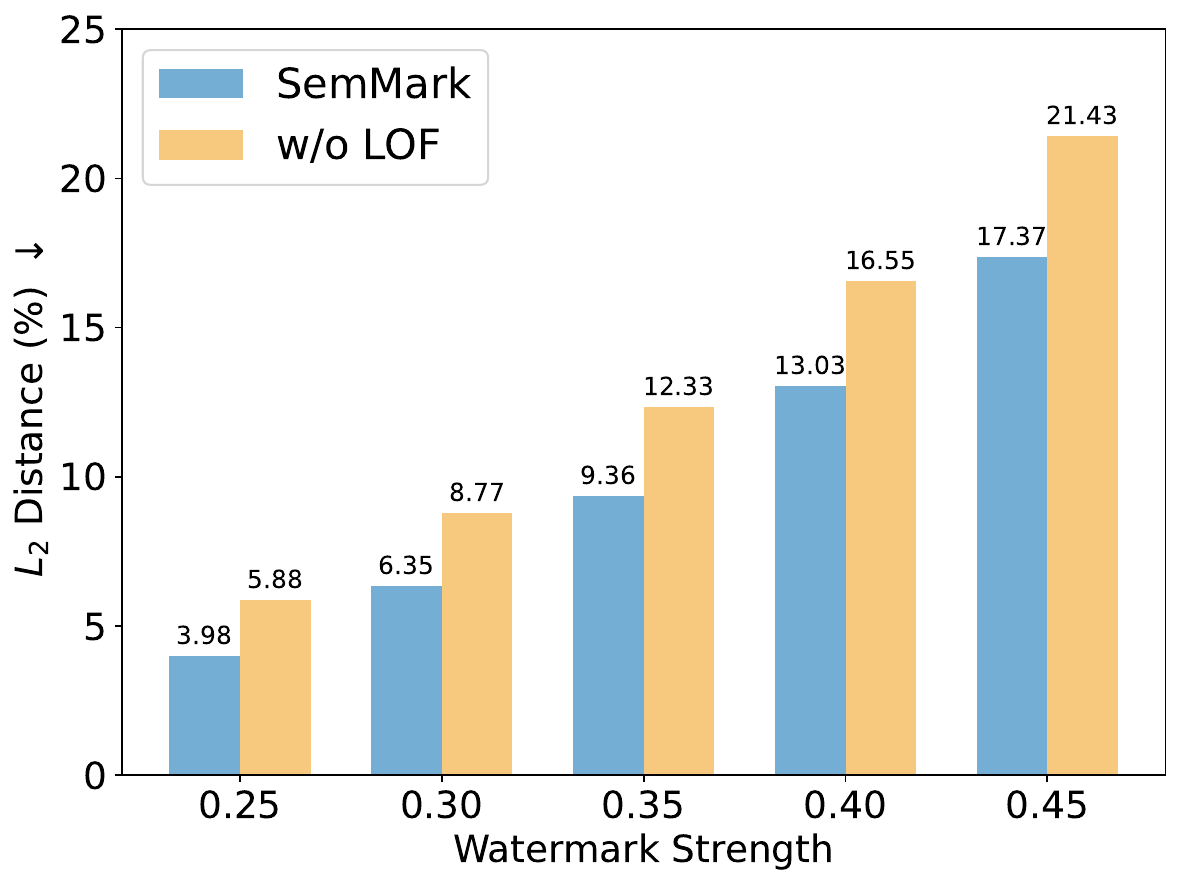}
  }
  \subfigure[Enron Spam] {
    \includegraphics[width=0.23\linewidth]{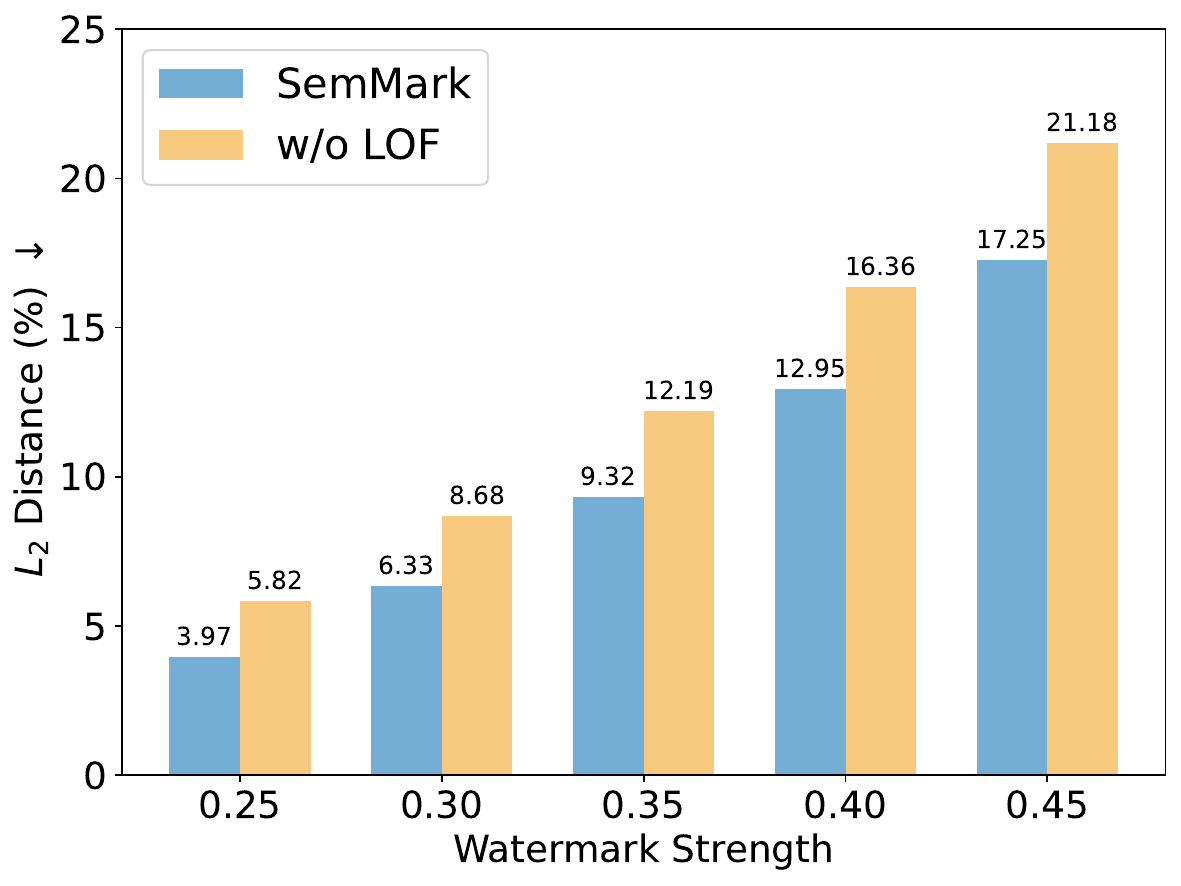}
  }
\caption{Similarity performance analysis of SemMark on the square of $L_2$ distance metrics.}
\label{fig: L2 result}
\end{figure*}

In Figures \ref{fig: cos result} and \ref{fig: L2 result}, we explore the similarity performance of SemMark under different watermark strengths. Specifically, Figures \ref{fig: cos result} and \ref{fig: L2 result} use cosine similarity and the square of $L_2$ distance metrics to evaluate the closeness of the watermarked embedding to the original embedding, where ``w/o LOF'' represents the use of a fixed hyperparameter watermark weight instead of adaptive watermark weights. 
As shown in Figure \ref{fig: cos result}, our method significantly improves cosine similarity, with an average improvement of \textbf{1.81\%}, and a maximum improvement of \textbf{5.09\%} on the more complex multi-classification MIND dataset. In Figure \ref{fig: L2 result}, our method reduces the square of $L_2$ distance by \textbf{3.62\%} on average, with a maximum reduction of \textbf{10.19\%} on the MIND dataset.
This demonstrates that the adaptive weighting mechanism can significantly alleviate the impact on the overall original embedding by assigning larger weights to outliers, thereby significantly improving the harmlessness. 

\end{document}